\documentclass[aip,jcp,reprint,superscriptaddress,footinbib,floatfix]{revtex4-2}

\usepackage{graphicx}
\usepackage{bm}
\usepackage{amssymb}
\usepackage{amsbsy}
\usepackage{amsmath}
\usepackage{mathtools}
\usepackage{xcolor}
\usepackage{stmaryrd}
\usepackage{mathrsfs}
\usepackage{tikz}
\usetikzlibrary{shapes}
\usepackage{cancel}
\usepackage[normalem]{ulem}
\usepackage{hyperref}
\usepackage{multirow}

\usepackage{titlesec}
\titlespacing\section{0pt}{12pt plus 4pt minus 2pt}{8pt plus 4pt minus 2pt}
\titlespacing\subsection{0pt}{12pt plus 4pt minus 2pt}{6pt plus 2pt minus 2pt}
\titlespacing\subsubsection{0pt}{12pt plus 4pt minus 2pt}{4pt plus 2pt minus 2pt}

\newcommand{\fref}[1]{Fig.~\ref{#1}}

\newcommand{\eref}[1]{Eq.~(\ref{#1})}
\newcommand{\sref}[1]{Sec.~\ref{#1}}

\makeatletter
\let\cat@comma@active\@empty
\makeatother

\makeatletter 
\gdef\@ptsize{2}
\let\@currsize\normalsize 
\makeatother 

\begin{document}
\title{Linking molecular timescales to linear viscoelastic response in dilute and semidilute unentangled wormlike micelle solutions }
\author{Avishek Kumar}
\affiliation{IITB-Monash Research Academy, Mumbai, Maharashtra 400076, India}
\affiliation{Department of Chemical Engineering, Indian Institute of Technology Bombay, Mumbai, Maharashtra 400076, India}
\affiliation{Department of Chemical and Biological Engineering, Monash University, Melbourne, VIC 3800, Australia}
\author{Rico F. Tabor}
\affiliation{School of Chemistry, Monash University, Melbourne, VIC 3800, Australia}
\author{P.Sunthar}
\affiliation{Department of Chemical Engineering, Indian Institute of Technology Bombay, Mumbai, Maharashtra 400076, India}
\author{J. Ravi Prakash}
\email{ravi.jagadeeshan@monash.edu}
\affiliation{Department of Chemical and Biological Engineering, Monash University, Melbourne, VIC 3800, Australia}
\homepage{https://ravijagadeeshan.ac}

\begin{abstract}

Unentangled wormlike micelle solutions relax stress through a dynamic interplay of reversible scission and intrachain relaxation involving a hierarchy of molecular timescales whose relationship to linear viscoelastic response remains incompletely resolved. A multiparticle mesoscopic Brownian dynamics framework has been developed in which “persistent worms”, represented by bead-spring chains with sticky ends, assemble to form wormlike micelles via reversible scission and fusion. Both linear and ring-like micelles are formed across the dilute and semidilute concentration regimes. Accurate predictions of dynamic properties are obtained through inclusion of hydrodynamic interactions using a Rotne–Prager–Yamakawa tensor. We identify and quantify characteristic timescales governing micellar dynamics, including bond lifetimes, self- and non-self-recombination times, breakage times of wormlike micelles of length $L$, relaxation times of various contributions to stress, and the longest relaxation time. The dependence of these timescales on sticker strength, concentration, micellar topology and hydrodynamic interactions is systematically established. The presence of ring micelles is found to moderately prolong recombination and breakage processes, while hydrodynamic interactions are shown to affect some of the timescales by reducing sticker mobility. When appropriately scaled, the dependence on mean length of the non-self-recombination and micelle breakage times collapse onto universal, sticker-independent master curves. Linear viscoelastic storage and loss moduli, computed using the Green–Kubo relation, exhibit distinctive features in the intermediate-frequency regime that are absent in homopolymer solutions. A clear connection is made between micellar timescales and these signatures in the dynamic moduli at various characteristic frequencies, providing a direct link between microscopic dynamics and macroscopic rheology in unentangled wormlike micellar solutions, in dilute and semidilute concentration regimes.

\end{abstract}

\maketitle

\section{\label{sec:intro} Introduction}

Under certain conditions, amphiphilic molecules self-assemble into long, elongated  polymer-like structures known as wormlike micelles. Unlike polymers, these supramolecular structures continuously break and recombine, leading to a broad distribution of lengths; they are therefore often referred to as ``living" or ``equilibrium" polymers \cite{Cates2006,Israel2011}. This dynamic nature gives rise to rich and tunable rheological behaviour, making wormlike micellar solutions important in applications such as personal care products, drag reduction, enhanced oil recovery, and drug delivery \cite{Sullivan2007,Nicolas2007,Dreiss2007,Dreiss2013}. Despite their widespread use, establishing a clear connection between the microscopic dynamics of micelles and their macroscopic viscoelastic response remains a central challenge.

The pioneering work of Cates \cite{Cates1987} developed a theoretical framework to describe the viscoelastic properties of entangled wormlike micelle solutions. In this model, stress relaxation is governed by a composite timescale involving both reptation and micellar scission. In the limit where the micelle breakage time is much shorter than the reptation time, the stress relaxation reduces to a single-mode Maxwell response. This behaviour was demonstarted experimentally using the Cole–Cole representation of the storage and loss moduli, which shows the expected semi-circular relationship over a wide range of frequencies for many surfactant systems \cite{Rehage1991,Shibaev2015}. These ideas have been highly successful in rationalizing the rheology of concentrated and entangled wormlike micellar systems \cite{Oelschlaeger2003}. 

A key assumption underlying this framework is that recombination is a mean-field process, in which newly created micellar ends lose memory of their original partners and recombine with other chain ends \cite{Cates1988}. Under this assumption, recombination kinetics are governed solely by the average density of chain ends, leading to exponential recombination-time statistics and a single characteristic relaxation time. O’Shaughnessy and Yu \cite{Shaughnessy1995} subsequently introduced a dimensionless parameter $X = t_h/t^*$ to characterize the competition between self-recombination and recombination with other chains, where $t^*$ is the characteristic self-recombination time of two ends created by the same scission event and $t_h$ is the diffusion time required for a chain end to encounter the nearest end belonging to a different chain. This framework identifies two universality classes of recombination dynamics: for $X \ll 1$, recombination with other chains occurs before self-recombination and mean-field (MF) kinetics are recovered, whereas for $X \gg 1$, recombination at short times is dominated by self-recombination. The latter regime is referred to as diffusion-controlled (DC) recombination and is characterized by algebraic recombination-time distributions at short times, followed by exponential decay at longer times. These distinct recombination kinetics give rise to anomalous stress relaxation at short times and deviations from the semicircular Cole–Cole response at short times, even though Maxwell-like behaviour is recovered at long times \cite{Shaughnessy1995}.

Mesoscopic simulations have provided direct access to the microscopic kinetics underlying recombination in wormlike micelles. Coarse-grained Brownian-dynamics simulations of unentangled equilibrium polymers by ~\citet{Huang2006}, and FENE-C–based simulations of wormlike micelles by Padding and Boek \cite{Padding2004,Padding2008}, explicitly incorporated reversible scission–recombination dynamics and enabled direct measurement of first-recombination statistics, effective rate constants, and mean micellar lifetimes. These studies confirmed the two recombination regimes predicted by theory, with power-law self-recombination behaviour at short times followed by exponential decay associated with mean-field non-self recombination at longer times. At long times, the mean micellar lifetime was found to scale inversely with the average micellar length, consistent with Cates’ mean-field kinetic model \cite{Cates1988,Huang2006}. These results establish the presence of two distinct kinetic timescales in unentangled wormlike micelles: a short-time self-recombination timescale and a longer non-self recombination timescale, whose interplay governs the recombination kinetics and the resulting stress-relaxation behaviour.

More recently, Koide \cite{Koide2023} employed dissipative particle dynamics simulations of nonionic surfactant solutions to investigate unentangled micellar scission and recombination dynamics. In contrast to earlier models, micelles are not constrained to linear chains and can undergo branching and loop formation. Their analysis showed that self-recombination exhibits a power-law probability density, with an exponent determined by the subdiffusive mean-squared displacement of surfactant molecules, whereas non-self recombination displays an exponential survival function consistent with mean-field kinetics. They further identified scaling laws for the non-self recombination time as functions of aggregation number, temperature, and surfactant volume fraction, highlighting the role of micellar morphology and size in governing recombination kinetics.

In addition to these kinetic timescales, simulations have shown that the long-time stress relaxation is governed by a single dominant terminal relaxation time. Padding and Boek \cite{Padding2008} derived an expression for the terminal relaxation time of breakable Rouse chains and showed that the long-time decay of the simulated stress-relaxation modulus of unentangled wormlike micelle solutions is well described by a single exponential with this relaxation time. More recently, Koide and Goto \cite{Koide2022} identified a longest micellar relaxation time by combining the rotational relaxation time and the micellar lifetime using Kaplan–Meier survival-analysis methods \cite{Kaplan1958}. This timescale governs the crossover in surfactant mean-squared displacements and becomes independent of aggregation number for sufficiently large micelles, establishing the existence of a single dominant terminal relaxation time despite micellar polydispersity.

Taken together, existing theoretical \cite{Cates1987,Cates1988,Shaughnessy1995} and mesoscopic simulation studies \cite{Huang2006,Padding2004,Padding2008,Koide2022,Koide2023} have provided important insights into recombination kinetics and stress relaxation in wormlike micelle solutions; however, several fundamental issues remain unresolved for unentangled systems. In particular, while different microscopic timescales—such as bond lifetimes \cite{Cates1987,Cates1988,Padding2004}, self- and non-self recombination times \cite{Shaughnessy1995,Padding2004,Huang2006,Padding2008,Koide2023}, length-dependent breakage times \cite{Huang2006,Koide2022}, and the terminal relaxation time \cite{Padding2004,Padding2008}—have been introduced or examined separately in prior work, their interrelations and collective roles within a single, consistent mesoscopic framework has not been systematically established. Moreover, although ring micelles have been reported in experiments \cite{Martin1999,Cates2001,Oelschlaeger2002,Zhu2004} on wormlike micellar solutions, prior theoretical and mesoscopic modeling approaches have typically restricted micellar topology to linear chains. In addition, they have often neglected hydrodynamic interactions, leaving the combined influence of topology (specifically linear versus ring micelles) and hydrodynamic coupling on recombination kinetics and stress relaxation largely unexplored. Finally, although the distinction between mean-field and diffusion-controlled recombination kinetics has been well characterized at the level of recombination time statistics, it remains unclear how this kinetic dichotomy manifests directly in the linear viscoelastic spectrum of polydisperse unentangled wormlike micellar solutions.

In recent work \cite{Kumar2025}, we introduced a mesoscopic Brownian dynamics framework in which wormlike micelles assemble from smaller, indivisible units, which were termed persistent worms. The model incorporated reversible scission–fusion kinetics, semiflexibility, and hydrodynamic interactions, and was shown to reproduce key static and rheological properties of unentangled dilute and semidilute wormlike micellar solutions. In particular, it captured the equilibrium length distributions of linear and ring micelles, as well as the frequency-dependent storage and loss moduli, and revealed qualitative rheological distinctions between wormlike micelles and homopolymer solutions. This study established the validity of the mesoscopic framework and its ability to describe the linear viscoelastic response; however, it did not systematically examine the hierarchy of microscopic timescales underlying stress relaxation.

The present work addresses this issue by extending the persistent worm mesoscopic framework to investigate the dynamics of dilute and unentangled semidilute wormlike micelles. The model combines reversible scission–fusion kinetics with Brownian dynamics simulations and naturally generates polydisperse ensembles of both linear and ring micelles. This enables a systematic examination of the characteristic timescales governing micellar dynamics and  how these timescales are influenced by hydrodynamic interactions and micellar topology.

The remainder of the paper is organized as follows. \sref{sec:goveqn} introduces the mesoscopic persistent worm framework employed in this study, along with the governing equations describing reversible scission and fusion and the Brownian dynamics simulation algorithm. Definitions of micellar length, effective concentration, and the stress tensor required for the analysis of dynamic and rheological properties are provided in \sref{sec:wlm}. \sref{sec:results} presents the results obtained from our simulations. In \sref{sec:cole_cole}, the linear viscoelastic response of wormlike micellar solutions is analyzed using storage and loss moduli and Cole–Cole representations. \sref{sec:timescales} focuses on the identification and quantification of molecular timescales governing micellar dynamics, and examines their dependence on concentration, sticker energy, micellar topology, and hydrodynamic interactions. In \sref{sec:mapping}, the molecular timescales are mapped onto features of the linear viscoelastic spectrum. The zero-shear rate viscosity and its concentration dependence in the dilute and unentangled semidilute regimes is analyzed in \sref{sec:vis_scaling}. \sref{sec:conclusion} summarizes the conclusions of the study.

\section{\label{sec:goveqn} The mesoscopic model and governing equations}

\noindent The mesoscopic model employed in this work is based on the persistent worm framework introduced in our recent study \cite{Kumar2025}. Here, we briefly summarize the aspects of the model that are directly relevant to the dynamic properties investigated in the present paper and refer the reader to Ref.~\cite{Kumar2025} for a detailed description of the framework and the validation studies against mean-field theory predictions for the static properties of wormlike micellar solutions \cite{Wittmer1998,Wittmer2000}.

\subsection{\label{sec:pw} The persistent worm as an indivisible unit}

Wormlike micelles are modeled as assemblies of persistent worms, which represent the shortest indivisible micellar units. Each persistent worm consists of $N_{\text{pw}}$ beads connected by Fraenkel springs with force law,  
\begin{equation}
    \bm{F}^{(\text{sp})} \left(\bm{r}_{\mu+1} - \bm{r}_{\mu}\right) = H \left[1 - \frac{b}{\left\vert\bm{r}_{\mu+1} - \bm{r}_{\mu}\right\vert}\right] \left(\bm{r}_{\mu+1} - \bm{r}_{\mu}\right),
\label{eq:Fraenkel}
\end{equation}
where $\bm{r}_{\mu}$ is the position vector of bead $\mu$, $H$ is the spring constant, and $b$ is the natural length of spring. The force acts along the connector vector $(\bm{r}_{\mu+1} - \bm{r}_{\mu})$, with magnitude $\lvert \bm{r}_{\mu+1} - \bm{r}_{\mu}\rvert$. With an appropriate choice of the magnitude of the spring constant $H$, the Fraenkel spring mimics a rigid rod of length $b$~\cite{Indranil2022,Amit2026}. The natural contour length of a persistent worm is therefore $\ell_{\text{pw}} = (N_{\text{pw}}-1)b$.

The terminal beads of each persistent worm are designated as sticky beads. Pairwise association of sticky beads permits reversible fusion and scission of persistent worms, leading to the formation of polydisperse linear and ring wormlike micelles. Sticker functionality is restricted to unity throughout this work, thereby preventing branching.

If $n_{\text{pw}}^{\text{T}}$ persistent worms are present in a simulation box of volume $V$, the total number of beads is $N^{\text{T}} = N_{\text{pw}} \times n_{\text{pw}}^{\text{T}}$, and the monomer concentration is $c = N^{\text{T}}/V$. Here, $c$ accounts for the monomers present in both linear and ring wormlike micelles.

\subsubsection{\label{sec:SDK} The pairwise bead-bead interaction potential}

The interaction between beads $\mu$ and $\nu$ separated by a distance $r_{\mu\nu}= \left\vert\bm{r}_{\mu} - \bm{r}_{\nu}\right\vert$, is described using the Soddemann-D\"unweg-Kremer (SDK) potential~\cite{SDK2001}, 
{\small
\begin{align}\label{eq:SDK} 
\frac{U_{\mu\nu}^{\text{SDK}}}{k_{\text{B}} T} = \left\{
\begin{array}{l l}
 \hspace{-1mm} 4 \left[ \left( \dfrac{\sigma}{  r_{\mu\nu}} \right)^{12} - \left( \dfrac{\sigma}{  r_{\mu\nu}} \right)^6 + \dfrac{1}{4} \right] - \epsilon \, ;& r_{\mu\nu}\leq 2^{1/6}\sigma \\ [15pt]
\hspace{-1mm} \dfrac{1}{2}\,  \epsilon  \left[ \cos \left(\alpha \left( \dfrac{ r_{\mu\nu}}{\sigma} \right)^2+ \beta\right) - 1 \right] ;& \hspace{-5mm} 2^{1/6}\sigma \leq  r_{\mu\nu} \leq r_{\rm c} \\ [15pt]
\hspace{-1mm} 0. &   r_{\mu\nu} \geq  r_{\rm c}
\end{array}\right.
\end{align}
}
where $r_{\mu\nu} = \vert\bm{r}_{\mu} - \bm{r}_{\nu}\vert$. Backbone–backbone and backbone–sticker interactions are purely repulsive ($\epsilon_{bb}=0$), sticker–sticker interactions are attractive with well depth $\epsilon_{st} \ge 1$. The interaction energies $\epsilon_{bb}$ and $\epsilon_{st}$ are expressed in units of $k_BT$, and all lengths are non-dimensionalized using $l_H=\sqrt{{k_{\text{B}}T}/{H}}$, with $\sigma/l_H=1$. The attractive interaction acts as a coarse-grained surrogate for the micellar end-cap energy \cite{Kumar2025}.

\noindent Reversible scission and fusion of micelles are implemented using a Monte Carlo scheme that satisfies detailed balance. When two stickers lie within the cutoff distance (i.e. $2^{1/6}\sigma \le r \le r_c$), a change in bonding energy is attempted with acceptance probability: $\min \left[1,\exp\left(- \Delta E / k_{\text{B}}T \right)\right]$, where,
\begin{equation}
\Delta E(r)=\frac12\left[\cos\left(\alpha\left(\frac{r}{\sigma}\right)^2+\beta\right)-1\right]\epsilon_{st}.
\end{equation}
Bond formation and bond breakage are attempted once per Brownian dynamics timestep. Although fusion has been treated as an activated process in some earlier studies \cite{Padding2008}, no explicit activation barrier is included here. Note that the model admits the possibility of semiflexible wormlike micelles through the inclusion of a bending potential within the Monte Carlo scheme; however, the present study is restricted to flexible wormlike micelles \cite{Kumar2025}.

\subsubsection{\label{sec:BD} Euler integration algorithm}

The position of each bead, $\bm{r}_{\mu}(t)$ $(\mu = 1,2,3, \ldots, N^{\text{T}})$, in the simulation box evolves according to a first-order Euler integration scheme for the numerical solution of the It\^o stochastic differential equation~\cite{JainPRE2012},
\begin{equation}\label{eq:BD}
    \bm{r}_{\mu}(t+\Delta t) = \bm{r}_{\mu}(t) + \frac{\Delta t}{4} \sum_{\nu=1}^{N^{\text{T}}} \bm{D}_{\mu\nu} \cdot \bm{F}_\nu +  \frac{1}{\sqrt{2}} \sum_{\nu=1}^{N^{\text{T}}}\bm{B}_{\mu\nu} \cdot \bm{\Delta W}_\nu
    \end{equation}
Here time scales are nondimensionalized using $\lambda_H={\zeta}/{4H}$, where $\zeta=6\pi \eta_{\text{s}} a$ is the Stokes friction coefficient of a bead of radius $a$ in a solvent of viscosity $\eta_{\text{s}}$. $\Delta\bm W_\nu$ is a non-dimensional Wiener process with zero mean and variance $\Delta t$. The non-dimensional diffusion tensor is given by $\bm D_{\mu\nu} = \delta_{\mu\nu} \bm{\delta} + \bm{\Omega}_{\mu\nu}$, where $\delta_{\mu\nu}$ is the Kronecker delta, $\bm{\delta}$ is the unit tensor, and $\bm{\Omega}_{\mu\nu}$ is the hydrodynamic interaction tensor. The $(\mu,\nu)$-th block of block matrices $\mathcal{D}$ and $\mathcal{B}$ is represented by $\bm{D}_{\mu\nu }$ and $\bm{B}_{\mu\nu }$, respectively. These matrices consist of $N^{\text{T}} \times N^{\text{T}}$ blocks, each of dimension $3 \times 3$. The matrix $\mathcal{B}$ satisfies the decomposition rule, $\mathcal{B} \cdot {\mathcal{B}}^{\top} = \mathcal{D}$. Hydrodynamic interactions are modeled using the regularized Rotne–Prager–Yamakawa (RPY) tensor:
\begin{equation}
{\bm{\Omega}_{\mu \nu}} = {\bm{\Omega}} ( {\bm{r}_{\mu}} - {\bm{r}_{\nu}} )
\end{equation}
where 
\begin{equation}
\bm{\Omega}(\bm{r}) =  {\Omega_1\, \bm{\delta} +\Omega_2 \, \frac{\bm{r r}}{{r}^{2}}} \, ; \,\, \text{with} \, \left\vert \bm{r} \right\vert = r
\end{equation}
and
\begin{equation*}
\Omega_1 = \begin{cases} \dfrac{3\sqrt{\pi}}{4} \dfrac{h^*}{r}\left({1+\dfrac{2\pi}{3}\dfrac{{h^*}^2}{{r}^{2}}}\right) & \text{for} \quad r\ge2\sqrt{\pi}h^* \\
 1- \dfrac{9}{32} \dfrac{r}{h^*\sqrt{\pi}} & \text{for} \quad r\leq 2\sqrt{\pi}h^* 
\end{cases}
\end{equation*}
\vspace{5pt}
\begin{equation*}
\Omega_2 = \begin{cases} \dfrac{3\sqrt{\pi}}{4} \dfrac{h^*}{r} \left({1-\dfrac{2\pi}{3}\dfrac{{h^*}^2}{{r}^{2}}}\right) & \text{for} \quad r\ge2\sqrt{\pi}h^* \\
 \dfrac{3}{32} \dfrac{r}{h^*\sqrt{\pi}} & \text{for} \quad r\leq 2\sqrt{\pi}h^* 
\end{cases}
\end{equation*}
The hydrodynamic interaction parameter $h^* = a/(\sqrt{\pi k_BT/H})$ is the dimensionless bead radius. 

The net nondimensional force $\bm{F}_\nu$ force acting on bead $\nu$ in \eref{eq:BD} is a sum of the nondimensional spring force, $\bm{F}_\nu^{(\text{sp})} = \bm{F}^{(\text{sp})}(\bm{Q}_\nu) - \bm{F}^{(\text{sp})}(\bm{Q}_{\nu-1})$ (derived from 
 \eref{eq:Fraenkel}) and the force due to the SDK potential, $\bm{F}_\nu^{\text{SDK}}$ (derived from \eref{eq:SDK}),
\begin{equation}\label{netforce}
    \bm{F}_\nu =  \bm{F}_\nu^{(\text{sp})} + \bm{F}_\nu^{\text{SDK}} 
\end{equation}
The subsequent section describes the simulation protocol used to integrate \eref{eq:BD}.

\subsubsection{\label{sec:simdet} Simulation Details}

Brownian dynamics simulations were performed using the HOOMD-Blue simulation toolkit~\cite{Nlist2019,HOOMD2020}. Hydrodynamic interactions were incorporated through the positively split Ewald (PSE) method, originally developed for colloidal suspensions~\cite{PSE2017} and later adapted for polymer solutions~\cite{Robe2024}. The latter implementation was used in the present study.

Each simulation was conducted in three stages. In the pre-equilibration stage, only backbone monomers were present. This was followed by an equilibration stage, during which the sticker strength was increased stepwise until the system reached a steady state. Finally, data were collected during a production stage. The duration of each stage was expressed in terms of the longest polymer relaxation time, estimated from a homopolymer chain with length equal to the mean micelle length. Since the mean micelle length depends on both the concentration of persistent worms and the sticker strength, preliminary simulations were first carried out until static properties reached a stationary state. The mean micelle length was then estimated from these runs. For simulations including hydrodynamic interactions, the Zimm relaxation time corresponding to the mean micelle length was used, whereas for simulations without hydrodynamic interactions, the Rouse relaxation time was adopted~\cite{Bird1987}. Pre-equilibration was typically carried out for 5–10 relaxation times, followed by equilibration for 20 relaxation times, and a production run of approximately $\tau_{run} = 10$ relaxation times. The non-dimensional timestep was fixed at $\Delta t = 10^{-3}$ throughout after ensuring timestep convergence. Ensemble averages and statistical uncertainties were obtained from 1000 independent simulation runs ($N_{\text{run}}$). For simulations without hydrodynamic interactions, $h^*$ was set to 0, while simulations with hydrodynamic interactions were conducted with $h^* = 0.2$.

    \begin{table}[b]
    \begin{center}
           \caption{Typical parameter values used in the Brownian dynamics simulations \label{parametervalues}}         
     \begin{tabular}{llllll}
     	     \hline
            {} & Parameter & & Symbol  &  & Values \\
            \hline
            1 & Backbone interaction strength  & & $\text{$\epsilon$}_{bb}$   &  & 0 \\
            2 & Sticker strength  & & $\text{$\epsilon$}_{st}$    & & 2 to 10 \\
            3 & Simulation box size  & & $L_{\text{box}} $    & & 24, 40 \\
            4 & Rest length of the spring  & & $b$  & & 3\\
            5 & Hydrodynamic interaction parameter  & & $h^*$  & & 0, 0.2 \\
            6 & Integration time step  & & $\Delta t$ &  & 0.001 \\
            7 & Number of beads in a persistent worm  & & $N_{\text{pw}}$   & &  3 \\
            8 & Persistent worm length & & $\ell_{\text{pw}}$   & & 6 \\
            9 & Independent simulation instances  & & $N_{\text{run}}$  & & 1000 \\
            10 & Scaled monomer concentration  & & $c/c^*_{\text{pw}}$  & & 0.2 to 1 \\
            \hline
           \end{tabular}
           \end{center}
    \end{table}

Because wormlike micelles are inherently polydisperse, their overlap concentration is not known \textit{a priori}. For the purpose of planning simulations, we therefore define a scaled concentration $c/c_{\text{pw}}^*$, where $c_{\text{pw}}^*$ is the overlap concentration of a system consisting purely of persistent worms (i.e., chains with $N_{\mathrm{pw}}$ beads with no sticky ends), defined as
\begin{equation}\label{cpw}
c_{\text{pw}}^* = \frac{N_{\text{pw}}}{\left(\tfrac{4\pi}{3}\right) R_{g_0,\text{pw}}^3}.
\end{equation}
Here $R_{g_0,\text{pw}}$ is the radius of gyration of a persistent worm in the dilute limit, obtained from simulations of bead–spring chains with purely repulsive interactions and without hydrodynamic interactions. While $c_{\text{pw}}^*$ does not distinguish between dilute and semidilute regimes, it provides a convenient reference for non-dimensionalizing the monomer concentration. In the simulations, the sticker strength $\epsilon_{st}$ and the concentration of persistent worms $c$ are prescribed input parameters, whereas the effective monomer concentration $c^{\text{eff}}$ (see \eref{eq:ceff} below) is an output of the simulation and depends on both $c/c_{\text{pw}}^*$ and $\epsilon_{st}$.

To examine the role of ring formation on micellar dynamics, two versions of the algorithm were employed: one allowing both linear and ring micelles (labelled ``With Rings (WR)'' in the figure legends), and another excluding ring formation (labelled ``No Rings (NR)'').
    
Table~\ref{parametervalues} summarizes the typical parameter values used in the Brownian dynamics simulations reported in this work. These parameters are consistent with those employed in our earlier study \cite{Kumar2025}.  We now proceed to define the static and dynamic properties of wormlike micelles that are required for the analysis of molecular timescales and linear viscoelastic response.

\subsection{\label{sec:wlm} Property definition of wormlike micelles assembled from persistent worms}

\subsubsection{\label{sec:lcd} Lengths and concentrations}

The definitions of the static and dynamic properties of wormlike micelles have been described in detail in our earlier work \cite{Kumar2025}. Here, we provide a concise summary of the quantities directly relevant to the analysis of dynamic properties presented in this study.

The contour length $L$ of a wormlike micelle (linear or ring), formed by the assembly of $m_{\text{pw}}^{\text{L}}$ persistent worms, is given by,
\begin{equation}
L = m_{\text{pw}}^{\text{L}}  \ell_{\text{pw}} = m_{\text{pw}}^{\text{L}} (N_{\text{pw}} - 1) b
\label{eq:wlmlength}
\end{equation}
Since micelles grow and shrink by the fusion and scission of a persistent worm length ($\ell_{\text{pw}}$), the contour length $L$ is a discrete quantity.

If $n^{\text{lin}}_{\text{L}}$ and $n^{\text{R}}_{\text{L}}$ denote the number of linear and ring micelles of length $L$, respectively, the total number of micelles ($\mathcal{N}_{\text{wlm}}$) in the simulation box is:
\begin{align}
& \mathcal{N}^{\text{lin}}_{\text{wlm}}   = \sum_{L_{\text{min}}}^{ L_{\text{max}}}  n^{\text{lin}}_{\text{L}} (L) ,  \,  \, \text{and} \, \,  
\mathcal{N}^{\text{R}}_{\text{wlm}} = \sum_{L_{\text{min}}}^{ L_{\text{max}}}  n^{\text{R}}_{\text{L}} (L) \nonumber \\
 \text{and}, \quad & \mathcal{N}_{\text{wlm}}= \mathcal{N}^{\text{lin}}_{\text{wlm}} + \mathcal{N}^{\text{R}}_{\text{wlm}} 
\label{eq:Nwlm}
\end{align}
where the limits $L_{\text{min}}$ and $L_{\text{max}}$ represent the shortest and longest possible length of a micelle, respectively \cite{Kumar2025}. Since sticky beads pair when persistent worms assemble into micelles, the \textit{effective} number of monomers in a linear micelle of length $L$ is:
\begin{equation}
        N_{\text{L,lin}}^{\text{eff}} =  (N_{\text{pw}} - 1) \,  m_{\text{pw,lin}}^{\text{L}} +1      
 \label{eq:Nefflin}   
\end{equation}
while for a ring micelle it is:
\begin{equation}
        N_{\text{L,R}}^{\text{eff}}   = (N_{\text{pw}} - 1)  \, m_{\text{pw,R}}^{\text{L}}
\label{eq:NeffR}   
\end{equation}
where $m_{\text{pw,lin}}^{\text{L}}$ and $m_{\text{pw,R}}^{\text{L}}$ is the number of persistent worms in a linear and ring wormlike micelle, respectively. 

The total effective number of monomers in linear and ring micelles in the simulation box are, respectively,
\begin{equation}
N_{\text{T,lin}}^{\text{eff}} = \sum_{L_{\text{min}}}^{ L_{\text{max}}}  n^{\text{lin}}_{\text{L}} (L) N_{\text{L,lin}}^{\text{eff}} ,  \,  \, \text{and} \, \, 
N_{\text{T,R}}^{\text{eff}} = \sum_{L_{\text{min}}}^{ L_{\text{max}}}  n^{\text{R}}_{\text{L}} (L) N_{\text{L,R}}^{\text{eff}} 
\label{eq:Nefftotal}
\end{equation}
leading to the effective concentrations:
\begin{equation}
c^{\text{eff}} = \frac{N_{\text{T,lin}}^{\text{eff}}}{V}  , \,  \, \text{and} \, \, c_{\text{R}}^{\text{eff}} = \frac{N_{\text{T,R}}^{\text{eff}}}{V} 
\label{eq:ceff}
\end{equation}
For simplicity, the subscript ``lin” is dropped in reporting $c^{\text{eff}}$, since most results are expressed in terms of the linear micelle concentration.

The probability of finding linear or ring micelles with contour length between $L$ and $L + \Delta L$ is defined as: 
\begin{align}
\psi_{\text{lin}} (L) \Delta L &  = \frac{  n^{\text{lin}}_{\text{L} + \Delta \text{L}}    - n^{\text{lin}}_{\text{L}} }{\mathcal{N}^{\text{lin}}_{\text{wlm}}} , 
 \text{and} \\ \nonumber 
 \psi_{\text{R}} (L) \Delta L & = \frac{  n^{\text{R}}_{\text{L} + \Delta \text{L}}    - n^{\text{R}}_{\text{L}} }{\mathcal{N}^{\text{R}}_{\text{wlm}}}
\label{eq:psiL}
\end{align}
The corresponding mean lengths of linear and ring wormlike micelle are defined, respectively, by,
\begin{equation}
\bar{L} = \sum_{L_{\text{min}}}^{ L_{\text{max}}}  L \, \psi_{\text{lin}} (L) \Delta L , \,  \, \text{and} \, \, \bar{L}_{\text{R}} = \sum_{L_{\text{min}}}^{ L_{\text{max}}}  L \, \psi_{\text{R}} (L) \Delta L 
\label{eq:meanL}
\end{equation}
Here, the subscript ``lin” has again been dropped from $\bar{L}$ for simplicity. 

As mentioned earlier wormlike micellar solutions are intrinsically polydisperse, a unique overlap concentration cannot be defined a priori. In our earlier work~\cite{Kumar2025}, we discussed the procedure for determining the overlap concentration $c^*$ for a fixed sticker energy $\epsilon_{st}$, following the approach of ~\citet{Wittmer1998} and we refer the reader to Ref. \cite{Kumar2025} for details.

\subsubsection{\label{sec:stress_dyn} Stress tensor and dynamic moduli}

The nondimensional polymeric stress tensor for a polydisperse solution of linear and ring wormlike micelles is computed using the Kramers–Kirkwood expression,

\begin{equation}\label{Kramers}
\bm{\tau}^{\text{p}} = \frac{1}{\mathcal{N}_{\text{wlm}}} \left\langle \sum_{\alpha=1}^{\mathcal{N}_{\text{wlm}}} \sum_{\nu=1}^{N_b^{\text{L}}} \left(\bm{r}_\nu^{(\alpha)}-\bm{r}_c^{(\alpha)} \right) \bm{F}_{\alpha\nu} \right\rangle
\end{equation}
where the stress is nondimensionalised by $\mathcal{N}_{\text{wlm}} \, (k_{\text{B}} T/V)$, $N_b^{\text{L}} = m_{\text{pw}}^{\text{L}} N_{\text{pw}}$ is the number of beads in a micelle of length $L$, and $\bm{r}_{\text{c}}^{(\alpha)} = \left( 1/N_{\text{L}}^{\text{eff},\alpha}\right)   \sum_{\mu=1}^{N_{\text{L}}^{\text{eff},\alpha} } \bm{r}_\mu$ is the center of mass of micelle $\alpha$, where $N_{\text{L}}^{\text{eff},\alpha}$ denotes the \textit{effective} number of monomers in the $\alpha$-th micelle. 

\noindent The total force on bead $\nu$ in micelle $\alpha$ is: 
\begin{equation}\label{totalforce}
\bm{F}_{\alpha\nu} = \sum_{\beta=1}^{\mathcal{N}_{\text{wlm}}} \sum_{\substack{\mu=1 \\ \mu\ne\nu}}^{N_b^{\text{L}}} \bm{F}_{\alpha\nu,\beta\mu}^{\text{SDK}} + \sum_{\substack{\mu=1 \\ \mu\ne\nu}}^{N_b^{\text{L}}} \bm{F}_{\alpha\nu,\alpha\mu}^{\text{(sp)}} 
\end{equation}
where the summations explicitly account for pairwise bead–bead interactions and intra-micellar spring forces.

\noindent The dimensionless net stress tensor $\bm{S}$ in a simulation box for a single trajectory in \eref{Kramers} is,
\begin{multline}
\label{tensor_S}
\bm{S} = \sum_{\alpha=1}^{\mathcal{N}_{\text{wlm}}} \sum_{\nu=1}^{N_b^{\text{L}}} \left( \bm{r}_\nu^{(\alpha)}-\bm{r}_c^{(\alpha)} \right) \bm{F}_{\alpha\nu} = \\
\frac{1}{2} \sum_{\nu=1}^{N^{\text{T}}}\sum_{\substack{\mu=1  \mu\ne\nu}}^{N^{\text{T}}} \bm{r}_{\nu\mu} \bm{F}_{\nu\mu}^{\text{SDK}} + \\\sum_{\alpha=1}^{\mathcal{N}_{\text{wlm}}} \sum_{i=1}^{N_s^{\text{L}}} \bm{Q}_{i}^{(\alpha)} \bm{F}^{{\text{(s)}}}
\left( \bm{Q}_{i}^{(\alpha)} \right)
\end{multline}

where $N_s^{\text{L}}$ is the number of springs in a wormlike micelle of length $L$, independent of its architecture. The tensor $\bm{S}$ is used to evaluate the dimensionless shear relaxation modulus, which under isotropic equilibrium conditions is given by:
\begin{equation}\label{Gtiso}
    G(t) = \frac{1}{3} \Big [  G_{xy}(t) + G_{xz}(t) + G_{yz}(t) \Big ]
\end{equation}
with individual components $G_{ij}(t)$ given by the Green–Kubo relation,
\begin{equation}\label{stressauto}
    G_{ij}(t) =\frac{1}{\mathcal{N}_{\text{wlm}}} \, \Big\langle \! S_{ij}(0) \, S_{ij}(t) \! \Big\rangle 
\end{equation} 
The relaxation modulus is fitted to a sum of exponentials as: $G(t) = \sum_{k=1}^{n} a_k \exp(-t/\tau_k)$, with fitting parameters $a_k$ and $\tau_k$, and $n$ denoting the number of exponentials.
Finally, the storage ($G^\prime(\omega)$) and loss ($G^{\prime\prime}(\omega)$) moduli are obtained via Fourier transformation of $G(t)$:
\begin{align}  \label{Gprime}
    G^\prime (\omega) = \int_0^\infty \!\! d (\omega t) \, G(t) \sin(\omega t)\\[10pt] \nonumber
    G^{\prime\prime} (\omega) =  \int_0^\infty \!\! d (\omega t) \, G(t) \cos(\omega t)
\end{align}

Using the simulation framework and property definitions described above, we compute the linear viscoelastic response of wormlike micellar solutions in the dilute and unentangled semidilute regimes. In the following section, the viscoelastic behaviour is analyzed using the Cole–Cole representation to highlight relaxation features unique to wormlike micelles and to motivate the identification of the underlying microscopic timescales.

\section{\label{sec:results} Results and Discussion}

\subsection{\label{sec:cole_cole} Cole-Cole plots}

We first compare the Cole–Cole response of monodisperse homopolymer solutions with that of wormlike micelles which allows us to identify qualitative differences introduced by micellar kinetics. We then examine how the Cole–Cole response of wormlike micelles varies with sticker energy and concentration.

\begin{figure*}[t]
\begin{center}
\begin{tabular}{cc}
\includegraphics[width=8.25cm]{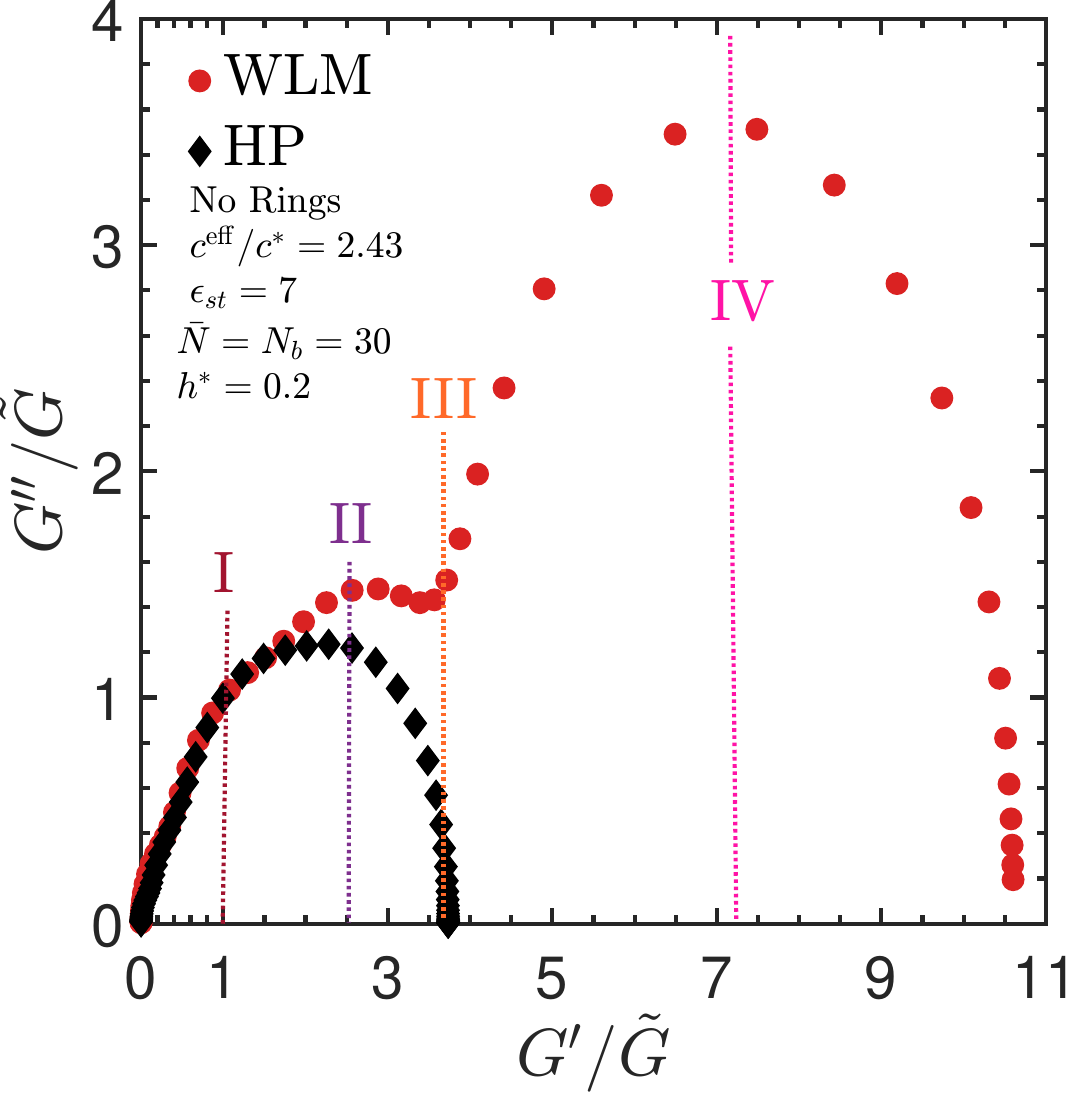} &
\includegraphics[width=8.5cm]{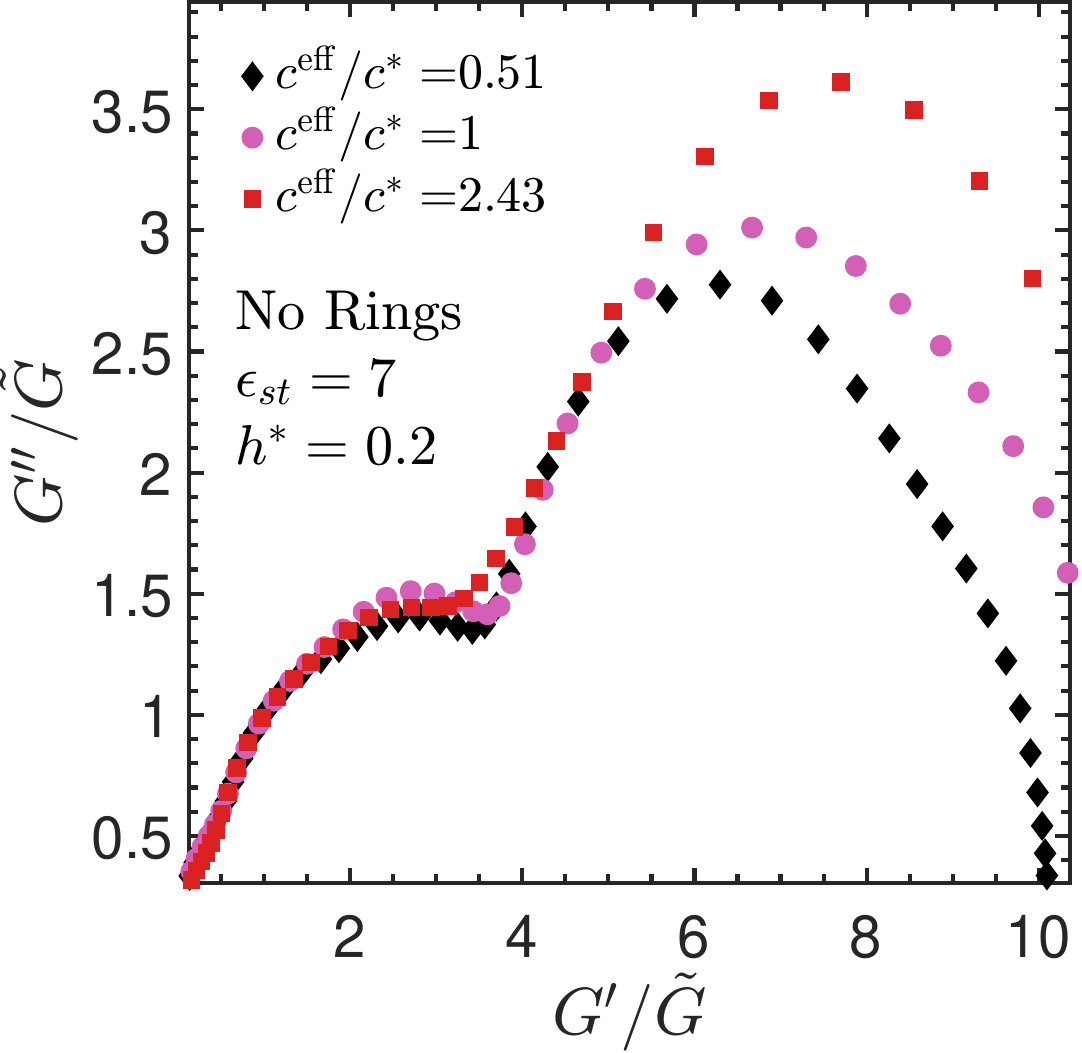} \\
(a)  & (b) \\[10pt]
\includegraphics[width=8.45cm]{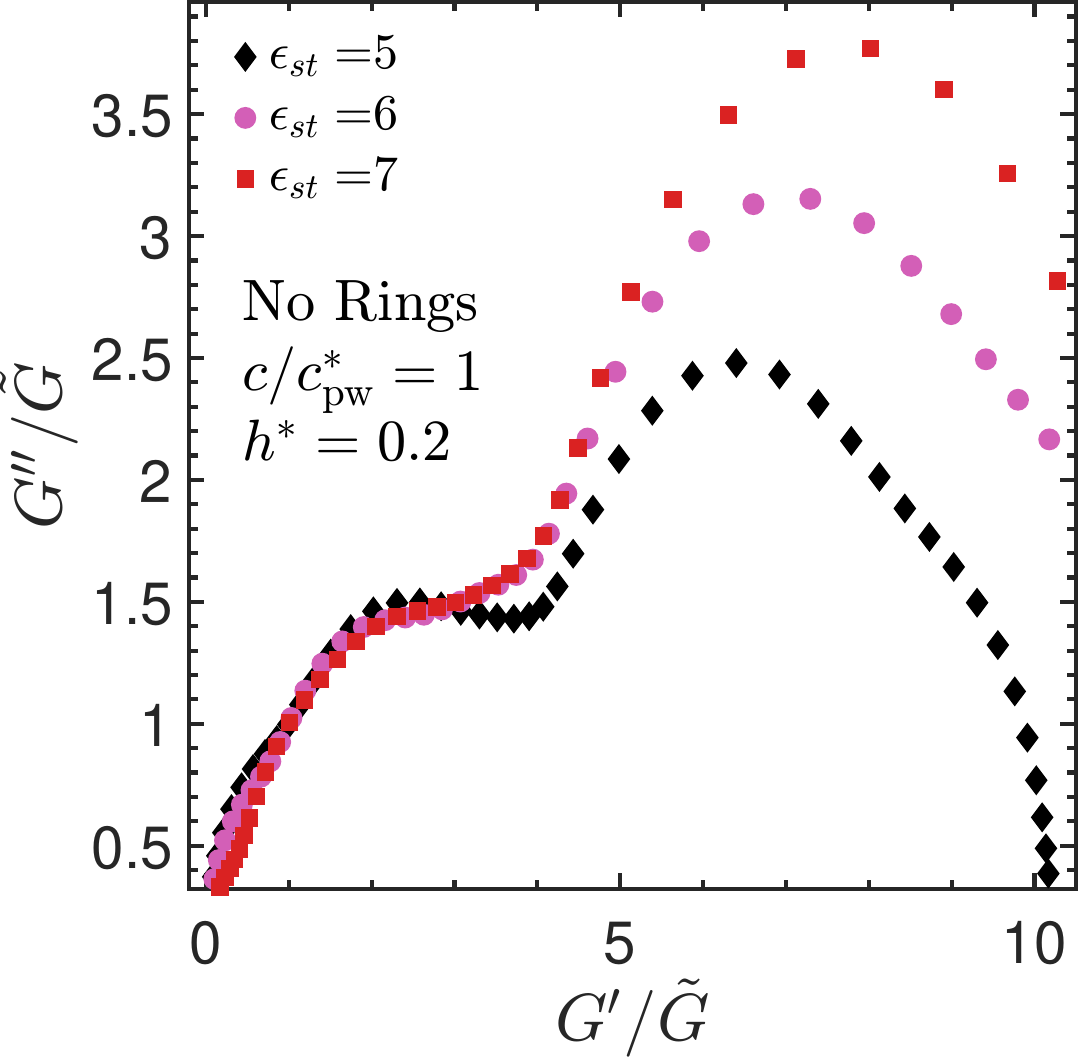} &
\includegraphics[width=8.5cm]{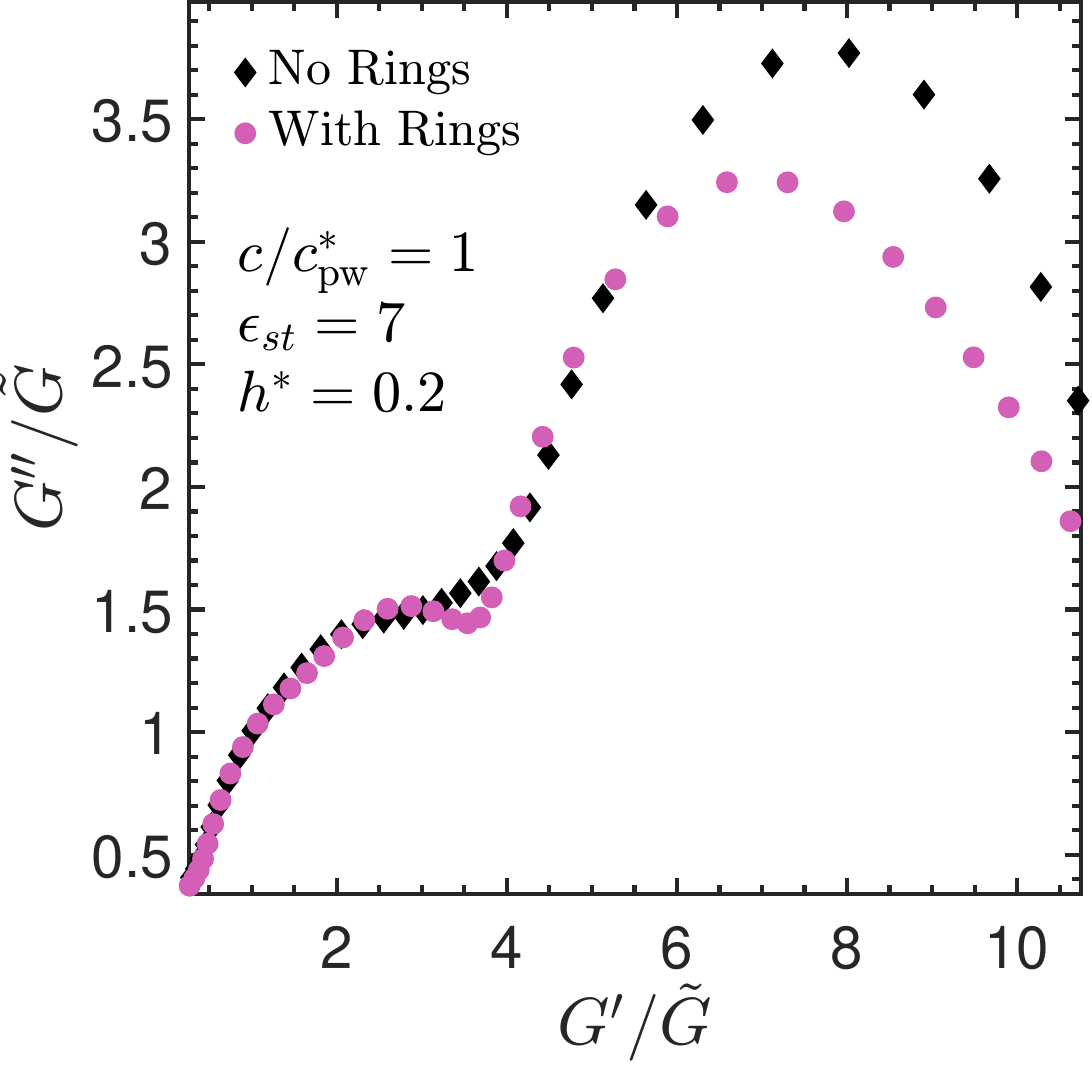} \\
(c)  & (d) 
\end{tabular}
\end{center}
    \vspace{-10pt}
\caption{Cole–Cole representations of the linear viscoelastic response of unentangled wormlike micellar solutions under different conditions. (a) Comparison between a monodisperse homopolymer solution (HP) at concentration $c/c^*=2.43$ with $N_b=30$ beads per chain and a wormlike micellar (WLM) solution with effective concentration $c^{\text{eff}}/c^*=2.43$ and sticker energy $\epsilon_{st}=7$, for the ``No Rings" case, under athermal solvent conditions with hydrodynamic interaction strength ($h^*=0.2$). (b) Effect of increasing wormlike micellar concentration at fixed sticker energy $\epsilon_{st}$ for the ``No Rings" case. (c) Effect of increasing sticker energy at fixed persistent worm concentration $c/c^*_{\text{pw}}=1$ for the ``No Rings" case. (d) Comparison between systems without rings (``No Rings") and with rings (``With Rings") at fixed sticker energy $\epsilon_{st}=7$ and persistent worm cocnetration $c/c^*_{\text{pw}}=1$. In panels (a)–(d), $G^\prime$ and $G^{\prime\prime}$ are normalised by $\tilde{G}$, defined as the value of the moduli at the crossover frequency $\omega=\omega_c^*$, where $G^\prime = G^{\prime\prime}=\tilde{G}$.}
\label{fig1}
    \vspace{-10pt}
\end{figure*}

\subsubsection{\label{sec:cole_cole_wlm} Cole–Cole Analysis of Homopolymer and Wormlike Micellar Solutions}

In our earlier work \cite{Kumar2025}, we compared the storage ($G^{\prime}$) and loss ($G^{\prime\prime}$) moduli of monodisperse homopolymer solutions with those of wormlike micellar solutions. We showed that when the monomer concentration in the homopolymer solution matches the effective concentration of a wormlike micellar solution, and the polymer chain length is equal to the mean micellar length, the longest nondimensional relaxation time is identical in both systems, where the longest relaxation time is given by: $\tau_1=\tau_{G^\prime=G^{\prime\prime}}^{\text{WLM}} = \tau_{G^\prime=G^{\prime\prime}}^{\text{HP}} = 1/\omega_c^*$. 

Here, we revisit the same systems but focus instead on their Cole–Cole representations. The wormlike micellar solution considered has a sticker energy $\epsilon_{st}=7$ and an effective concentration $c^{\text{eff}}/c^* = 2.43$, with a mean micellar length $\bar{L}=87$, corresponding to a mean number of beads $\bar{N}=\bar{L}/b+1 = 30$. For comparison, a monodisperse homopolymer solution with $N_b=\bar{N}=30$ was simulated at a semidilute concentration $c/c^*=2.43$, under athermal solvent conditions with the strenth of hydrodynamic interactions given by $h^*=0.2$. A clear distinction between the two solutions is observed in \fref{fig1}(a), as discussed below. To identify the longest relaxation time in both systems, the storage ($G^{\prime}$) and loss ($G^{\prime\prime}$) moduli in the Cole–Cole plot are non-dimensionalized by $\tilde{G}$, defined as the value of the moduli at the crossover frequency $\omega_c^*$, where $\tilde{G} = G^{\prime} = G^{\prime\prime}$. The dotted vertical line marked I in \fref{fig1}(a) corresponds to this longest relaxation time for both solutions.

In the Cole–Cole representation, increasing values of $G^{\prime}$ along the $x$-axis correspond to increasing frequency. The monodisperse homopolymer solution exhibits a smooth and monotonic curve. At high frequencies, the storage moduli $G^\prime$ approaches a constant value characteristic of an elastic solid, while the loss modulus $G^{\prime\prime}$ tends to zero as polymer chains no longer contribute to the dynamic viscosity. Consequently, in the Cole–Cole plot, the value of $G^{\prime\prime}$ approaches zero. At low frequencies, the homopolymer solution exhibits the terminal behaviour, with $G^{\prime}$ scaling as $\omega^2$ and $G^{\prime\prime}$ scaling linearly with $\omega$. In this regime, the system behaves as a viscoelastic liquid, with the loss modulus dominating for frequencies lower than that corresponding to the longest relaxation time. For a single-mode Maxwell fluid, the Cole–Cole plot forms a perfect semicircle, with its peak occurring at the frequency $\omega_c^*$ where $G^{\prime} = G^{\prime\prime}$, with the relaxation time ($1/\omega_c^*$) obtained from the crossover point directly representing the single characteristic relaxation time of the system. In contrast, as shown in \fref{fig1}(a), the simulated homopolymer solution at finite concentration relaxes through multiple modes rather than a single exponential decay, and the Cole–Cole curve therefore deviates from a perfect semicircle.

The wormlike micellar solution, however, exhibits two pronounced maxima (marked II and IV in \fref{fig1}(a)) and an apparent minimum (marked III) at distinct locations along the Cole–Cole curve. These features are absent in homopolymer solutions and reflect additional relaxation processes arising from reversible scission, recombination, and intramicellar stress relaxation. The detailed origin of these features and their associated timescales are examined in the subsequent sections.

Similar qualitative viscoelastic behaviour is observed in \fref{fig1}(b)-(d) under different conditions. In \fref{fig1}(b), the concentration of wormlike micelles is increased while keeping the sticker energy fixed at $\epsilon_{st}=7$ for the ``No Rings" case. In \fref{fig1}(c), the sticker energy is increased at fixed persistent worm concentration ($c/c^*_{\mathrm{pw}}=1$). In \fref{fig1}(d), results obtained with and without ring micelles are compared. 

Across all cases, a sequence of qualitatively similar characteristic features is observed in the Cole–Cole plots, indicating that the underlying hierarchy of relaxation processes is robust. Quantitative differences are primarily observed in the magnitude of $G^{\prime\prime}/\tilde{G}$, which increases with increasing concentration, increasing sticker energy, and in the absence of rings, along with the disappearance of the minimum between the two maxima in some cases. The former trend arises because $G^{\prime\prime}$ is nondimensionalized by $\mathcal{N}_{\text{wlm}} \, (k_{\text{B}} T/V)$. For a fixed concentration of persistent worms, increasing the sticker energy or effective micellar concentration, promotes the formation of longer micelles. As a result, the mean micellar length increases and the total number of micelles $\mathcal{N}_{\text{wlm}}$ in the simulation box decreases. 

The emergence of these additional qualitative features in wormlike micellar solutions forms the central focus of the present work. In the following section, each characteristic frequency is analyzed in detail, its connection to specific microscopic processes is established, and the corresponding timescales are extracted. We further examine how hydrodynamic interactions and micellar topology, including the presence or absence of rings, influence these timescales.

\subsection{\label{sec:timescales} Molecular timescales governing linear viscoelastic behaviour of wormlike micelles}

\subsubsection{\label{sec:tbond} Bond breakage time} 

\begin{figure}[tb]
\centering
\includegraphics[width=8.5cm]{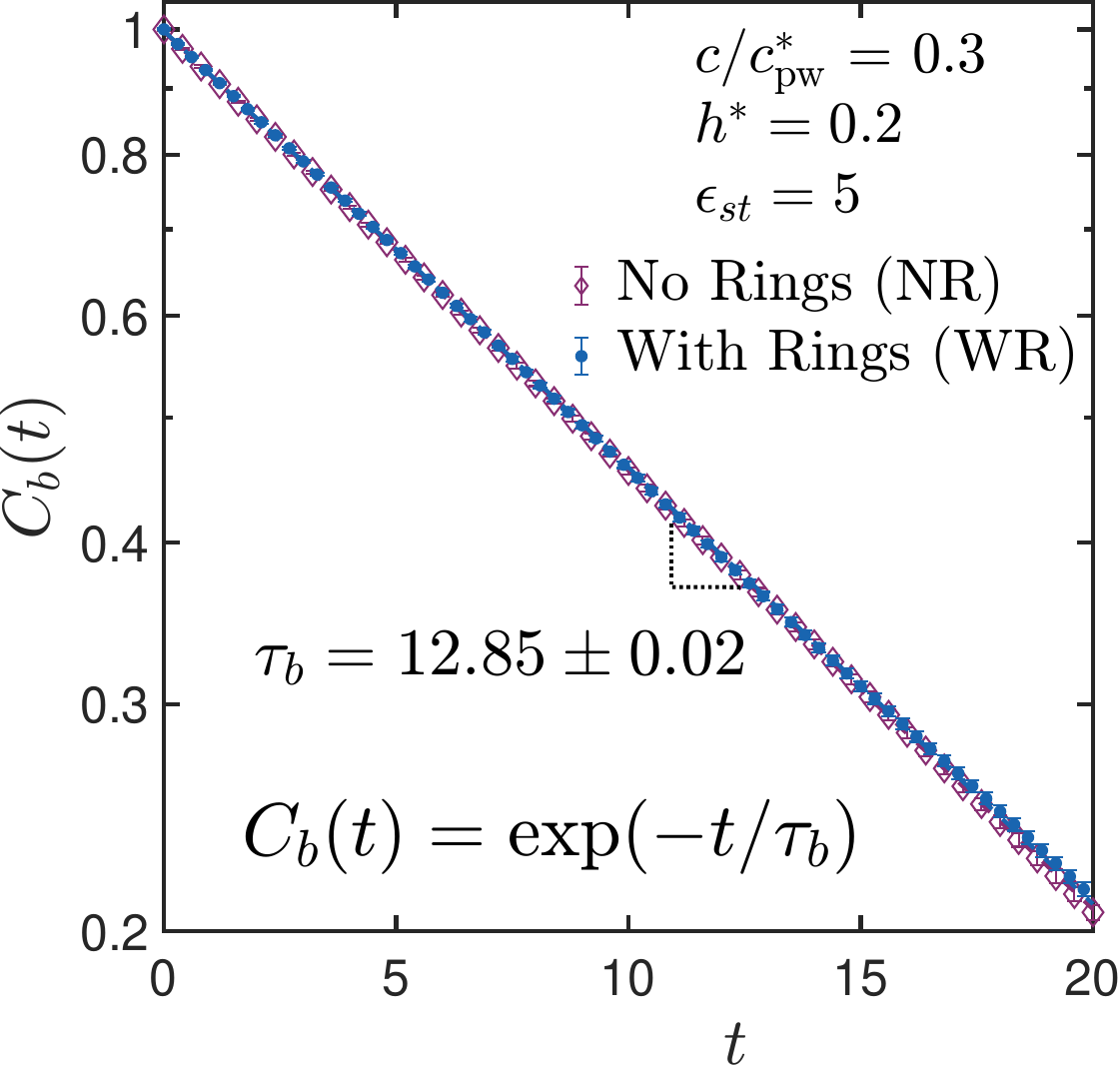}  
\caption{Bond autocorrelation function $C_b(t)$ for wormlike micelle solutions with and without rings at $\epsilon_{st}=5$ and $c/c_{\mathrm{pw}}^* = 0.3$.}
\label{fig2}
    \vspace{-10pt}
\end{figure}

The bond breakage time ($\tau_b$) is defined as the average lifetime of a bond formed between a given pair of stickers. It is obtained from the decay of the bond autocorrelation function~\cite{Gowers2015,Robe2024}, defined as:

\begin{equation}
    C_b(t) = \frac{\Big\langle \sum_{i=1}^{N_{\mathrm{st}}^{\text{T}}}\sum_{j=1}^{N_{\mathrm{st}}^{\text{T}}} \mathcal{M}_{ij}(t_0)\,\mathcal{M}_{ij}(t_0+t) \Big\rangle}{\Big\langle \sum_{i=1}^{N_{\mathrm{st}}^{\text{T}}}\sum_{j=1}^{N_{\mathrm{st}}^{\text{T}}} \mathcal{M}_{ij}^2(t_0)\Big\rangle}
\end{equation}
Here, $\mathcal{M}_{ij}(t)$ is a binary bonding matrix of dimension $N_{\mathrm{st}}^{\text{T}} \times N_{\mathrm{st}}^{\text{T}}$, where $\mathcal{M}_{ij}(t)=1$ if stickers $i$ and $j$ are bonded at time $t$, and $\mathcal{M}_{ij}(t)=0$ otherwise. The product $\mathcal{M}_{ij}(t_0)\mathcal{M}_{ij}(t_0+t)$ equals unity if the same pair of stickers is bonded at both times $t_0$ and $t_0+t$, and is zero otherwise. The double summation runs over all stickers  $N_{\mathrm{st}}^{\text{T}}$ in the simulation, while the angular brackets $\langle \cdots \rangle$ denote averaging over all reference times $t_0$ along the simulation trajectory (ensemble-time average). Following \citet{Gowers2015}, a continuous bond lifetime definition is employed here, whereby a bond contributes to the correlation function only if the same pair of stickers remains continuously bonded over the entire time interval $t$ without intermediate breakage.

The bond autocorrelation function, $C_b(t)$, obtained from simulations, decays exponentially as: $
C_b(t) = \,\exp\!\left(-t/\tau_b\right),$
where the characteristic decay time $\tau_b$ represents the bond breakage time. An example of this exponential fit is shown in \fref{fig2} for wormlike micellar solutions with and without rings at $c/c_{\mathrm{pw}}^* = 0.3$ and $\epsilon_{\mathrm{st}}=5$.

\begin{figure*}[t]
\begin{center}
\begin{tabular}{cc}
\includegraphics[width=8.5cm]{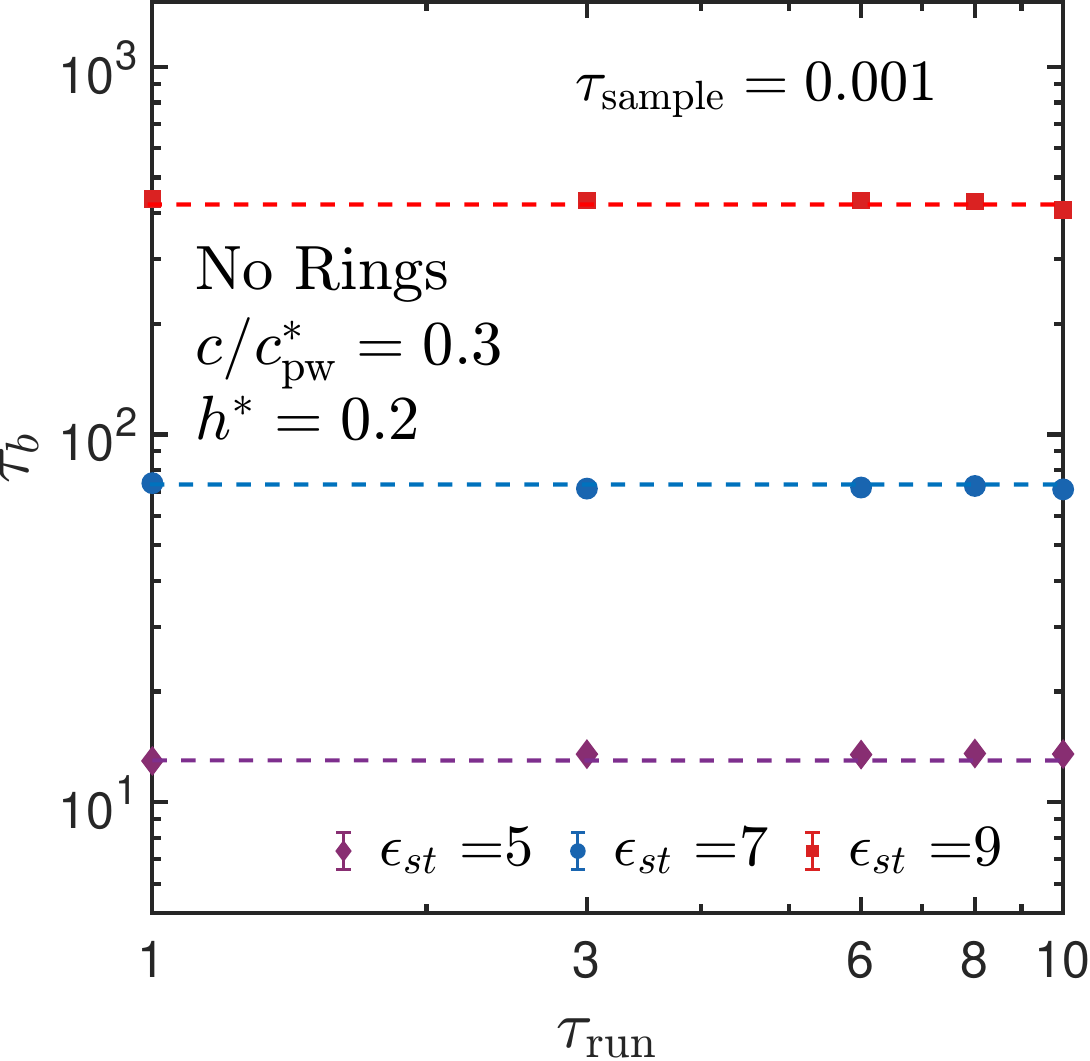} &
\includegraphics[width=8.5cm]{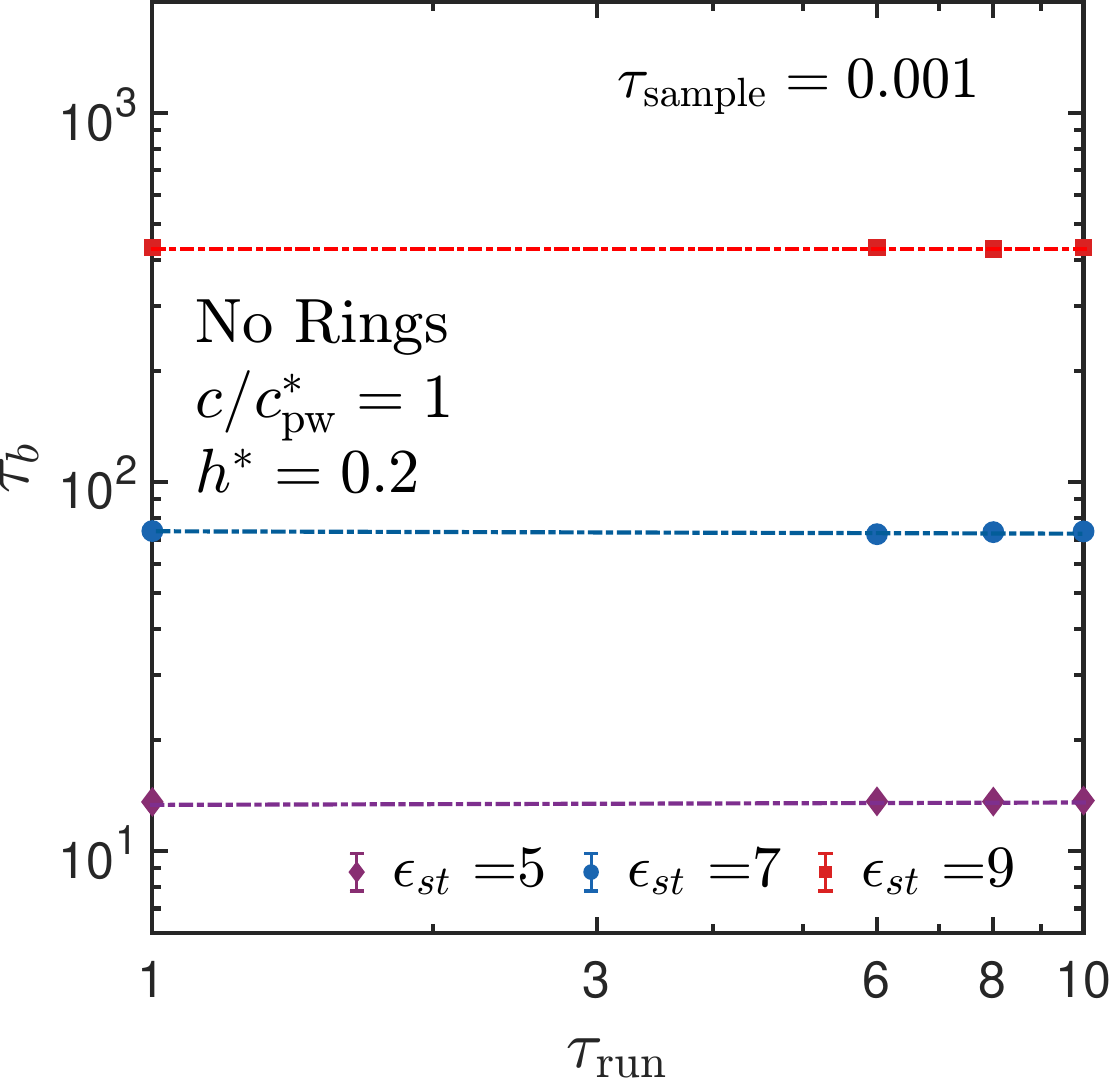} \\
(a)  & (b) \\[10pt]
\includegraphics[width=8.5cm]{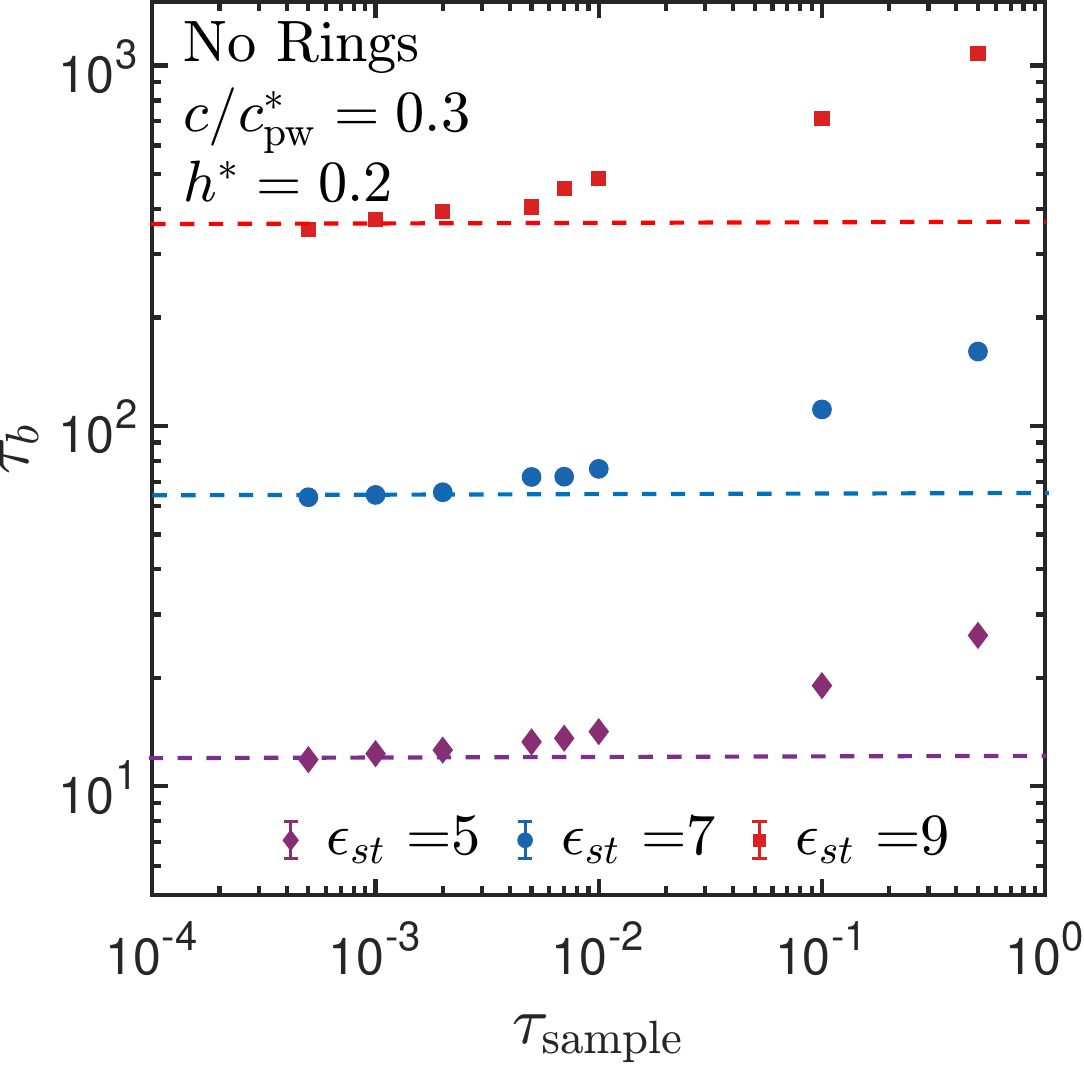} &
\includegraphics[width=8.5cm]{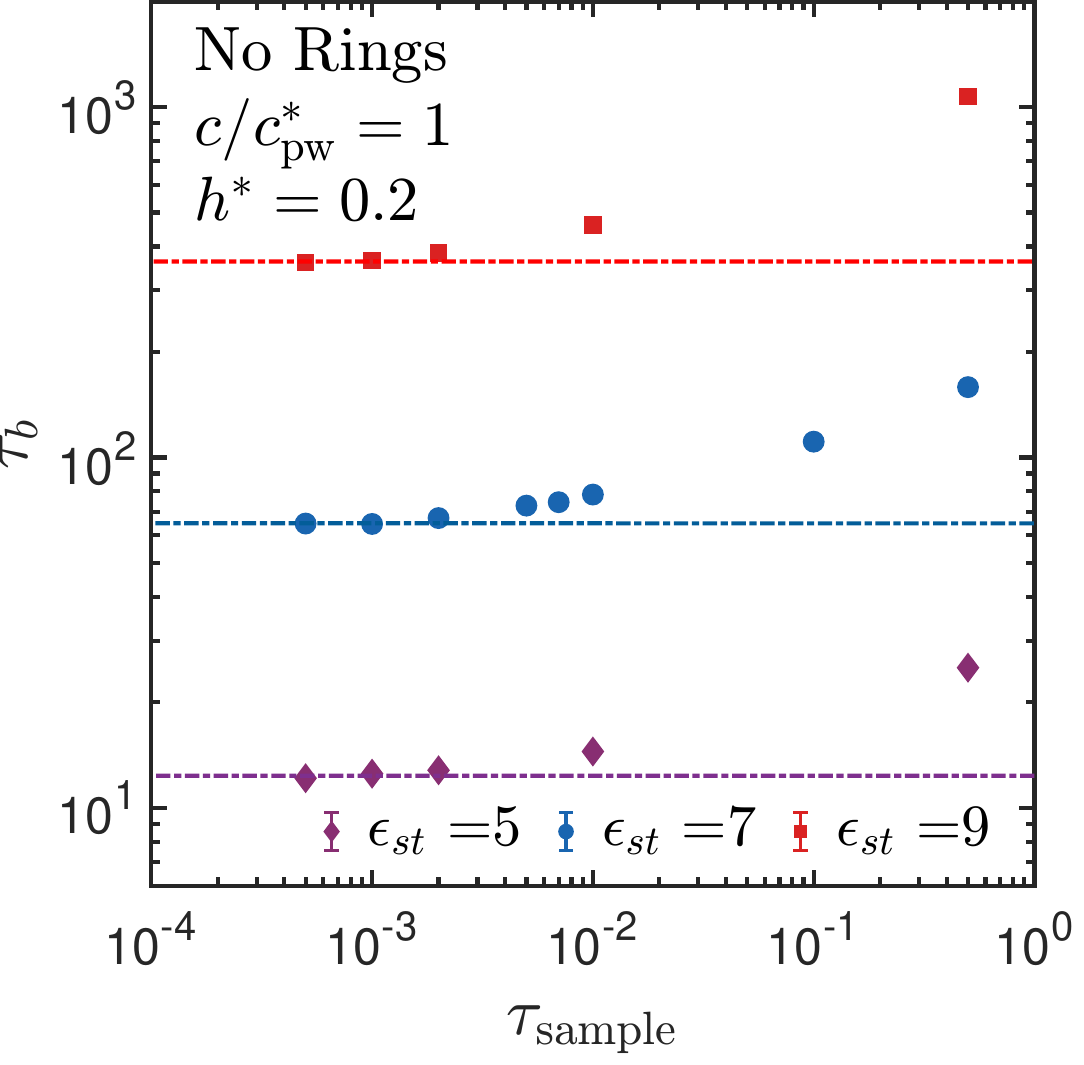} \\
(c)  & (d) 
\end{tabular}
\end{center}
    \vspace{-10pt}
\caption{Dependence of the bond breakage time $\tau_b$ on the total production run time $\tau_{\mathrm{run}}$ in (a) the dilute regime ($c/c_{\mathrm{pw}}^* = 0.3$) and (b) the semidilute regime ($c/c_{\mathrm{pw}}^* = 1$) for $\epsilon_{st}=5, 7,$ and $9$, with nondimensional sampling interval $\tau_{\mathrm{sample}} = 0.001$. Dependence of $\tau_b$ on the nondimensional sampling interval $\tau_{\mathrm{sample}}$ in (c) the dilute regime ($c/c_{\mathrm{pw}}^* = 0.3$) and (d) the semidilute regime ($c/c_{\mathrm{pw}}^* = 1$) for $\epsilon_{st}=5, 7,$ and $9$.}
\label{fig3}
    \vspace{-10pt}
\end{figure*}

We first examine the robustness of the estimation of $\tau_b$ with respect to simulation parameters. The bond breakage time is found to be independent of the total production run time $\tau_{\mathrm{run}}$ (see \fref{fig3}(a) and (b)) in both dilute ($c/c_{\mathrm{pw}}^* = 0.3$) and semidilute regimes ($c/c_{\mathrm{pw}}^* = 1$), confirming that the simulations are sufficiently long to obtain reliable bond statistics. In contrast, $\tau_b$ is sensitive to the sampling interval $\tau_{\mathrm{sample}}$, which determines how frequently configurations are stored during the production run (see \fref{fig3}(c) and (d)). Because $\tau_b$ is extracted from stored configurations during post-processing, an excessively large sampling interval may miss short-lived bonding events and therefore lead to an overestimation of $\tau_b$. We find that a nondimensional sampling interval $\tau_{\mathrm{sample}} \lesssim 0.002$ yields converged estimates of $\tau_b$, while smaller sampling intervals provide negligible improvement at the cost of significantly increased data storage.

\begin{figure*}[t]
\begin{center}
\begin{tabular}{ccc}
\includegraphics[width=5.45cm]{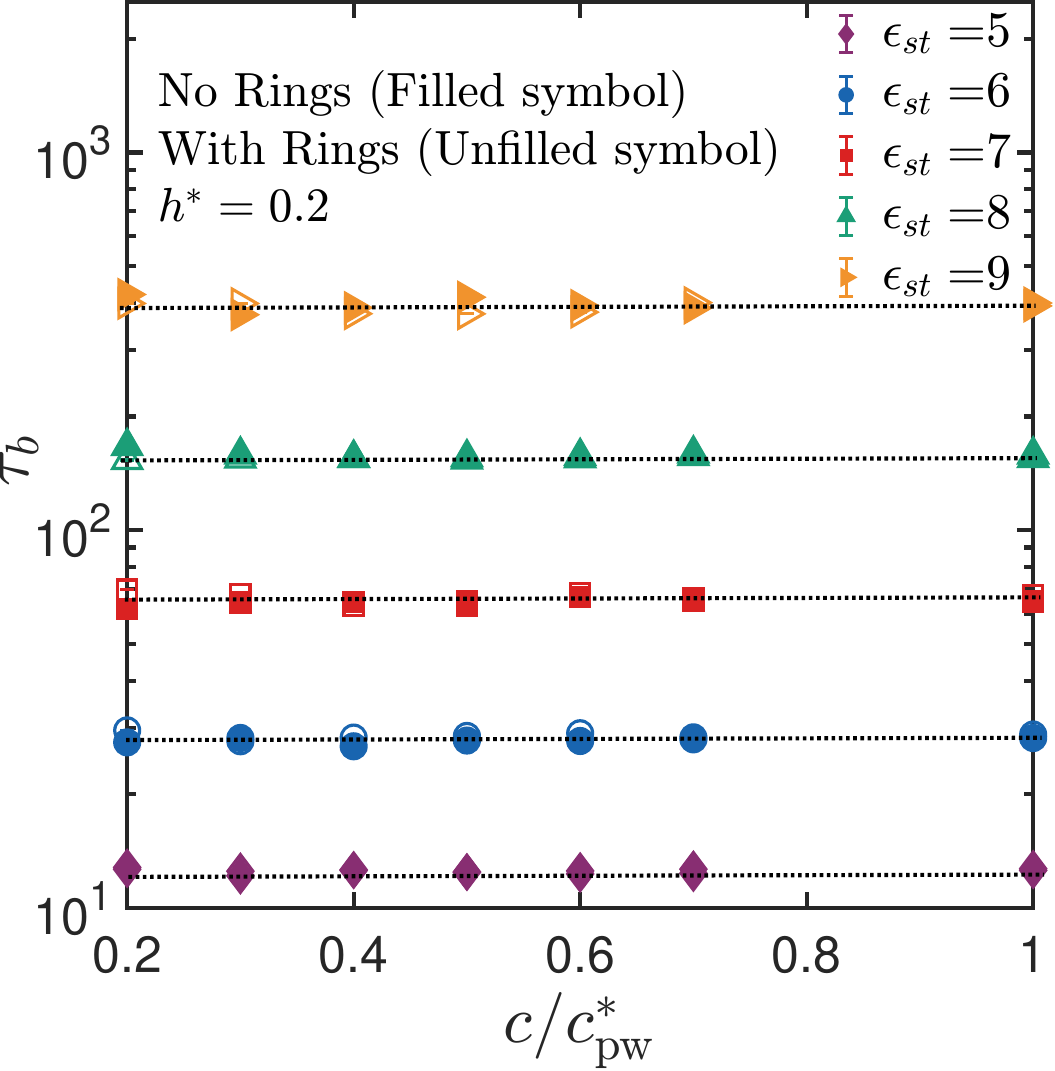} &
\includegraphics[width=5.6cm]{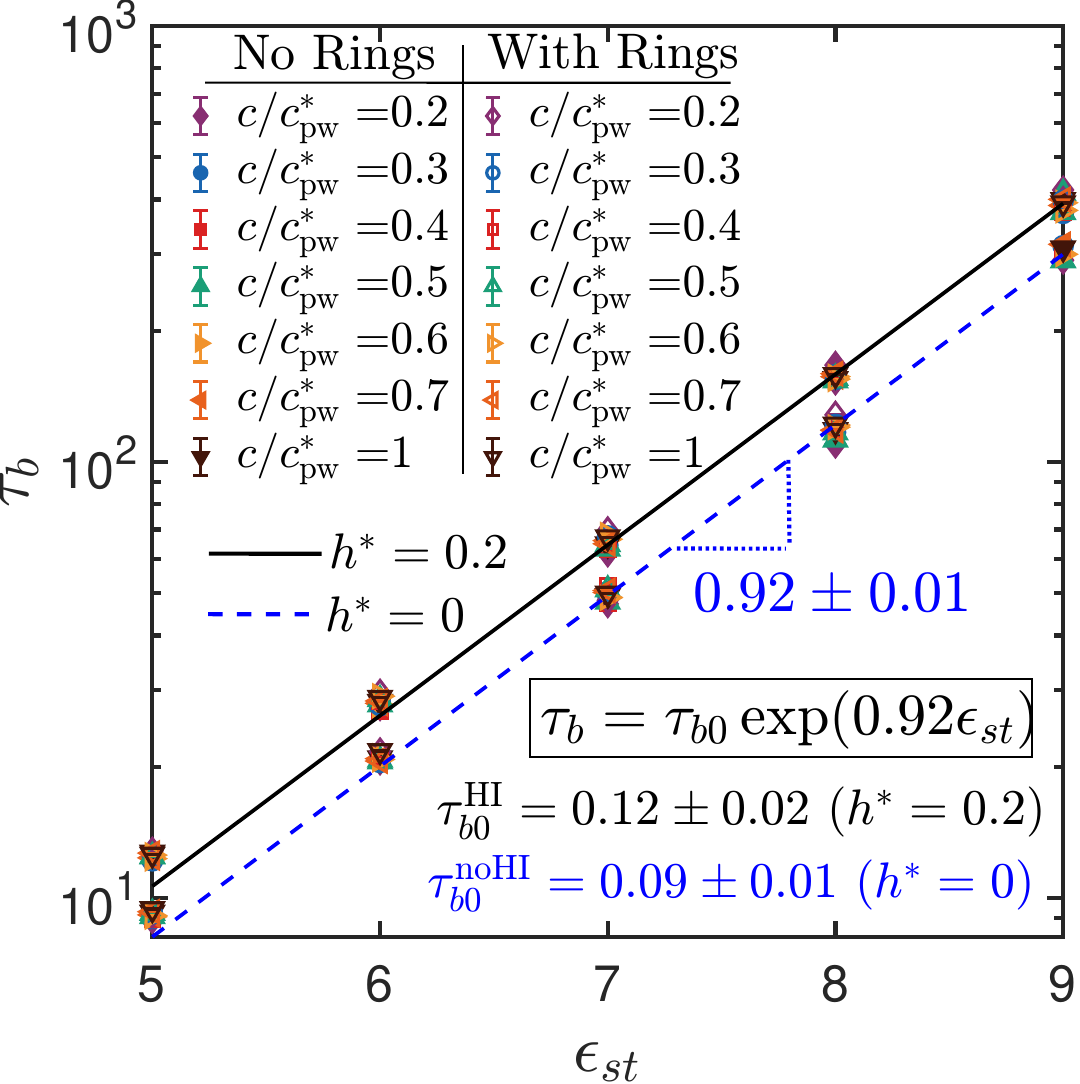} &
\includegraphics[width=5.7cm]{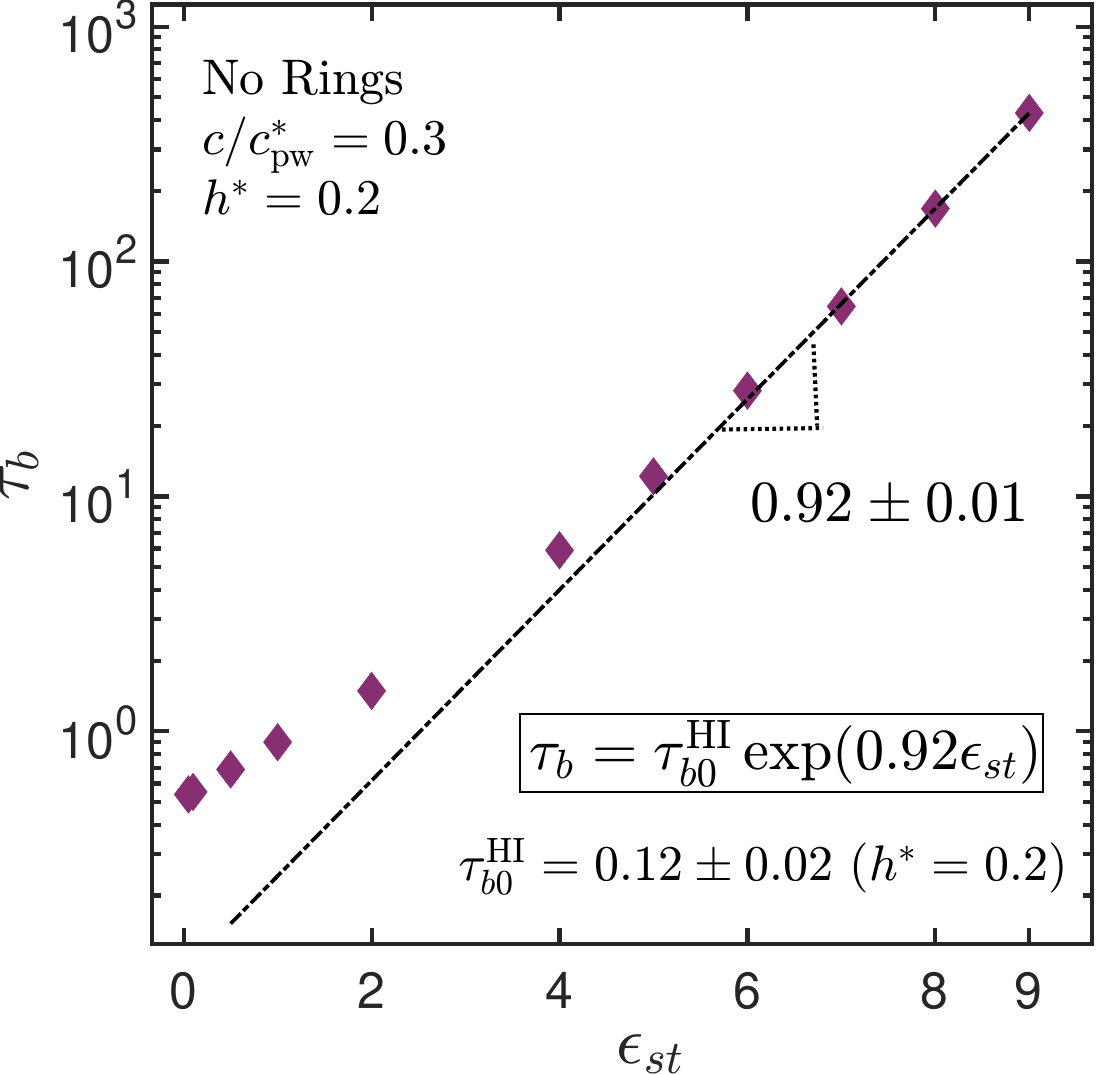} \\[-5pt] 
(a)  & (b) & (c)\\
\end{tabular}
\vspace{-10pt}
\end{center}
\caption{(a) Effect of bond breakage time $\tau_b$ on the concentration of persistent worm ($c/c_{\mathrm{pw}}^*$), at different sticker energies ($\epsilon_{st}$). (b) Exponential dependence of bond breakage time $\tau_b$ on the sticker energy $\epsilon_{st}$ for various $c/c_{\mathrm{pw}}^*$, in the presence and absence of rings and hydrodynamic interactions. (c) Dependence of $\tau_b$ on $\epsilon_{st}$ for dilute wormlike micellar solutions ($c/c_{\mathrm{pw}}^* = 0.3$), in the absence of rings and in the presence of hydrodynamic interactions ($h^*=0.2$), illustrating that exponential scaling is observed only for $\epsilon_{st} > 5$. }.
\label{fig4}
    \vspace{-10pt}
\end{figure*}

Since the bond breakage time represents an average over the lifetimes of all sticker–sticker bonds in the system, it is effectively independent of micellar length and micellar concentration (see \fref{fig4}(a)). In contrast, $\tau_b$ exhibits a clear exponential dependence on the sticker interaction strength $\epsilon_{st}$ for sufficiently large sticker energies ($\epsilon_{st} \ge 5$), indicating that bond dissociation is governed primarily by the local sticker–sticker interaction (see \fref{fig4}(b)). This behaviour is consistent with earlier studies \cite{Stukalin2013,Mordvinkin2021,Katashima2022}.

Since bond breakage time exhibits weak dependence on concentration, it is plotted in \fref{fig4}(c) as a function of sticker energy for a dilute wormlike micellar solution ($c/c_{\mathrm{pw}}^* = 0.3$) to highlight its behavior at lower sticker energies. The exponential scaling of $\tau_b$ emerges only at sufficiently large interaction energies ($\epsilon_{st} \ge 5$), where bond lifetimes become increasingly controlled by the sticker interaction energy. At lower sticker strengths, bonded configurations are short-lived, leading to deviations from exponential (Arrhenius-like) behaviour.

\begin{figure*}[t]
\begin{center}
\begin{tabular}{cc}
\includegraphics[width=8cm]{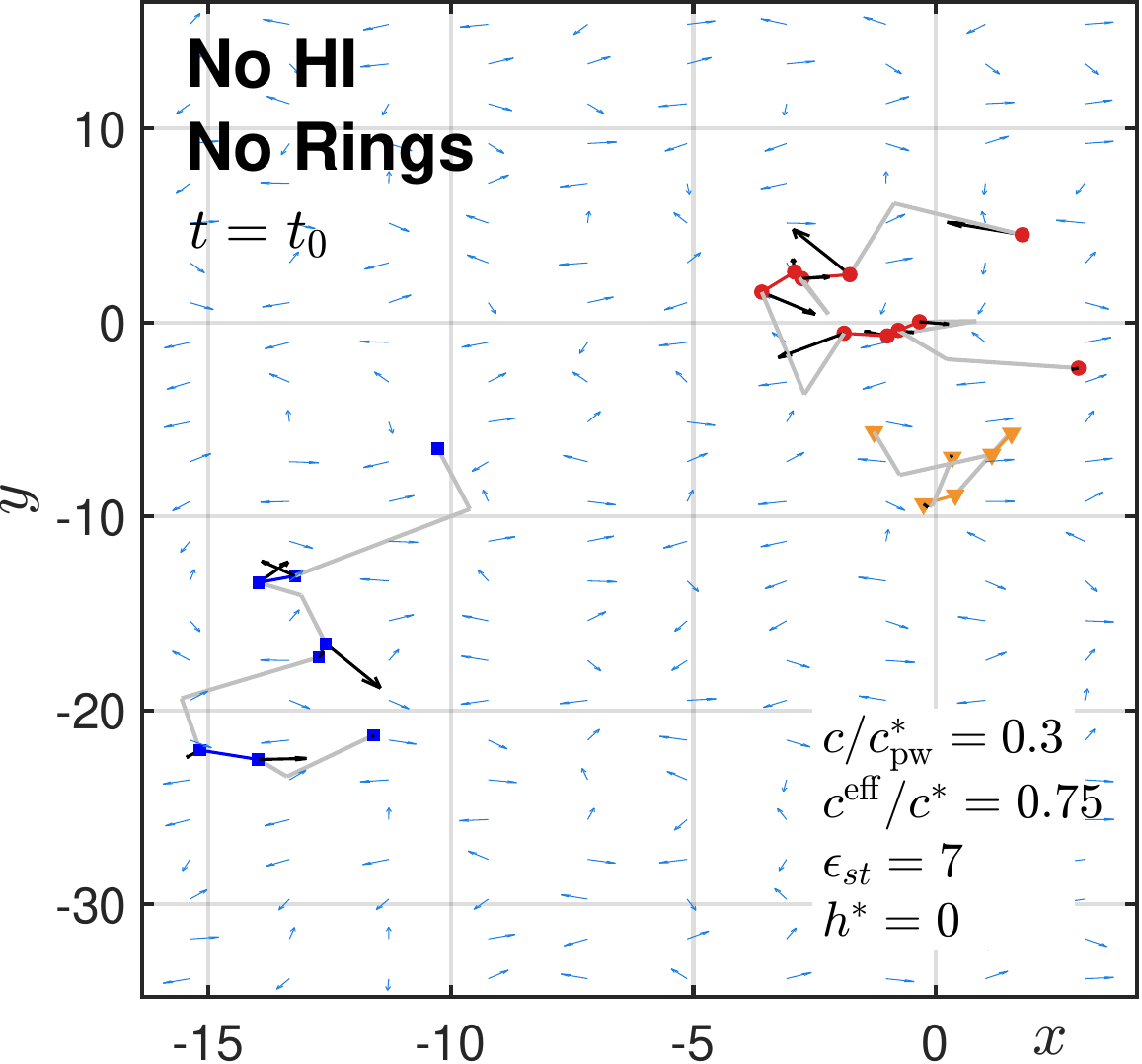} &
\includegraphics[width=8cm]{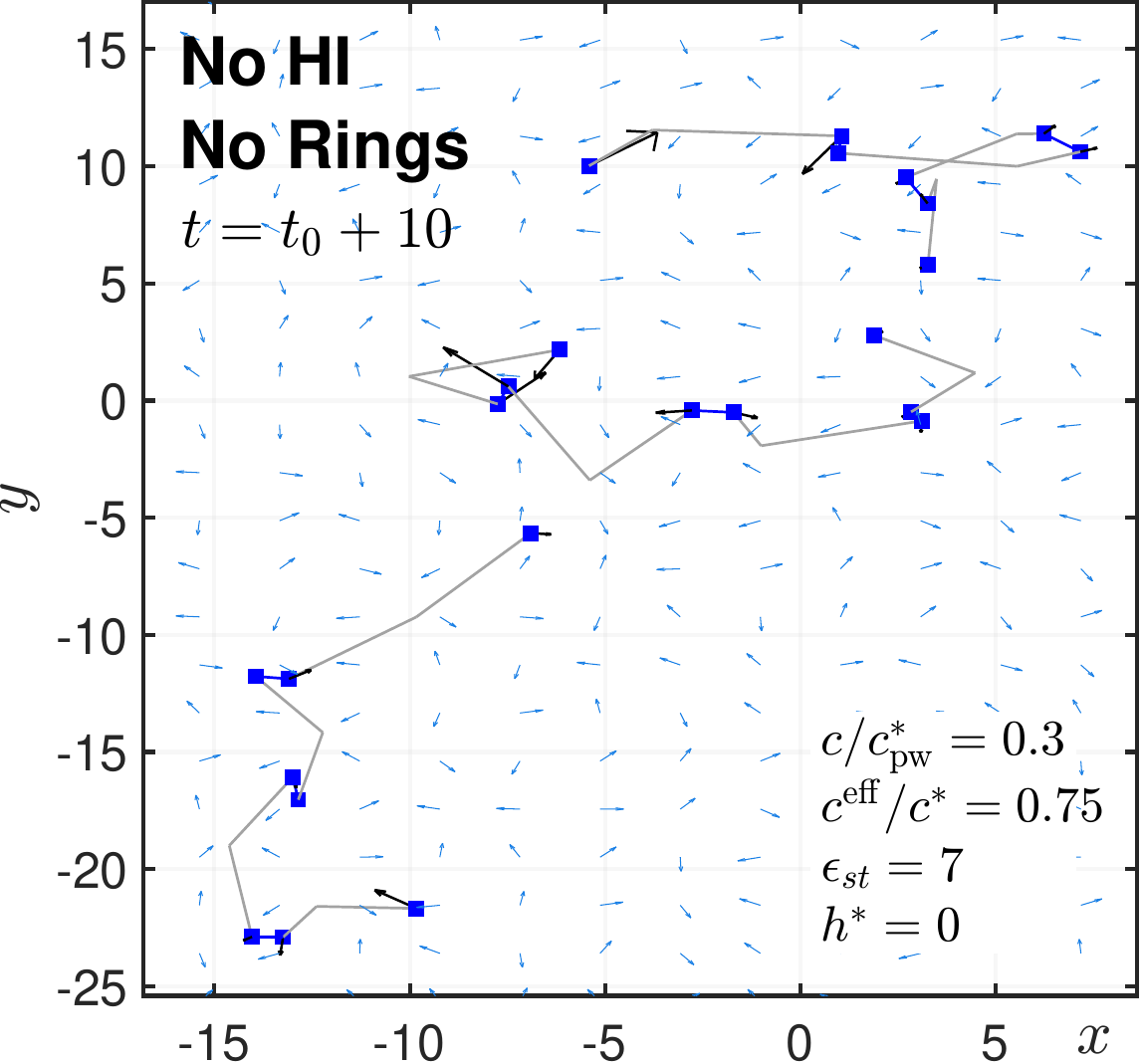} \\
(a)  & (b) \\[10pt]
\includegraphics[width=8cm]{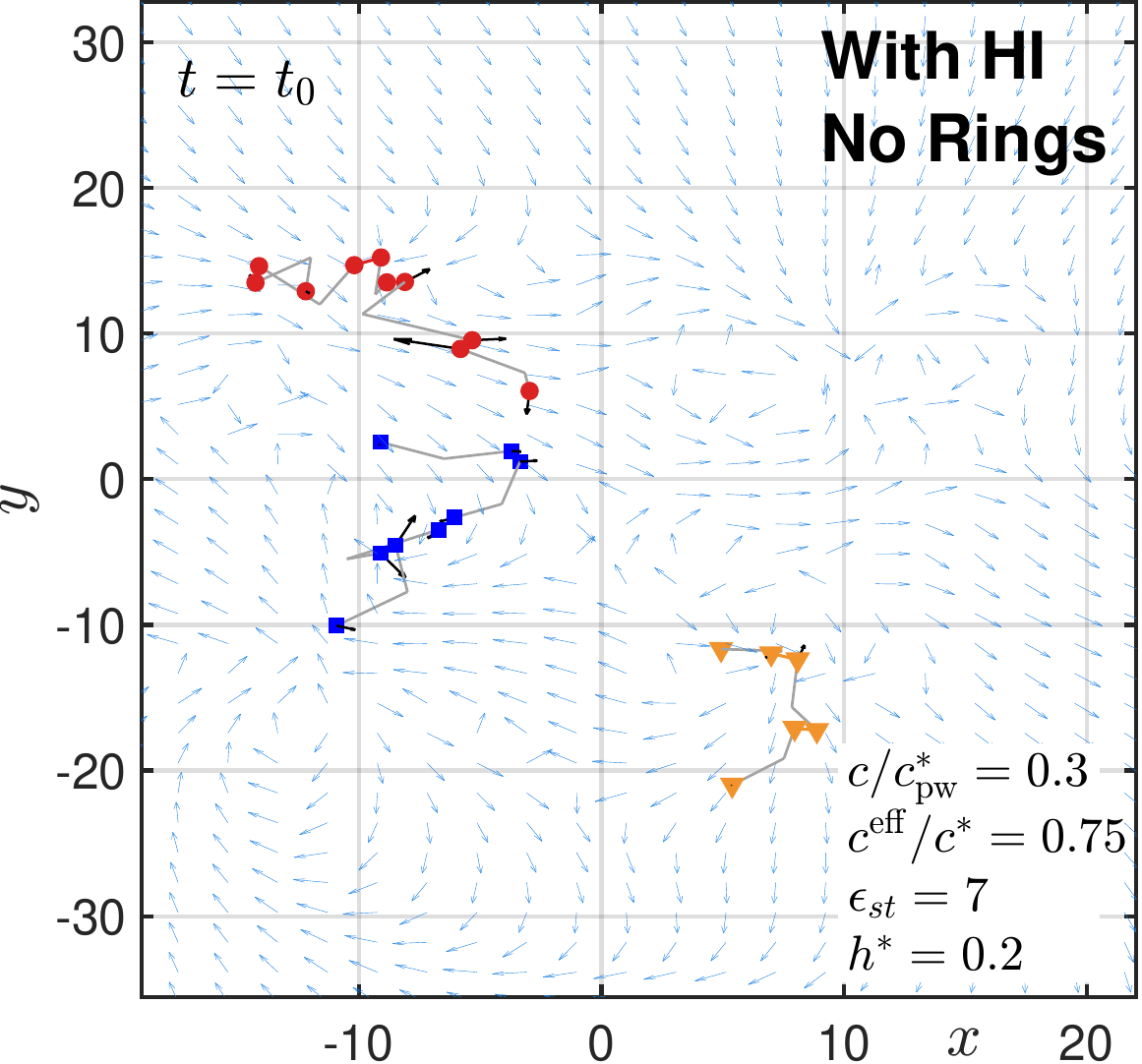} &
\includegraphics[width=8cm]{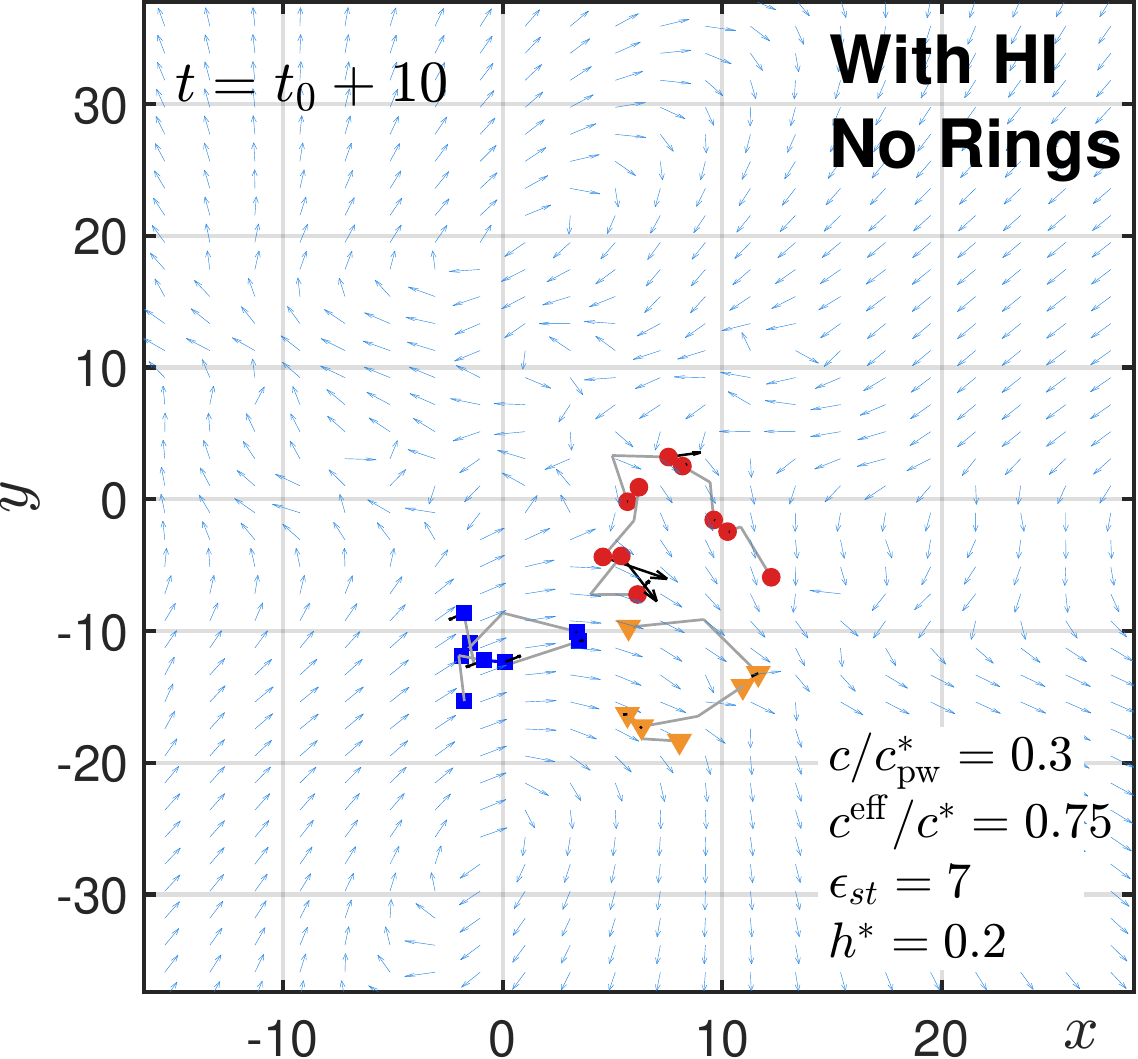} \\
(c)  & (d) 
\end{tabular}
\end{center}
    \vspace{-10pt}
\caption{Instantaneous velocity fields around bonded stickers illustrating the role of hydrodynamic interactions in the absence of rings. Panels (a) and (b) show the velocity field in the absence of hydrodynamic interactions at dimensionless times $t_0$ and $t_0+10$, respectively, where the blue arrows represent a schematic random solvent velocity field illustrating thermal fluctuations. Panels (c) and (d) show the corresponding disturbance velocity field in the presence of hydrodynamic interactions at the same times. The black arrows denote the net forces acting on sticky monomers. Orange (triangle), blue (square), and red (circle) colored stickers correspond to micelles containing $m_{\mathrm{pw,lin}}^{\mathrm L}=3,4$, and $5$ persistent worms, respectively.}
\label{fig5}
    \vspace{-10pt}
\end{figure*}

Most notably, hydrodynamic interactions have a pronounced effect, significantly increasing the bond breakage time $\tau_b$ by stabilizing bonded stickers through long-ranged hydrodynamic coupling (see \fref{fig4}(a)). This effect is further illustrated in \fref{fig5}, which shows instantaneous velocity fields around monomers in the absence and presence of hydrodynamic interactions at different time snapshots. In the presence of hydrodynamic interactions, the nondimensional disturbance velocity at the $i$-th grid point is computed as
$\bm{v}_i^{\mathrm{HI}} = (1/4)\sum_{\nu=1}^{N^{\mathrm{T}}} \bm{D}_{i\nu} \cdot \bm{F}_{\nu}$, where $\bm{D}_{i\nu}$ is the nondimensional diffusion tensor corresponding to a hypothetical bead at the $i$-th grid point, and $\bm{F}_{\nu}$ is the net nondimensional force acting on bead $\nu$ (see \sref{sec:BD}). In the absence of hydrodynamic interactions, the nondimensional velocity of bead $\nu$ reduces to $\bm{v}_{\nu}^{\mathrm{no\text{-}HI}} = (1/4) \, \bm{F}_{\nu}$. Note that in the free-draining approximation, forces acting on beads do not generate fluid disturbances in the surrounding solvent, and therefore the velocity at grid points away from the beads is zero.

To illustrate solvent motion in the absence of hydrodynamic interactions, a schematic random velocity field is shown in \fref{fig5}(a) and (b), representing thermal fluctuations of the solvent. In an explicit solvent description, the velocity of solvent particles at thermal equilibrium follows the Maxwellian distribution about the local fluid velocity \cite{Bird1987}. From the equipartition theorem, each Cartesian component satisfies $(1/2) m_s \langle v_\alpha^2 \rangle = (1/2) k_{\text{B}} T$ which yields a Gaussian distribution with variance $\langle v_\alpha^2 \rangle = k_{\text{B}} T/m_s$, where $m_s$ is the solvent particle mass.

In the present work, however, the solvent is treated implicitly within an overdamped Brownian dynamics framework, and solvent velocities are not explicitly defined. The velocity field shown in \fref{fig5}(a) and (b) is therefore purely illustrative and does not correspond to actual solvent dynamics. The velocity components are sampled from a Gaussian distribution with zero mean and are subsequently normalized by their standard deviation, $\sqrt{\langle v_\alpha^2 \rangle}$, such that the resulting nondimensional distribution has unit variance, $\langle v_\alpha^{*2} \rangle = 1$. This normalization is introduced solely for visualization clarity and does not influence the bead dynamics. In contrast, in the presence of hydrodynamic interactions (see \fref{fig5}(c) and (d)), forces acting on beads generate disturbance flows that extend into the surrounding solvent through hydrodynamic coupling.

The net forces acting on the sticky monomers are represented by black arrows. To clearly visualize the velocity field around bonded stickers, only a subset of micelles in the wormlike micellar solution is shown in \fref{fig5}. The orange (triangles), blue (squares), and red (circles) colored stickers correspond to micelles containing $m_{\mathrm{pw,lin}}^{\mathrm{L}} = 3, 4$, and $5$ persistent worms, respectively.

In the absence of hydrodynamic interactions (see \fref{fig5}(a) and (b)), the velocity field consists only of random solvent fluctuations, and no disturbance flow is generated by forces on the bead. Consequently, the bonded configuration is less stable and the stickers detach more readily. In contrast, in the presence of hydrodynamic interactions (\fref{fig5}(c) and (d)), the flow field exhibits correlated motion around bonded stickers at dimensionless times $t_0$ and $t_0 + 10$, indicating that hydrodynamic interactions generate local flow patterns that advect bonded partners towards each other. These correlated flows effectively stabilize the bonded state and delay bond breakage, thereby increasing the average bond lifetime. This effect is also evident from the evolution of micellar configurations. In the absence of hydrodynamic interactions, the micelle containing $m_{\mathrm{pw,lin}}^{\mathrm{L}} = 5$ persistent worms (red circles) breaks between the snapshots at $t_0$ and $t_0 + 10$ (see \fref{fig5}(a) and (b)). The resulting fragments are subsequently rearranged to form three micelles, each with $m_{\mathrm{pw,lin}}^{\mathrm{L}} = 4$ (blue squares), such that the total number of persistent worms in the solution is conserved. In contrast, in the presence of hydrodynamic interactions, the bonded configuration remains intact over the same time interval.

We also observe from \fref{fig1} and \fref{fig4}(b) that micellar topology, i.e., the presence (labelled as ``WR'') or absence (labelled as ``NR'') of rings in wormlike micellar solutions, does not significantly affect the bond breakage time.

\subsubsection{\label{sec:recombtime} Recombination timescale}

In this section, we extract the timescale associated with the \textit{first recombination event} of a sticker that has just become unbound. When a sticker unbinds due to scission at time $t_1$, it remains unbound for a certain duration before recombining for the first time at time $t_2$. The interval $\tau_{\mathrm{rec}} = t_2 - t_1$, is defined as the ``recombination time". Upon scission, the sticker may recombine with the same partner from which it detached, a process termed ``self-recombination", with the corresponding timescale referred to as the ``self-recombination time ($\tau_s$)". Alternatively, the sticker may diffuse away and recombine with a different partner; this timescale is referred to as the ``non-self recombination time ($\tau_{ns}$)".

\begin{figure*}[t]
\begin{center}
\begin{tabular}{cc}
\includegraphics[width=8.45cm]{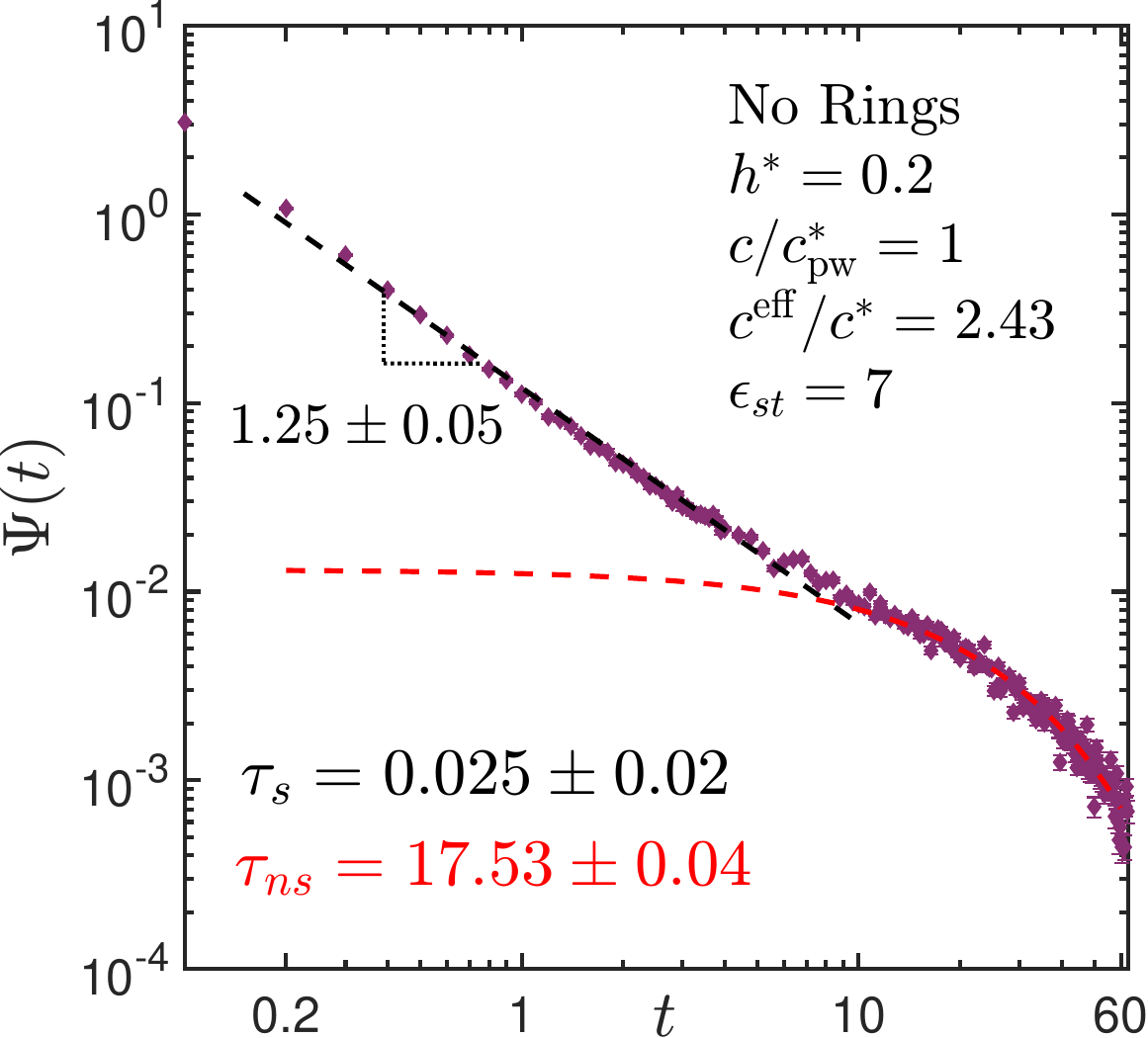} &
\includegraphics[width=8.5cm]{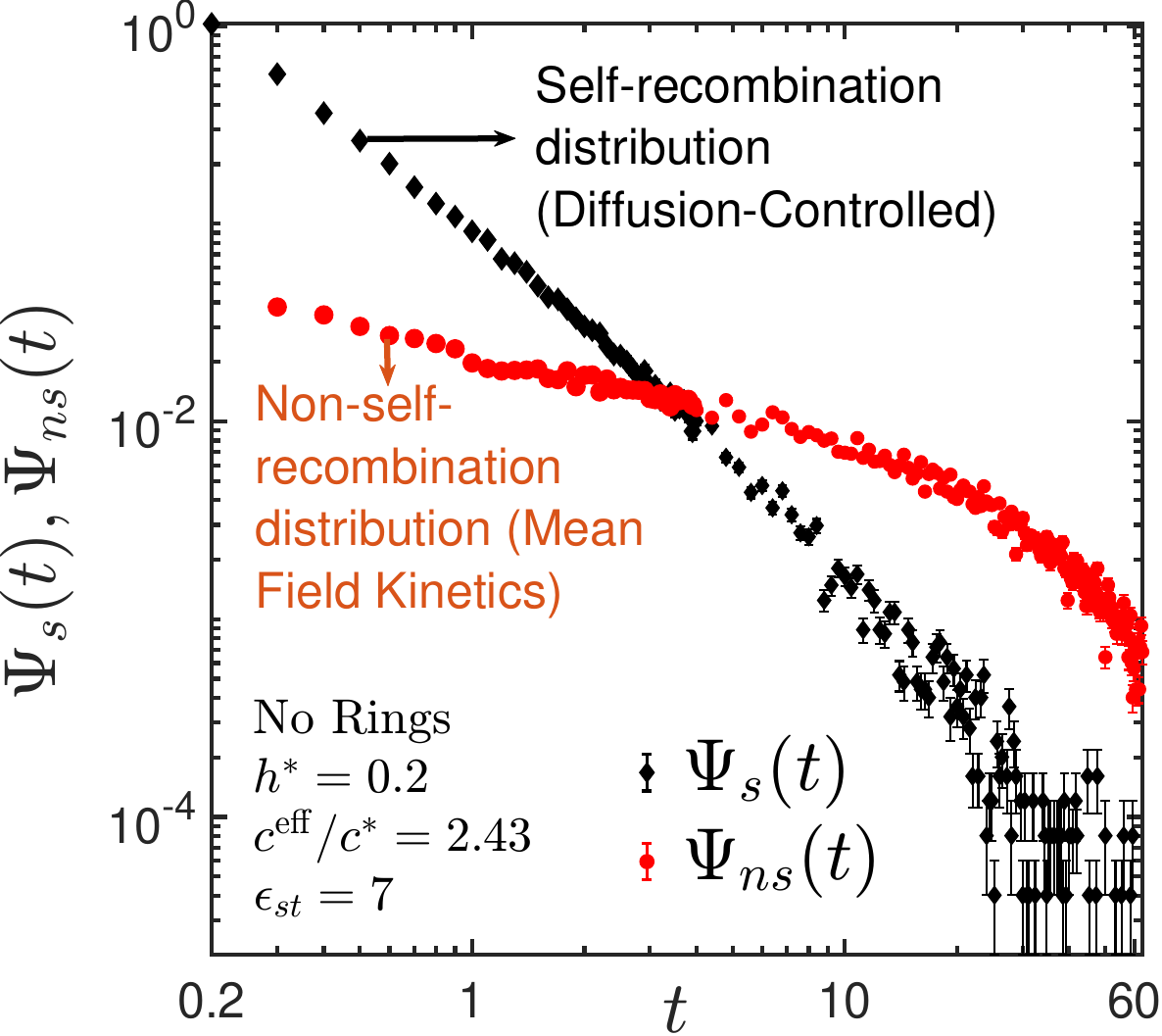} \\[-5pt] 
(a)  & (b) \\
\end{tabular}
\vspace{-10pt}
\end{center}  
\caption{(a) Recombination probability distribution $\Psi(t)$ for wormlike micellar solutions at $c^{\mathrm{eff}}/c^* = 2.43$, $c/c_{\mathrm{pw}}^* = 1$, and $\epsilon_{\mathrm{st}} = 7$ in the presence of hydrodynamic interactions ($h^* = 0.2$). At short times ($t \lesssim 10$), self-recombination dominates and the distribution exhibits a power-law decay, whereas at longer times non-self recombination becomes dominant and the distribution follows an exponential decay. (b) The self-recombination probability distribution $\Psi_{s}(t)$ and the non-self recombination probability distribution $\Psi_{ns}(t)$ are shown separately. The self recombination (diffusion-controlled kinetics)  exhibits a power-law decay, while the  non-self recombination (mean-field kinetics) follows an exponential decay throughout the simulation runtime.} 
\label{fig6}
\end{figure*}

From the simulation trajectories, we collect a set of recombination times $\left\{ \tau_{\mathrm{rec}} \right\}$ for individual stickers, where each value corresponds to the elapsed time between a scission event of a previously bound sticker at time $t_1$ and its subsequent first recombination at time $t_2$. A histogram of these recombination times is then constructed to obtain the probability density $\Psi(t)dt$, which represents the probability that a free sticker created at time $t=0$ undergoes its first recombination within the interval $[t,t+dt]$.

Previous studies \cite{Ryckaert2009,Ryckaert2006,Koide2023,Shaughnessy1995,Padding2004} have shown that sticker recombination proceeds through two distinct kinetic regimes. At short times, self-recombination dominates via diffusion-controlled (DC) kinetics, and the probability distribution exhibits an algebraic decay:
$\Psi(t) \approx \left(1/\tau_s\right)\left(\tau_s/t\right)^{5/4}, \quad \tau_s < t < \tau_{ns},$
where $\tau_s$ and $\tau_{ns}$ are the self-recombination and non-self recombination timescales, respectively. At longer times, the unbound sticker loses memory of its original partner and recombines with a new partner. This regime, referred to as mean-field (MF) kinetics, is characterized by an exponential probability distribution:
$\Psi(t) \approx \left(1/\tau_{ns}\right)\exp\left(-t/\tau_{ns}\right), \quad t > \tau_{ns}$. The relative magnitudes of $\tau_s$ and $\tau_{ns}$ determine the dominant recombination pathway. When $\tau_s \ll \tau_{ns}$, self-recombination occurs before the sticker diffuses far enough to encounter a new partner. Conversely, when $\tau_{ns} \ll \tau_s$, recombination with a new partner becomes the dominant mechanism.

\begin{figure*}[tbph]
\begin{center}
\begin{tabular}{cc}
\includegraphics[width=8.5cm]{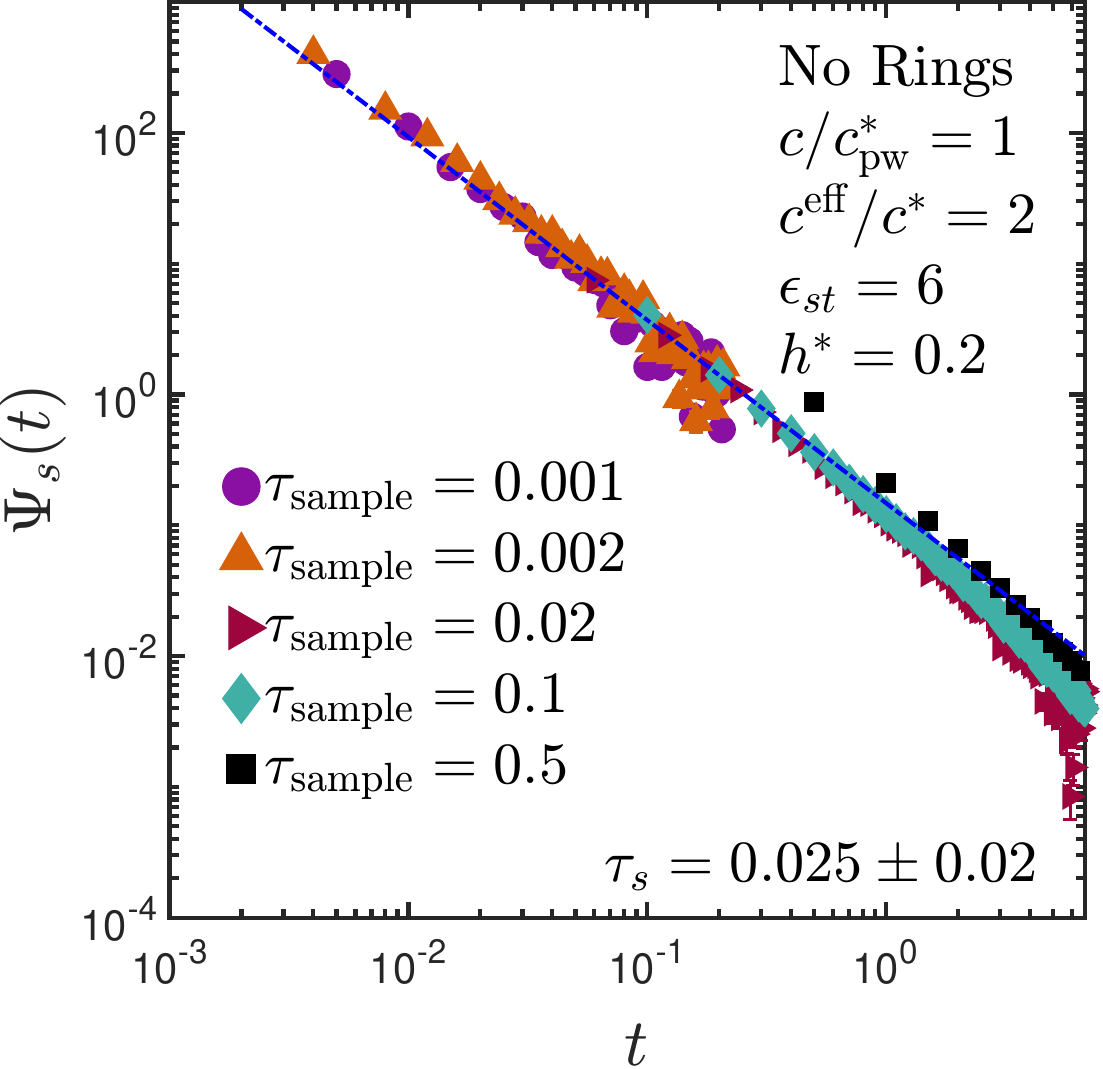} &
\includegraphics[width=8.5cm]{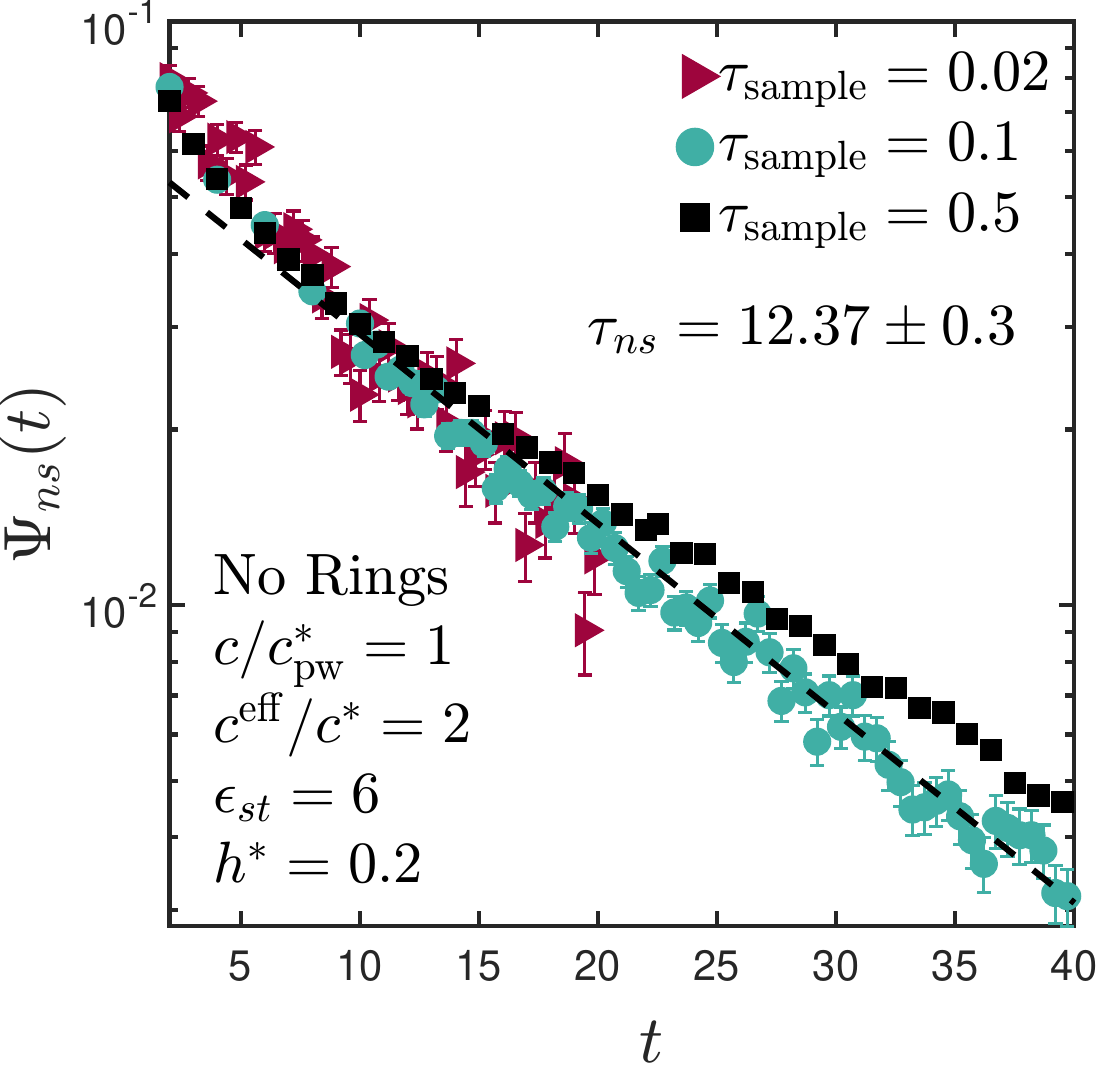} \\[-5pt] 
(a)  & (b) \\
\end{tabular}
\vspace{-10pt}
\end{center}  
\caption{Dependence of recombination probability distributions on the dimensionless sampling interval $\tau_{\text{sample}}$ for wormlike micellar solutions at $c/c_{\mathrm{pw}}^* = 1$, $c^{\mathrm{eff}}/c^* = 2$, and $\epsilon_{st} = 6$, in the presence of hydrodynamic interactions ($h^* = 0.2$) and in the absence of rings. (a) Self recombination probability distribution $\Psi_{s}(t)$ for different sampling intervals, plotted on a log–log scale. For $\tau_{\text{sample}} \le 0.1$, the distributions collapse onto a single curve, indicating that the measured probability distribution is independent of the sampling interval. (b) Non-self recombination probability distribution $\Psi_{ns}(t)$ for different sampling intervals, plotted on a semi-log scale. For large sampling intervals ($\tau_{\text{sample}} \ge 0.5$), short-lived recombination events are missed, which distorts the measured distribution and leads to an overestimation of recombination times.}
\label{fig7}
    \vspace{-10pt}
\end{figure*}

A representative illustration of the two kinetic regimes is presented in \fref{fig6}(a) for $c^{\mathrm{eff}}/c^* = 2.43$ and $\epsilon_{\mathrm{st}} = 7$. At short times ($t \lesssim 10$), the distribution exhibits a power-law decay corresponding to diffusion-controlled self-recombination. At longer times, a clear crossover to an exponential decay is observed, indicating the onset of mean-field recombination kinetics.

\begin{figure*}[t]
\begin{center}
\begin{tabular}{cc}
\includegraphics[width=8.35cm]{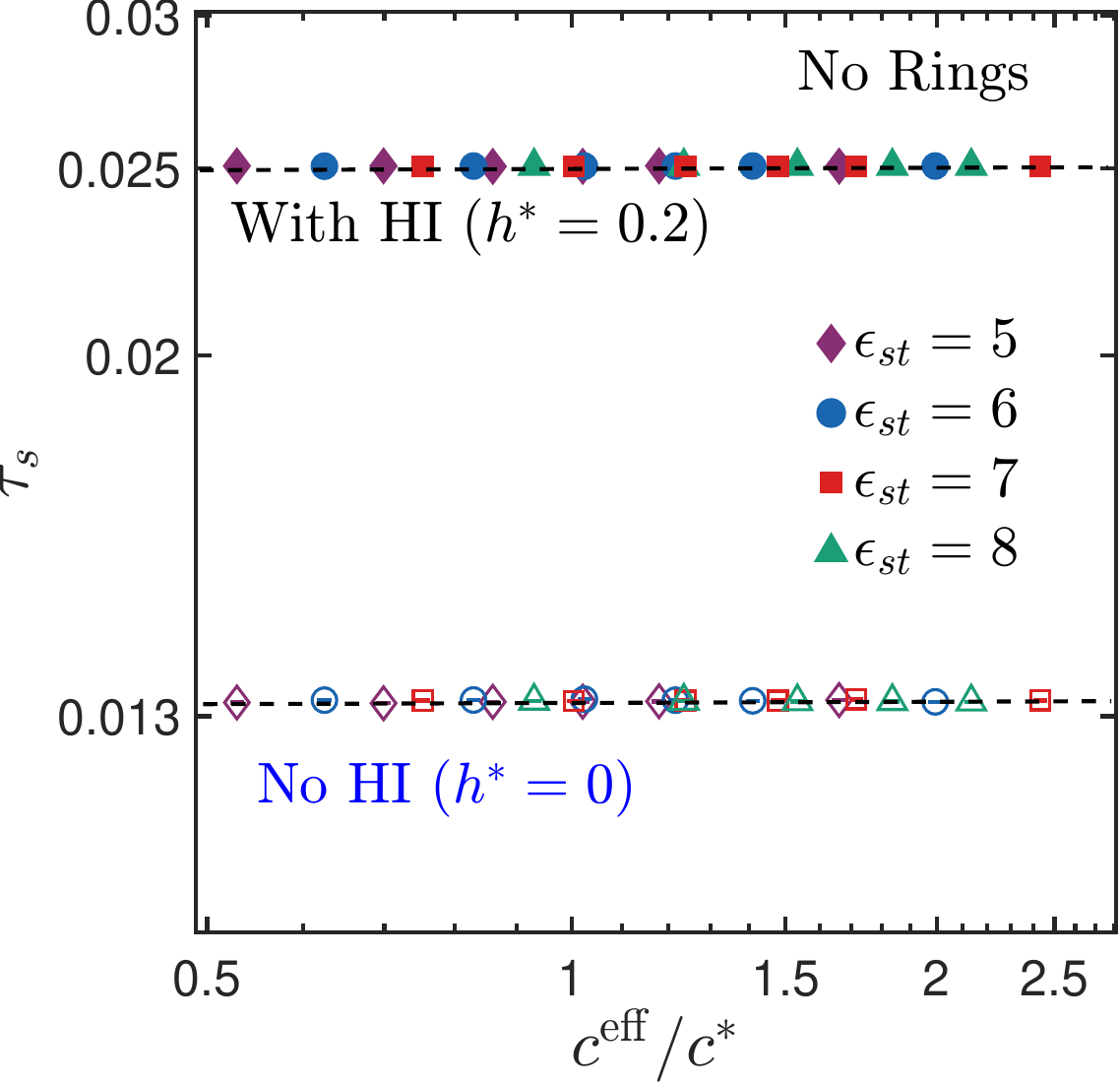} &
\includegraphics[width=8.5cm]{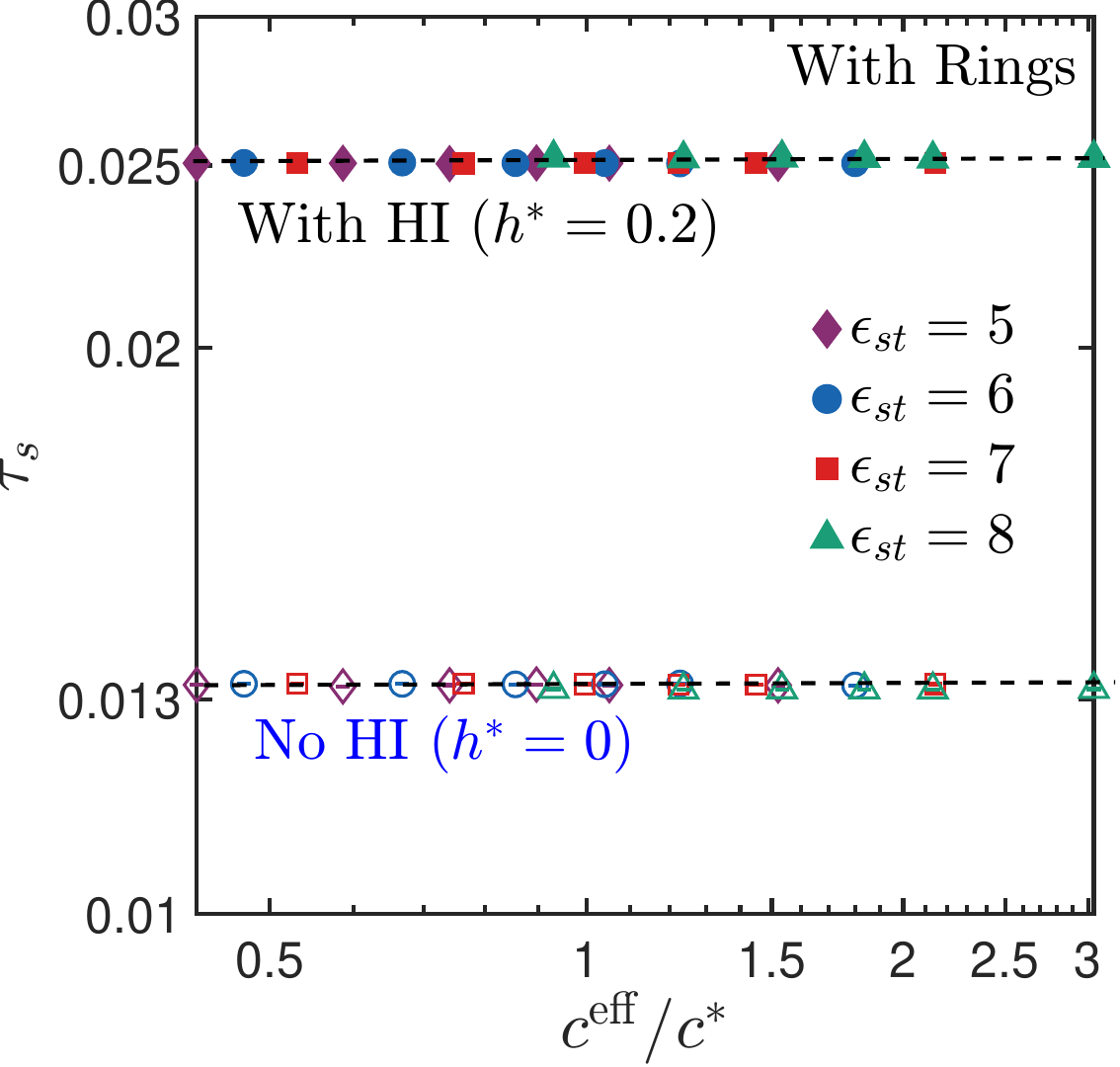} \\[-5pt] 
(a)  & (b) \\
\end{tabular}
\vspace{-10pt}
\end{center}
\caption{Self-recombination timescales as a function of wormlike micelle concentration, shown for simulations performed (a) without rings and (b) with rings, in both the absence and presence of hydrodynamic interactions.}\label{fig8}
\vspace{-10pt}
\end{figure*}

To illustrate the individual contributions of diffusion-controlled and mean-field recombination over the total simulation runtime (given by $\tau_{\text{run}}$, see \sref{sec:simdet}), these distributions are plotted separately in \fref{fig6}(b). For the diffusion-controlled regime, we collect recombination times $\left\{ \tau_{\mathrm{rec}} \right\}$ corresponding to events in which a sticker recombines with the same partner from which it detached following a scission event. The corresponding self-recombination probability density function $\Psi_{s}(t)dt$ is then computed. Similarly, the mean-field recombination distribution is obtained by collecting recombination times $\left\{ \tau_{\mathrm{rec}} \right\}$ for events in which the sticker diffuses away after scission and recombines with a different partner, yielding the non-self recombiation probability density function $\Psi_{ns}(t)dt$. We find that $\Psi_{s}(t)$ follows a power-law behaviour throughout the simulation runtime ($\tau_{\text{run}}$), whereas $\Psi_{ns}(t)$ exhibits a purely exponential decay.

The nondimensional sampling interval $\tau_{\text{sample}}$, defined in \sref{sec:tbond}, also influences the measured recombination probability distributions. This effect is illustrated in \fref{fig7}(a) and (b). For $\tau_{\text{sample}} \le 0.1$, the probability distributions $\Psi_s(t)$ collapse onto a single curve, indicating that the statistics of self-recombination (diffusion-controlled kinetics) have converged. However, for larger sampling intervals (e.g., $\tau_{\text{sample}} = 0.5$) the distribution deviates from this master curve. Similarly, \fref{fig7}(b) shows the effect of the sampling interval on the non-self recombination (mean-field) probability distribution $\Psi_{ns}(t)$. For large values of $\tau_{\text{sample}}$, short-lived recombination events are not resolved, which distorts the measured distribution and leads to an overestimation of the recombination times. Based on these observations, we select a nondimensional sampling interval of $\tau_{\text{sample}} = 0.1$, which yields converged probability distributions while avoiding the need to store prohibitively large datasets. The self-recombination ($\tau_s$) and non-self recombination ($\tau_{ns}$) timescales reported below are therefore obtained using this optimized sampling interval.

From our simulations, the self-recombination time ($\tau_s$) is found to be independent of both the concentration of persistent worms and the sticker energy. This behaviour is expected at such short times, where the free sticker remains in the immediate vicinity of its original partner and the dynamics are governed primarily by local thermal fluctuations and the short-range relative diffusion of the two unbonded stickers, rather than by the wormlike micelle concentration or the sticker interaction strength. In contrast, hydrodynamic interactions significantly affect the self-recombination time, as hydrodynamic coupling slows the relative motion of nearby beads and delays recombination, as shown in \fref{fig8}. The presence (see \fref{fig8}(a)) or absence (see \fref{fig8}(b)) of rings in wormlike micellar solution does not significantly affect the self-recombination time.

\begin{figure*}[t]
\begin{center}
\begin{tabular}{cc}
\includegraphics[width=8.35cm]{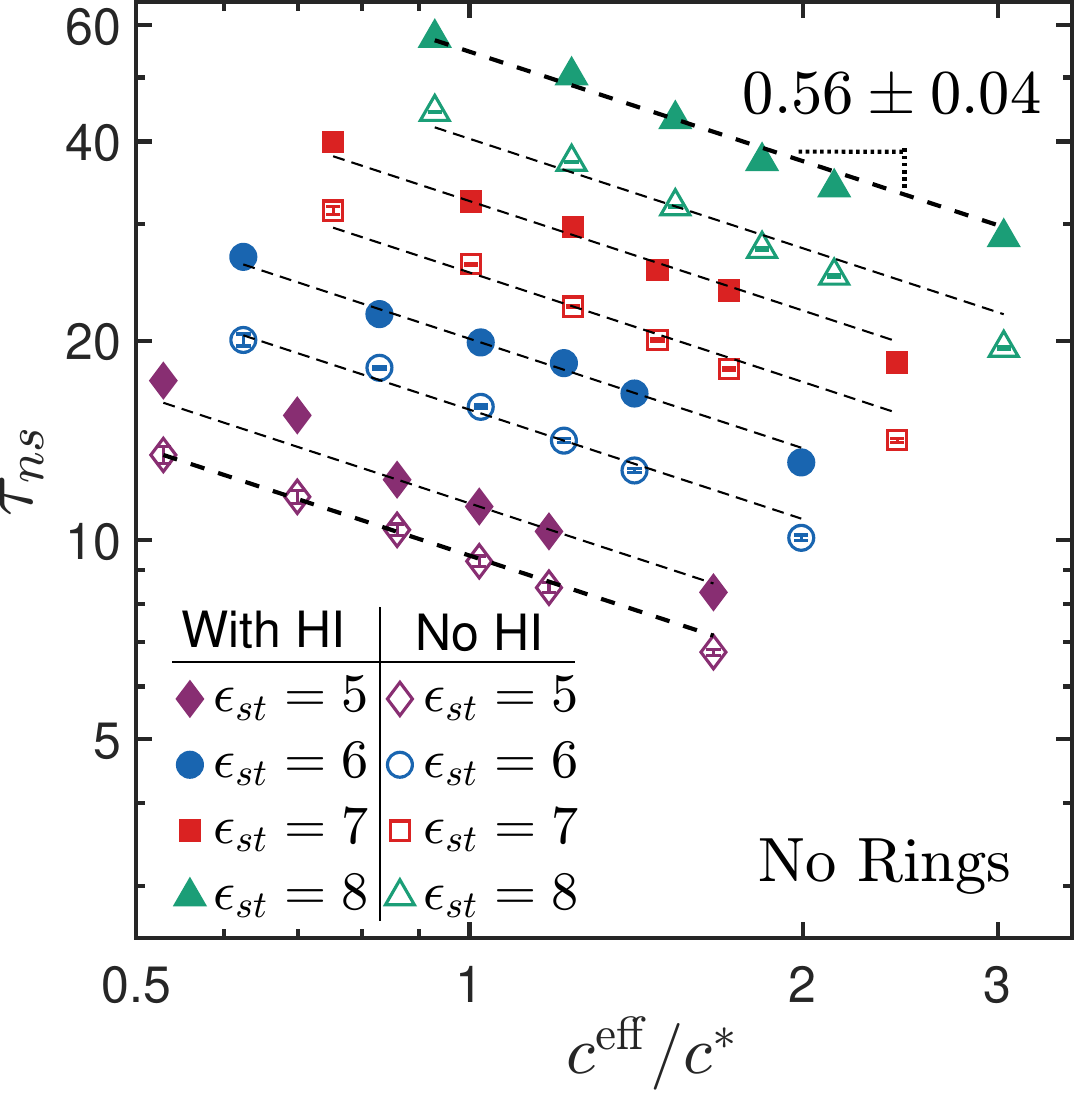} &
\includegraphics[width=8.5cm]{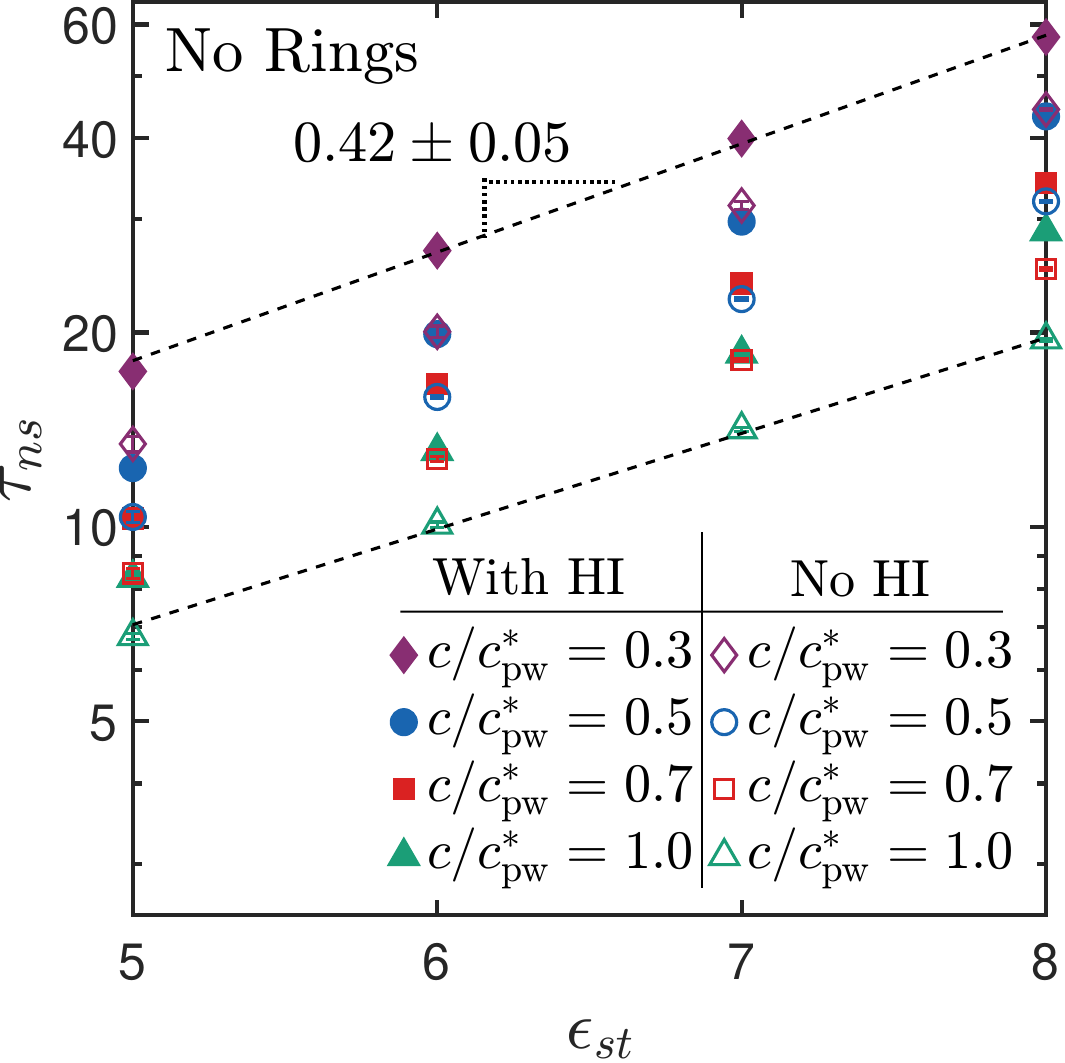} \\
(a)  & (b) \\[10pt]
\includegraphics[width=8.45cm]{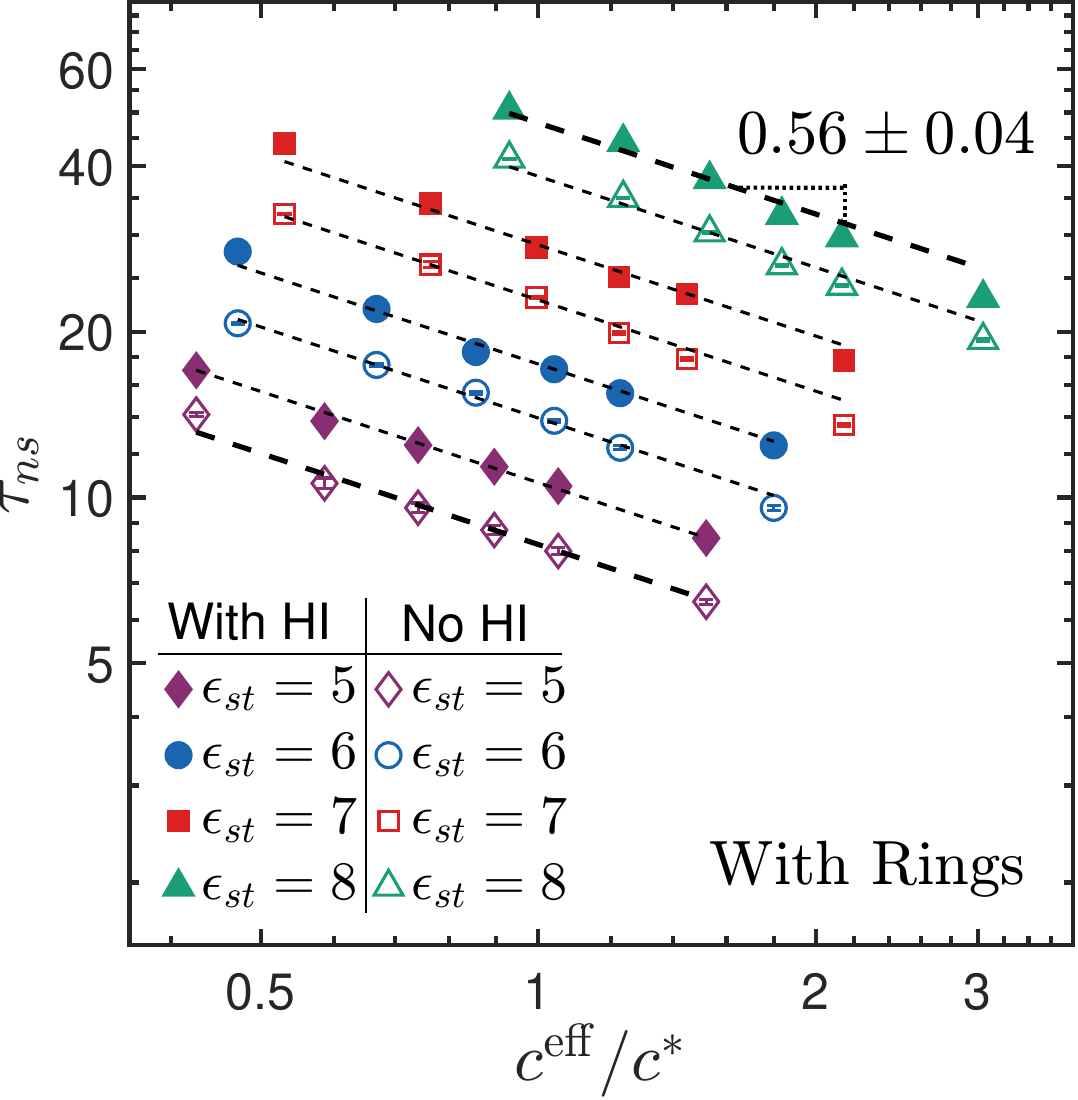} &
\includegraphics[width=8.5cm]{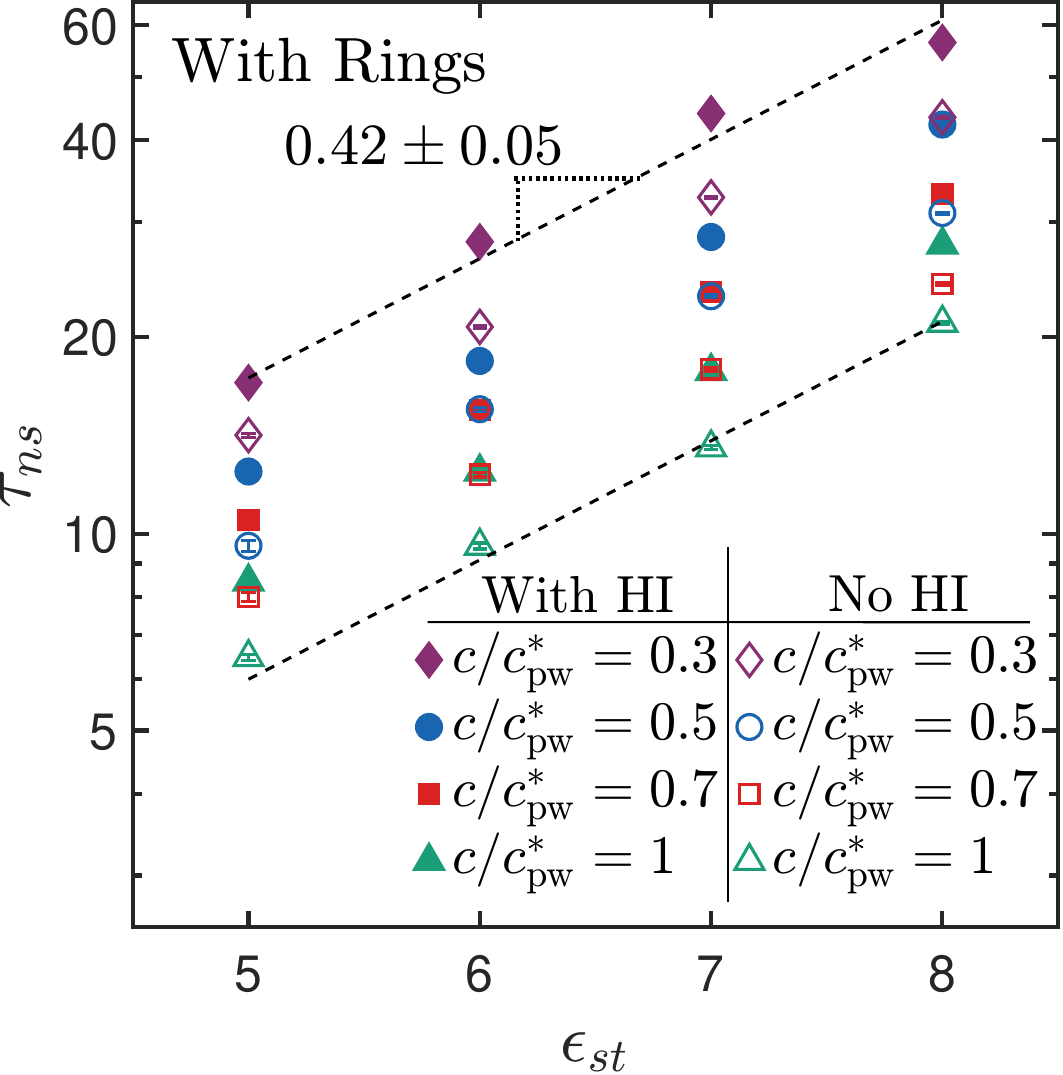} \\
(c)  & (d) 
\end{tabular}
\end{center}
    \vspace{-10pt}
\caption{Variation of the non-self recombination time with effective monomer concentration ($c^{\mathrm{eff}}/c^*$) in panels (a) and (c) and sticker energy ($\epsilon_{st}$) in panels (b) and (d). Panels (a) and (b) correspond to solutions without rings, while panels (c) and (d) correspond to solutions with rings, shown in the presence and absence of hydrodynamic interactions (HI).}\label{fig9}
    \vspace{-10pt}
\end{figure*}

The non-self recombination time ($\tau_{ns}$), in contrast, depends on the persistent worm concentration, sticker energy, and micellar topology, as shown in \fref{fig9}. Increasing the persistent worm concentration ($c^{\mathrm{eff}}/c^*$) reduces $\tau_{ns}$, since free stickers encounter new partners more frequently (as seen in \fref{fig9}(a) and (c)), while higher sticker energies ($\epsilon_{st}$) increases $\tau_{ns}$, as stronger bonding reduces the population of free stickers available for recombination (see \fref{fig9}(b) and (d)). In all cases, hydrodynamic interactions increase the non-self recombination time by delaying the recombination process. However, the slopes of $\tau_{ns}$ versus concentration curves of wormlike micelles solution and sticker energies (see \fref{fig9}) are independent of the presence or absence of hydrodynamic interactions. Furthermore, comparison of \fref{fig9}(c) and (d) with (a) and (b) shows that the presence of rings reduces the non-self recombination time. We next examine the scaling behavior of $\tau_{ns}$.

\begin{figure*}[tbh!]
\begin{center}
\begin{tabular}{cc}
\includegraphics[width=8.5cm]{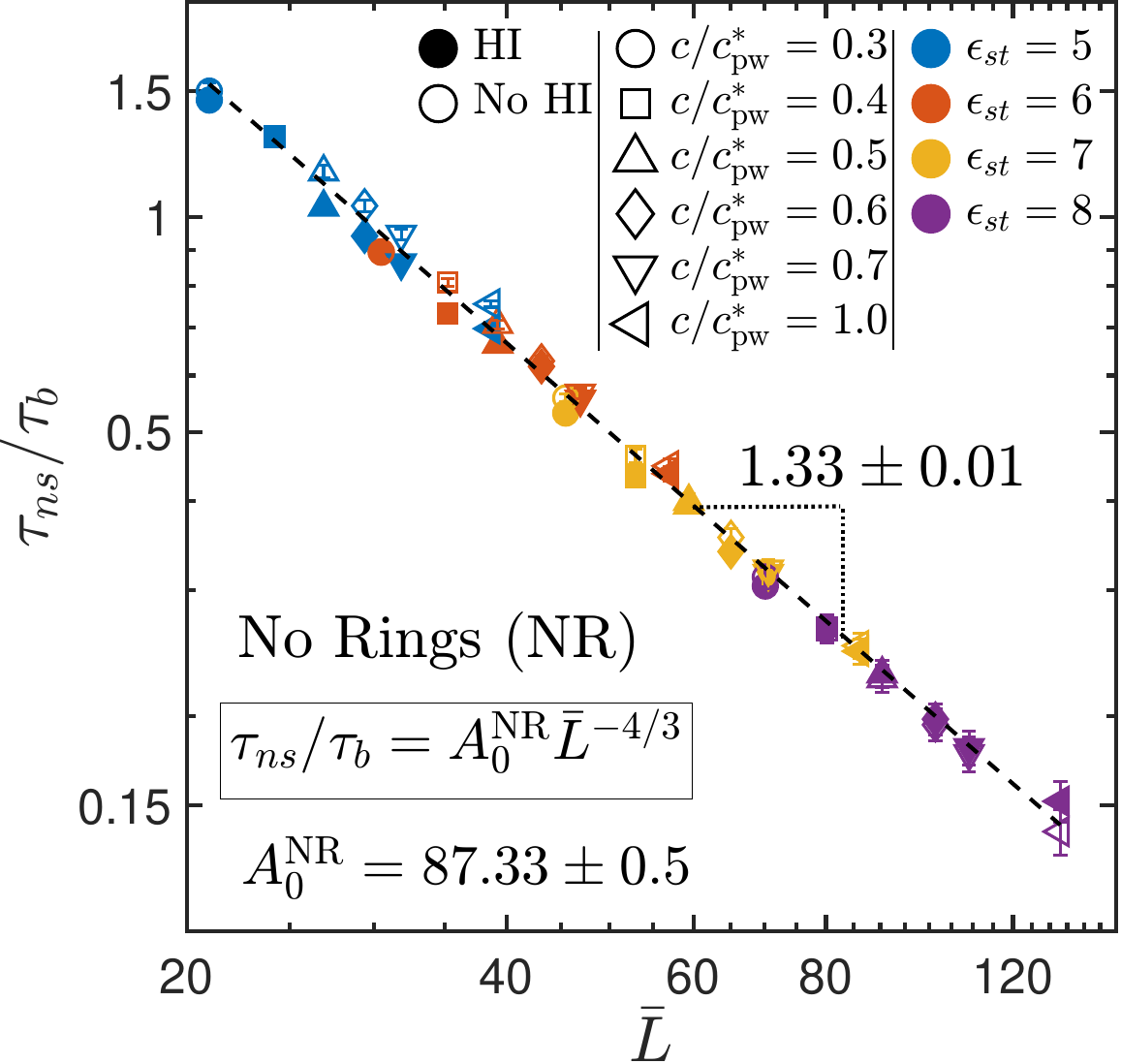} &
\includegraphics[width=8.5cm]{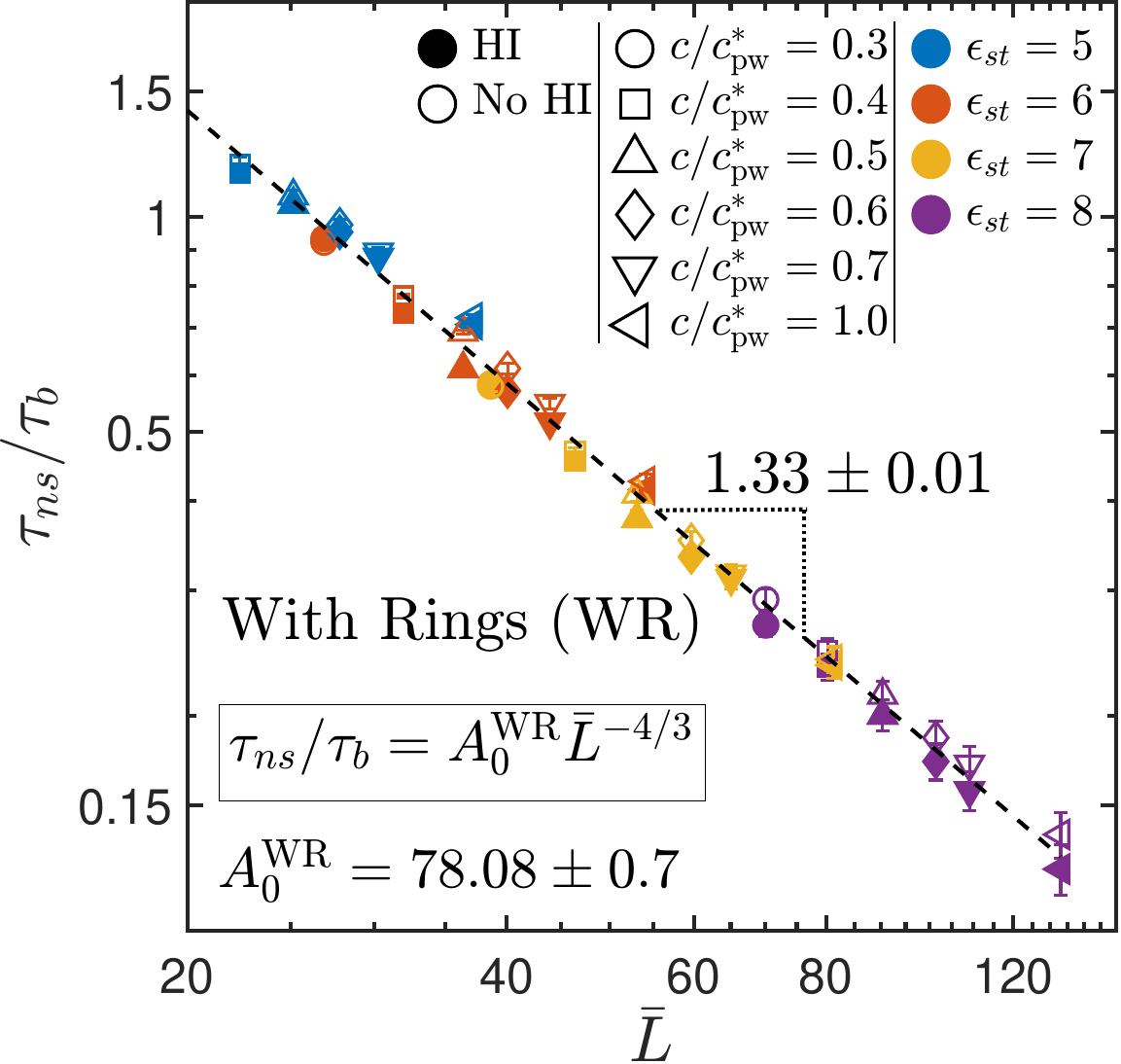} \\[-5pt] 
(a)  & (b) \\
\end{tabular}
\vspace{-10pt}
\end{center}
\caption{Variation of the non-self recombination time, $\tau_{ns}$, normalized by the bond breakage time $\tau_b$, with the mean length of linear micelles ($\bar{L}$) for solutions (a) without rings (NR) and (b) with rings (WR), shown for both the presence and absence of hydrodynamic interactions (HI). Filled and open symbols correspond to simulations with and without HI, respectively. Marker shapes represent the concentration $c/c^{*}_{\mathrm{pw}}$, with the same shape used consistently across all sticker energies, while colors denote the sticker energy $\epsilon_{st}$, with the same color applied consistently across all concentrations. The prefactor ratio is $A_0^{\mathrm{NR}} / A_0^{\mathrm{WR}} = 1.12$.}
\label{fig10}
    \vspace{-10pt}
\end{figure*}

\begin{figure*}[tbh!]
\begin{center}
\begin{tabular}{cc}
\includegraphics[width=8.5cm]{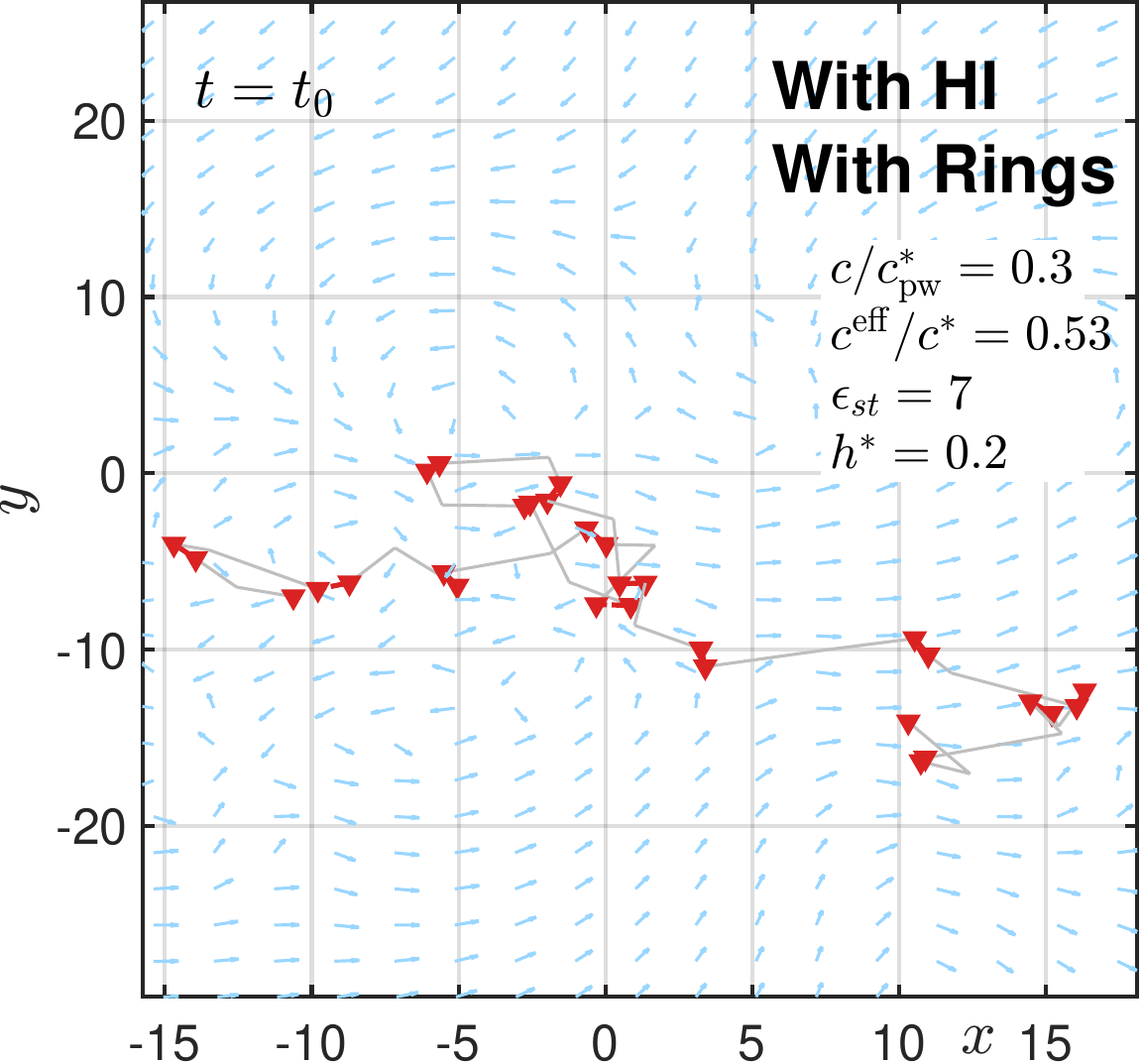} &
\includegraphics[width=8.45cm]{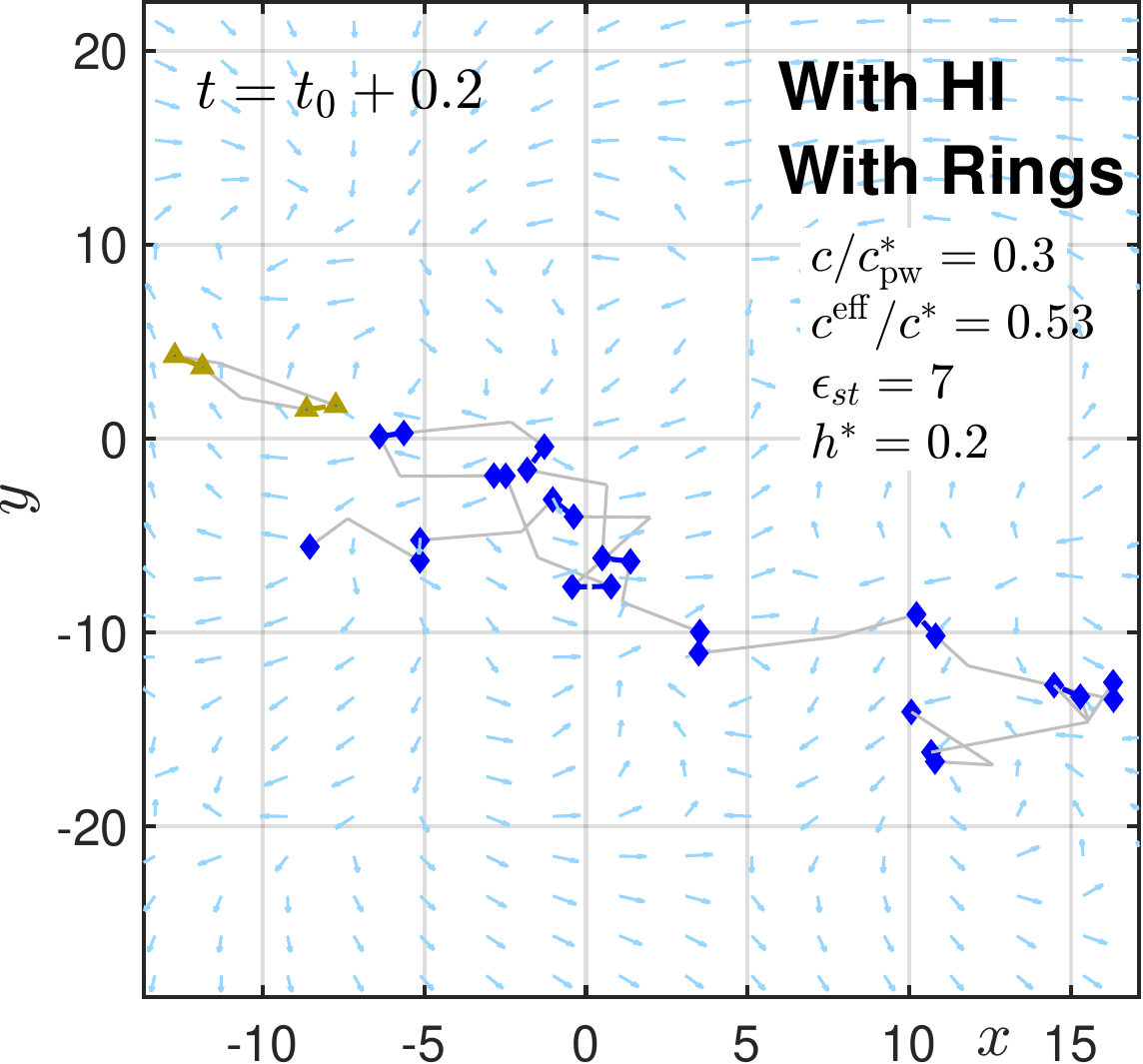} \\
(a)  & (b) \\[10pt]
\includegraphics[width=8.5cm]{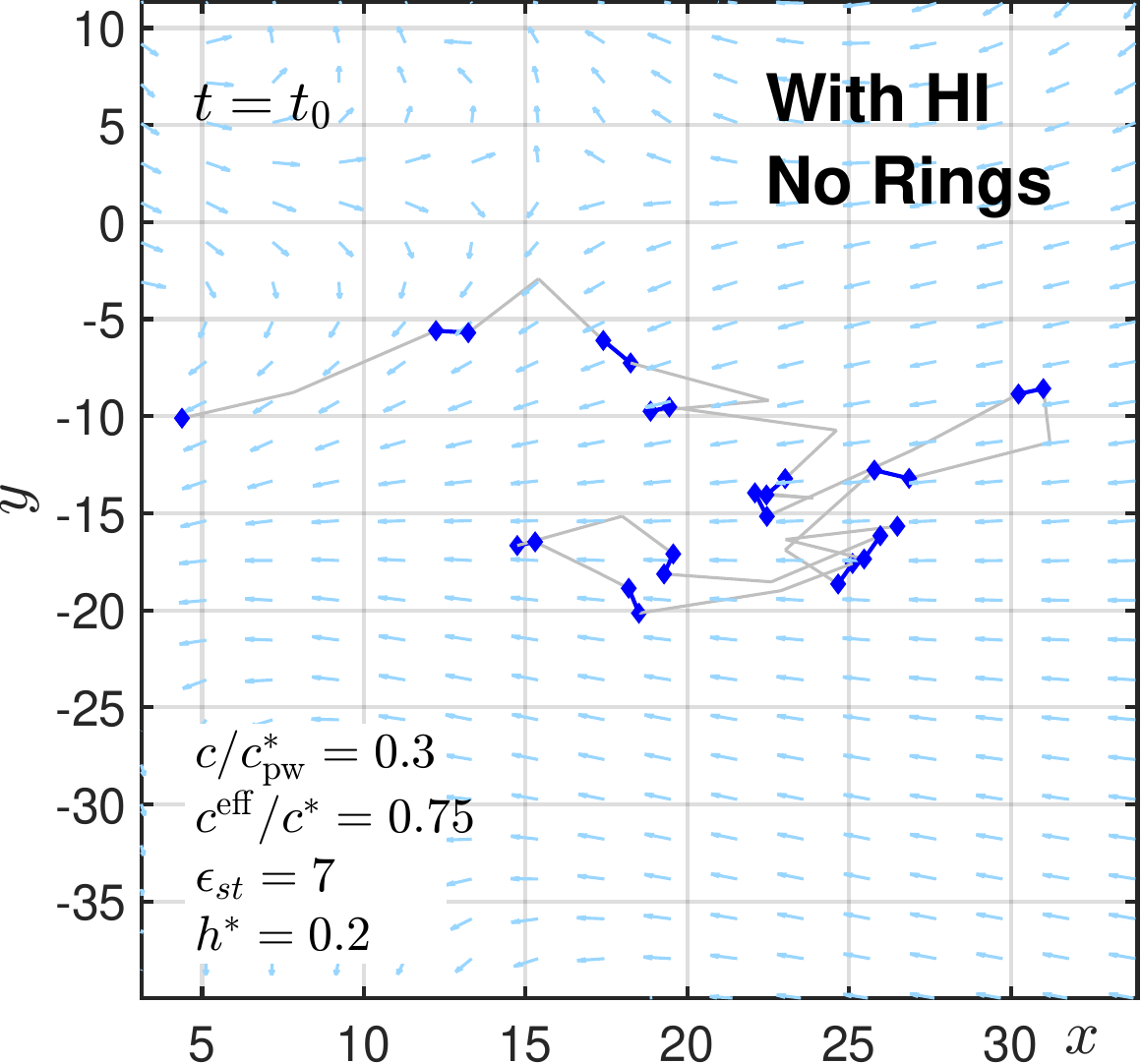} &
\includegraphics[width=8.5cm]{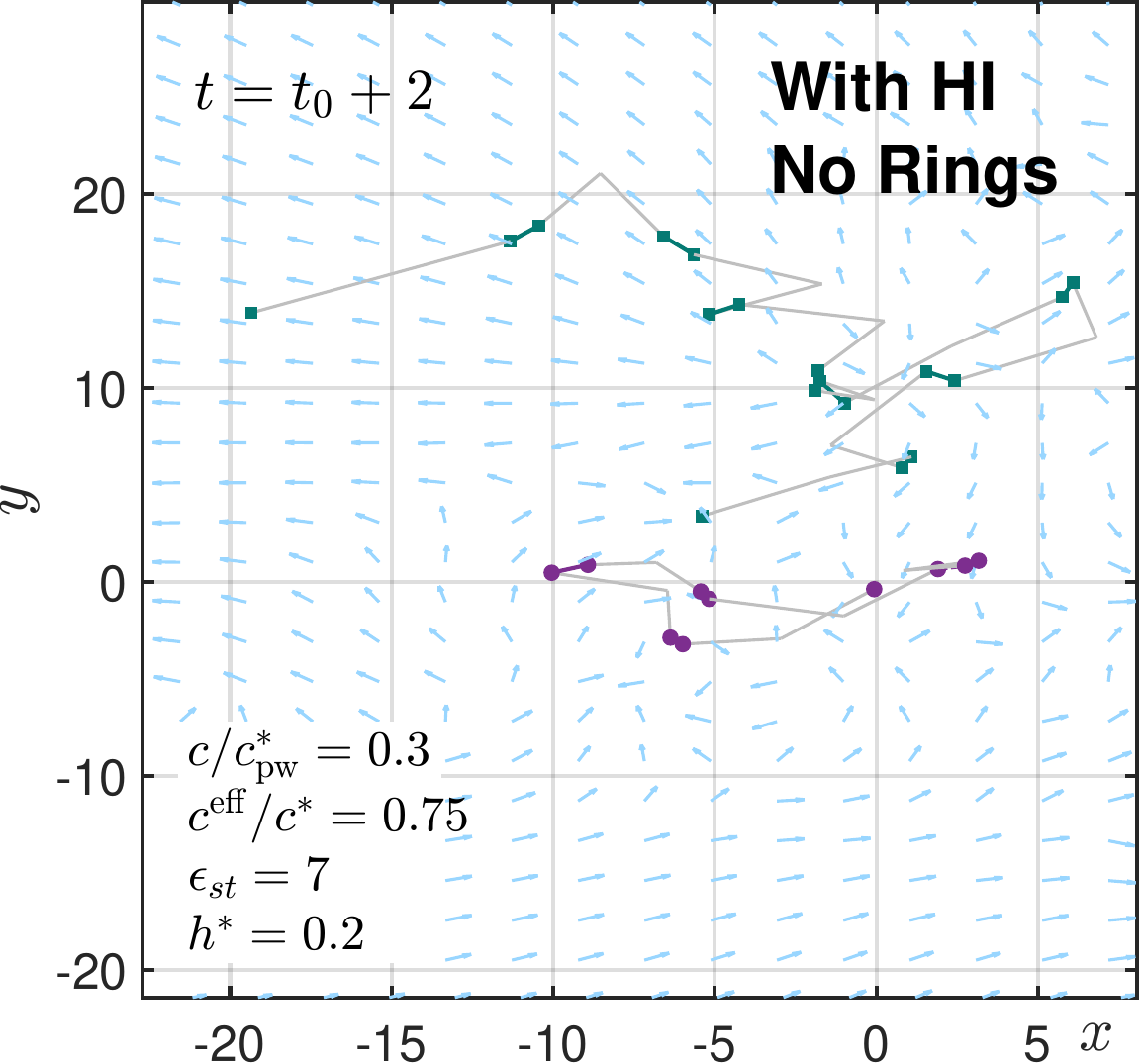} \\
(c)  & (d) 
\end{tabular}
\end{center}
    \vspace{-10pt}
\caption{Instantaneous velocity fields illustrating non-self recombination events in wormlike micellar solutions with hydrodynamic interactions. Panels (a) and (b) show the case with rings (WR) in the solutions, and panels (c) and (d) show the case without rings (NR). In (a) and (c), configurations are shown at time $t_0$, while in (b) and (d), configurations are shown at a later time. Only a subset of micelles is shown to highlight recombination events, while the remaining micelles in the solution are not displayed. Here purple circles, green squares, blue diamonds and red downward-pointing triangle, represents $m_{\mathrm{pw,lin}}^{\mathrm L} = 5,9,13$ and $15$ persistent worms in a linear micelles while olive green upward-pointing triangle represent $m_{\mathrm{pw,R}}^{\mathrm L} = 2$ persistent worms in a ring micelle.}
\label{fig11}
    \vspace{-10pt}
\end{figure*}

Remarkably, when $\tau_{ns}$ is nondimensionalized by the bond breakage time $\tau_b$ and plotted as a function of the mean micellar length $\bar{L}$, the data collapse onto a universal master curve of the form $\tau_{ns}/\tau_b = A_0 \bar{L}^{-4/3}$. This scaling is independent of both the effective concentration ($c^{\mathrm{eff}}/c^*$) and the sticker energy ($\epsilon_{st}$), since their influence is already captured through $\bar{L}$ and $\tau_b$. It is also independent of hydrodynamic interactions, whose effects are incorporated into $\tau_b$ (see \fref{fig10}). The prefactor $A_0$, however, depends on micellar topology. In the presence of rings (``WR'') in wormlike micellar solutions, $A_0^{\mathrm{WR}} = 78.08 \pm 0.7$, whereas in their absence (``NR''), $A_0^{\mathrm{NR}} = 87.33 \pm 0.5$ (see \fref{fig10}). The reduction in the non-self recombination time ($\tau_{ns}$) in the presence of rings can be attributed to the increased population of smaller rings in the solution. The formation of rings is favoured at lower concentrations of persistent worms and at higher sticker energies. Under these conditions, a larger fraction of micelles exist as small rings, since a free sticker on a linear micelle is more likely to encounter and bind with another sticker on the same chain, and when the sticker energy is high, the stickers remain bound for longer. As the concentration increases, the probability of ring formation decreases. At higher concentrations, stickers located at the ends of linear micelles are more likely to encounter stickers on neighbouring chains, promoting intermicellar association and the growth of longer linear micelles \cite{Kumar2025}. Consequently, the relative population of rings decreases with increasing concentration. The presence of small rings provide additional pathways for non-self recombination, increasing the probability of finding new partners. This is evident from \fref{fig9}(c) and (d), where the presence of rings reduces $\tau_{ns}$, indicating an increased probability of finding new partners. Similar trends are observed in \fref{fig10}(b).

\fref{fig11} shows instantaneous velocity fields for wormlike micellar solutions with and without rings, highlighting non-self recombination events. To emphasize these events, only a subset of micelles is shown, although the full system contains many more micelles. In the presence of rings (see \fref{fig11}(a) and (b)), a linear micelle with $m_{\mathrm{pw,lin}}^{\mathrm L}=15$ (red downward-pointing triangle) persistent worms at time $t_0$ breaks at $t_0+0.2$ into micelles of sizes $m_{\mathrm{pw,lin}}^{\mathrm L}=13$ (blue diamonds) and $m_{\mathrm{pw,R}}^{\mathrm L}=2$. The smaller micelle ($m_{\mathrm{pw,R}}^{\mathrm L}=2$) recombines with a new partner to form a ring (olive green upward-pointing triangle). In contrast, in the absence of rings (see \fref{fig11}(c) and (d)), a linear micelle with $m_{\mathrm{pw,lin}}^{\mathrm L}=13$ (blue diamonds) at time $t_0$ breaks into fragments of sizes $m_{\mathrm{pw,lin}}^{\mathrm L}=9$ (green squares) and $m_{\mathrm{pw,lin}}^{\mathrm L}=4$. The smaller fragment ($m_{\mathrm{pw,lin}}^{\mathrm L}=4$) diffuses through the solution and recombines at a later time with a single persistent worm (not shown in \fref{fig11}(d)) to form a longer linear micelle of size $m_{\mathrm{pw,lin}}^{\mathrm L}=5$ (purple circles) at $t_0+2$.

\subsubsection{\label{sec:taubL} Breakage time of a wormlike micelle of length $L$}

Two types of events terminate the life of a micelle of length $L$: (i) scission, where the micelle splits into shorter micelles, and (ii) recombination (self- or non-self), where the micelle merges with another micelle, resulting in a change in contour length. Over the duration of each simulation trajectory, from a nondimensional time $t=0$ to $t = \tau_{\mathrm{run}}$ (see \sref{sec:simdet}), termination events are monitored as follows: at a reference time $t_1$, a micelle with contour length $L$ and its associated set of bound sticker pairs is identified. As the simulation evolves, the micelle is kept track of to verify whether the same sticker pairs remain bound and whether the micelle length remains unchanged. If, at a later time $t_2$, either (i) a scission event occurs due to unbinding of a sticker pair or (ii) the micelle length changes due to recombination, the observation is terminated. The elapsed time $s^{\text{L}} = t_2 - t_1$ is then recorded as the survival time of that micelle of length $L$. This procedure is repeated for all micelles in the solution, with a new starting time $t_1$ for each micelle and termination at time $t_2$ when one of the above events occurs. It is clear that for any sufficiently large time interval, a number of termination events will be observed for micelles of length $L$, with varying survival times $s^{\text{L}}$. Here, we denote by $d^{\mathrm{L}}$ the number of scission events occurring within any particular time interval for micelles of length $L$, while $r^{\mathrm{L}}$ is used to denote the number of recombination events for micelles of length $L$ occurring within the same time interval. With this background, we discuss in this section how the characteristic equilibrium timescale for scission, defined as the breakage time $\tau_{br}^{\text{L}}$ for micelles of contour length $L$, can be determined.  

The nondimensional simulation time from $t=0$ to $t = \tau_{\mathrm{run}}$ is divided into  $p$ equal time intervals, $\Delta t_i$, each of duration $\tau_{\mathrm{sample}}$ (see \sref{sec:tbond}), with $i = 1, 2, \dots, p$, where $p = \tau_{\mathrm{run}}/\tau_{\mathrm{sample}}$. 
In each of these intervals, the following two data sets are constructed: (i) the number of scission events $\{d_i^{\mathrm{L}}\}$, and (ii) the number of recombination events $\{r_i^{\mathrm{L}}\}$, with the index $i$ denoting the particular time interval in which the data is collected (note that scission and recombination events are counted separately). Using these recorded events, the survival probability $\Sigma^{\text{L}}(s)$ of micelles of length $L$ up to time $s$ is computed using the Kaplan–Meier estimator~\cite{Kaplan1958,Koide2023,Koide2022}: 
 \begin{equation}
    \Sigma^{\text{L}}(s) = \prod_{k:\, \Delta t_k \leq s} \left(1 - \frac{d_k^{\text{L}}}{n_k^{\text{L}}}\right),
\label{survival_prob} 
\end{equation} 
where the product is taken over all time intervals $\Delta t_k$ up to time $s$. Here, $n_k^{\mathrm{L}} = n_{k-1}^{\mathrm{L}} - d_{k-1}^{\mathrm{L}} - r_{k-1}^{\mathrm{L}}$ denotes the number of micelles surviving up to the previous time interval $\Delta t_{k-1}$. The long-time decay of $\Sigma^{\mathrm{L}}(s)$ can be fitted to a single exponential, $\Sigma^{\mathrm{L}}(s) = A \exp(-s/\tau_{br}^{\mathrm{L}})$, from which the characteristic breakage time $\tau_{br}^{\mathrm{L}}$ of micelles of length $L$ can be extracted. Since the calculation of the characteristic breakage time is focussed on the long-time regime, it is important to examine the influence of the choice of sampling interval $\tau_{\text{sample}}$. The survival probability is found to be insensitive to the value of $\tau_{\text{sample}}$ for $\tau_{\text{sample}} \leq 0.1$, with the long-time decay converging to a unique curve. This behaviour is consistent with the sampling interval dependence observed for non-self recombination timescales at long times (see \fref{fig6}(b)). As a result, an optimized value of $\tau_{\text{sample}} = 0.1$ is used for computing the survival probability across different micelle lengths.

\begin{figure}[t]
\centering
\includegraphics[width=8.5cm]{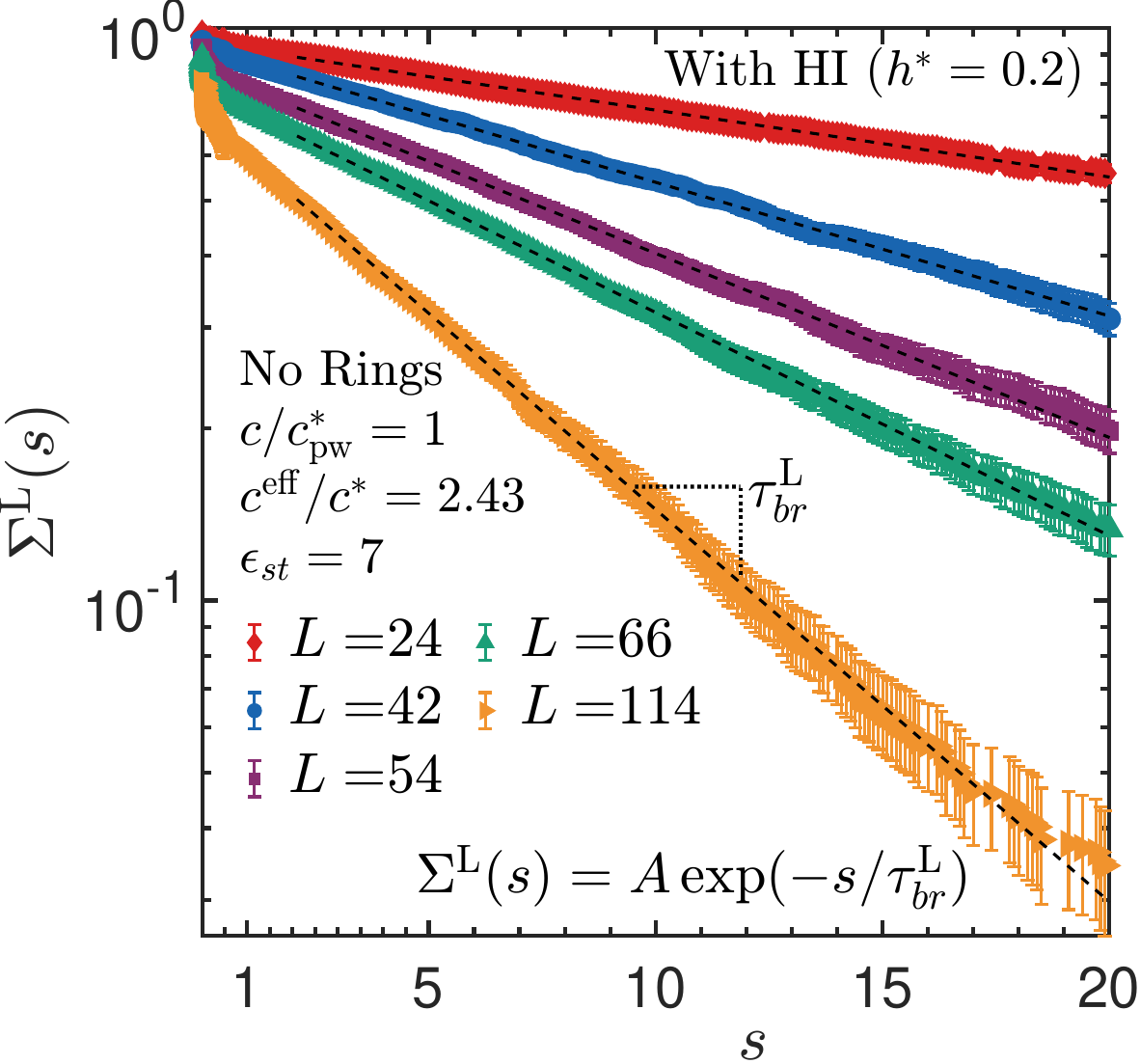}  
\vspace{-10pt}
\caption{Survival probability, $\Sigma^{\mathrm{L}}(s)$, of wormlike micelles with varying contour length $L$ at $c^{\mathrm{eff}}/c^* = 2.43$ and $\epsilon_{\mathrm{st}} = 7$, in the presence of hydrodynamic interactions and in the absence of rings. Different symbols and colours denote micelles of varying contour lengths $L$, with the corresponding number of persistent worms $m_{\mathrm{pw,lin}}^{\mathrm{L}}$ indicated as follows: $L=24$ (red diamonds, $m_{\mathrm{pw,lin}}^{\mathrm{L}}=4$), $L=42$ (blue circles, $m_{\mathrm{pw,lin}}^{\mathrm{L}}=7$), $L=54$ (purple squares, $m_{\mathrm{pw,lin}}^{\mathrm{L}}=9$), $L=66$ (up-triangles green, $m_{\mathrm{pw,lin}}^{\mathrm{L}}=11$), and $L=114$ (right-triangles orange, $m_{\mathrm{pw,lin}}^{\mathrm{L}}=19$).}
\label{fig12}
    \vspace{-10pt}
\end{figure}

\begin{table}[tbph]
\centering
\caption{Characteristic breakage times $\tau_{br}^{\mathrm{L}}$  for micelles of varying lengths $L$, shown in \fref{fig12}, in the absence of rings and in the presence of hydrodynamic interactions.}
\label{table:table2}
\setlength{\tabcolsep}{1.8em}
\renewcommand{\arraystretch}{1.3}
\vspace{10pt}
\begin{tabular}{ | c | c | c | }
\hline
\hline
{} & $L$ & $\tau_{br}^{\mathrm{L}}$  \\
\hline
1 & 24  & 37.07 $\pm$ 0.05  \\
\hline
2 & 42  & 18.69 $\pm$ 0.02  \\
\hline
3 & 54  & 13.45 $\pm$ 0.03 \\
\hline
4 & 66  & 10.94 $\pm$ 0.04  \\
\hline
5 & 114 & 6.31 $\pm$ 0.05  \\
\hline
\hline
\end{tabular}
\end{table}

\fref{fig12} shows representative survival probabilities $\Sigma^{\text{L}}(s)$ for micelles with varying contour lengths at $c^{\mathrm{eff}}/c^* = 2.43$ and $\epsilon_{st}=7$, along with the $95\%$ confidence interval using the exponential Greenwood formula \cite{Klein2012}. The  characteristic breakage timescale $\tau_{br}^{\mathrm{L}}$ for the different micelle lengths studied here, obtained from the slope of the long-time decay of $\Sigma^{\mathrm{L}}(s)$, is displayed in Table~\ref{table:table2}. The error bars increase for longer micelles due to their lower population~\cite{Kumar2025}.

\begin{figure*}[t]
\begin{center}
\begin{tabular}{cc}
\includegraphics[width=8.5cm]{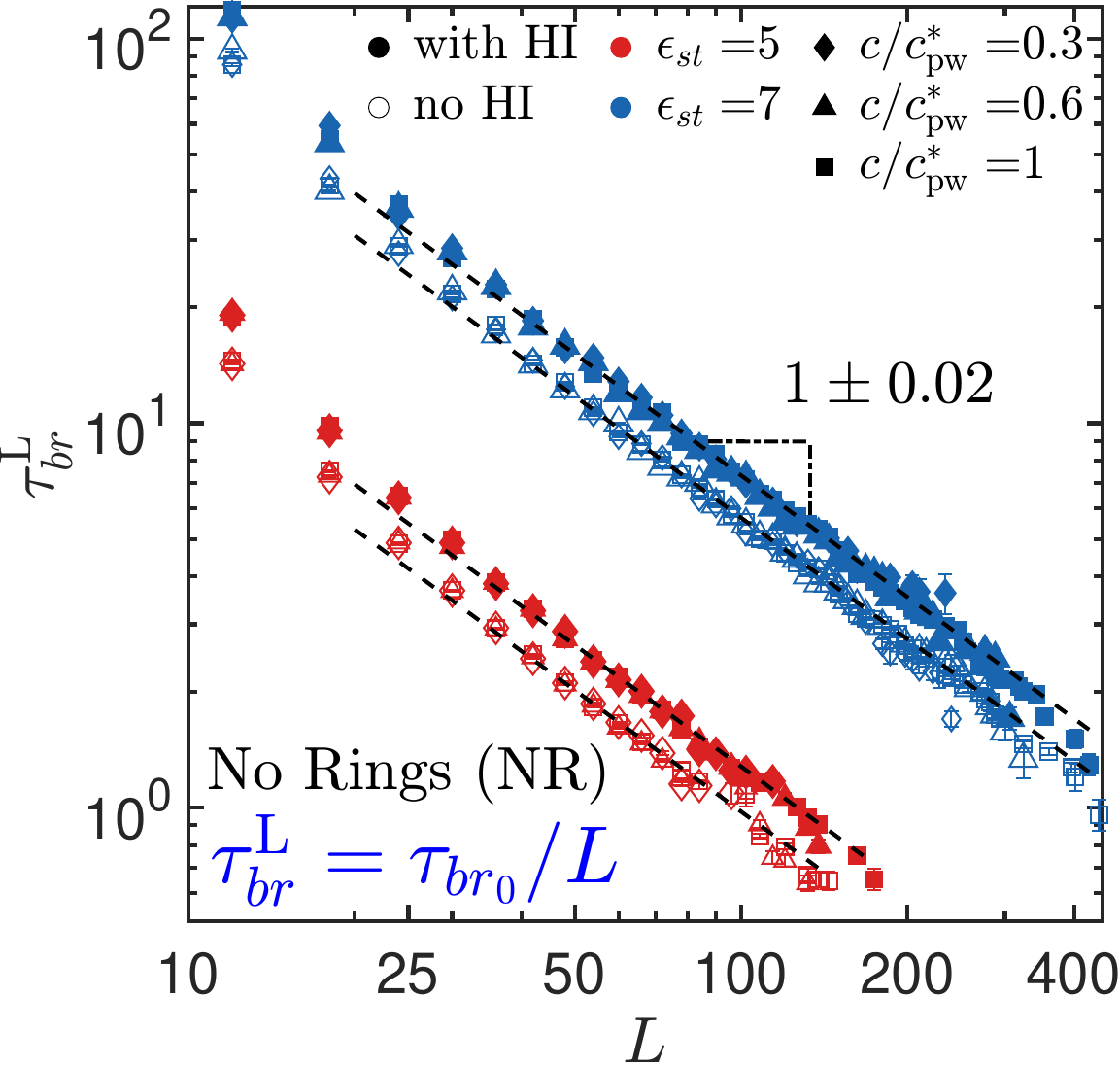} &
\includegraphics[width=8.5cm]{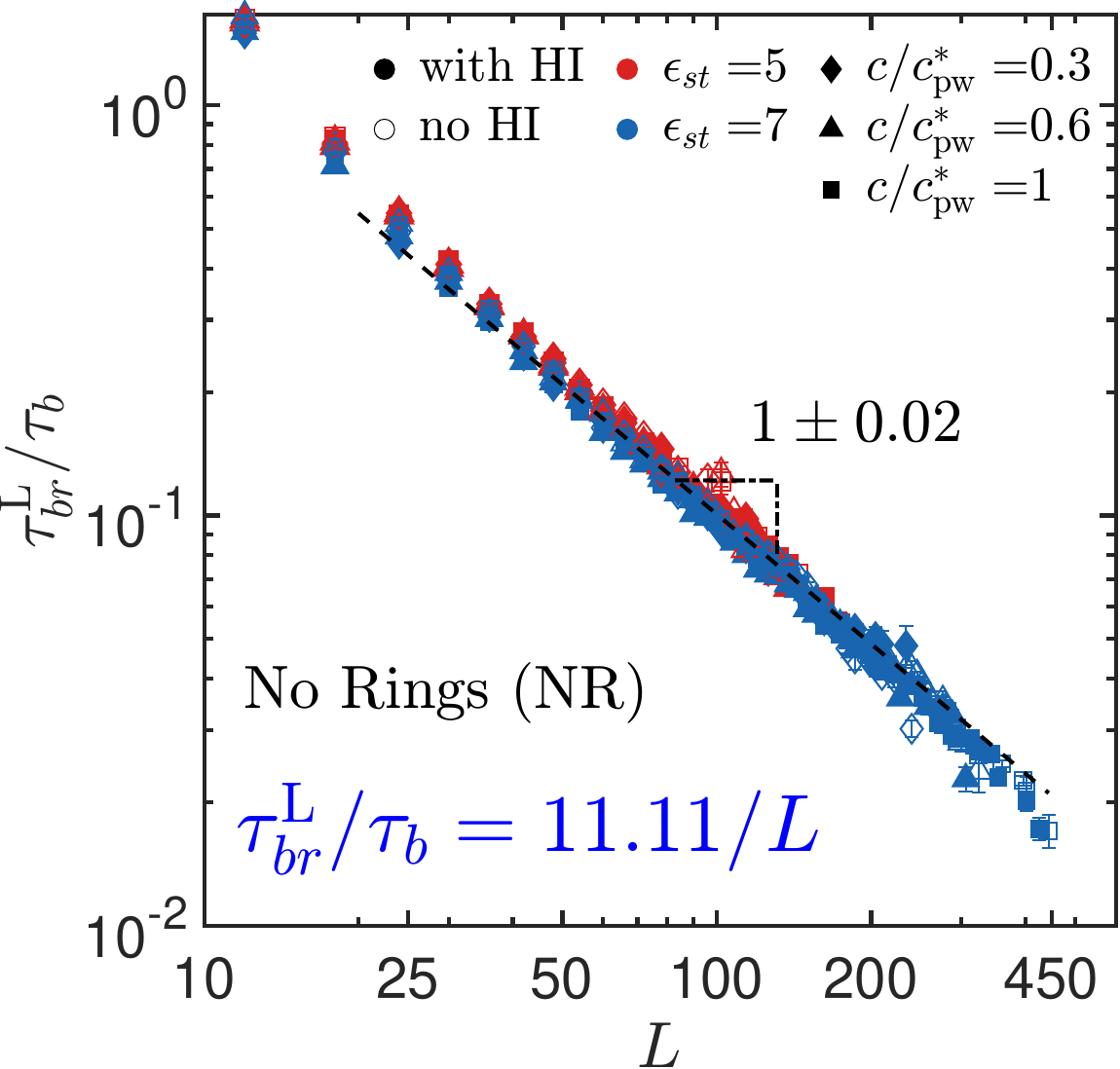} \\[-5pt] 
(a)  & (b) \\
\end{tabular}
\vspace{-10pt}
\end{center}
\caption{
(a) Breakage time $\tau_{br}^{\mathrm{L}}$ of wormlike micelles as a function of contour length $L$ for different concentrations of persistent worms ($c/c_{\mathrm{pw}}^*$), shown in the presence (``with HI'') and absence (``no HI'') of hydrodynamic interactions, for sticker energies $\epsilon_{st} = 5$ (red) and $7$ (blue) in micelle solutions without rings. The breakage time increases with increasing sticker energy for all $L$. (b) Normalized bond breakage time, $\tau_{br}^{\mathrm{L}}/\tau_b$, for wormlike micelles without rings at sticker energies $\epsilon_{st} = 5$ (red) and $7$ (blue), where $\tau_b$ is the bond breakage time defined in \sref{sec:tbond}. In both panels, solid symbols denote simulations with hydrodynamic interactions, while open symbols denote those without hydrodynamic interactions. Different symbols correspond to concentrations $c/c_{\mathrm{pw}}^*$: 0.3 (diamond), 0.6 (triangle), and 1 (square).}
\label{fig13}
\vspace{-5pt}
\end{figure*}

\fref{fig13}(a) shows the breakage time of wormlike micelles, $\tau_{br}^{\mathrm{L}}$, as a function of contour length. The breakage time exhibits a strong dependence on the sticker energy and a weak dependence on the concentration of persistent worms. As established in \sref{sec:tbond}, the bond lifetime $\tau_b$ varies exponentially with sticker energy, indicating that increasing sticker energy suppresses bond scission and consequently increases $\tau_{br}^{\mathrm{L}}$ for a given micelle length. 

Hydrodynamic interactions also play a significant role in the breakage dynamics. In the absence of hydrodynamic interactions, micelles break more rapidly, resulting in shorter $\tau_{br}^{\mathrm{L}}$. When hydrodynamic interactions are included, the breakage process is prolonged, as long-range hydrodynamic correlations damp local bond fluctuations and stabilize the micelle against scission. Regardless of whether hydrodynamic interactions are present or absent, the breakage time is inversely proportional to the micelle contour length ($L$), i.e., $\tau_{br}^{\text{L}} = A_0/L$, as displayed in \fref{fig13}(a), where $A_0$ is a constant. This scaling is consistent with previous studies \cite{Cates1987,Koide2022,Padding2004}. The breakage time corresponding to the mean micelle length $\bar{L}$ is therefore given by $\tau_{br}^{\bar{\text{L}}} = A_0/\bar{L}$ \cite{Cates1987,Padding2004}.

Furthermore, when the breakage time is normalized by the bond lifetime $\tau_b$ (see \sref{sec:tbond}), the data for micelles of different contour lengths collapse onto a single master curve (see \fref{fig13}(b)), independent of both concentration and sticker energy. This collapse indicates that the effects of sticker energy, concentration, and hydrodynamic interactions on the breakage dynamics are primarily mediated through $\tau_b$. The resulting scaling relation is given by: $\tau_{br}^{\text{L}}/\tau_b = C_0/L^{\alpha}$, with $C_0 = 11.11 \pm 0.42$ and $\alpha = 1.00 \pm 0.02$.

\subsubsection{\label{sec:tau1} Longest relaxation time of wormlike micelles}

The longest relaxation time of wormlike micelles, $\tau_1 = 1/\omega_{\left(G^\prime=G^{\prime\prime}\right)}$, is determined from the intersection of the storage and loss moduli (\eref{Gprime}) for different sticker energies and micelle concentrations. To identify the intersection point, we compute $\tan\delta = G^\prime/G^{\prime\prime}$ and plot it as a function of frequency $\omega$. The frequency at which $\tan\delta = 1$ defines $1/\tau_1$.

\begin{figure*}[tbph]
\begin{center}
\begin{tabular}{cc}
\includegraphics[width=8.5cm]{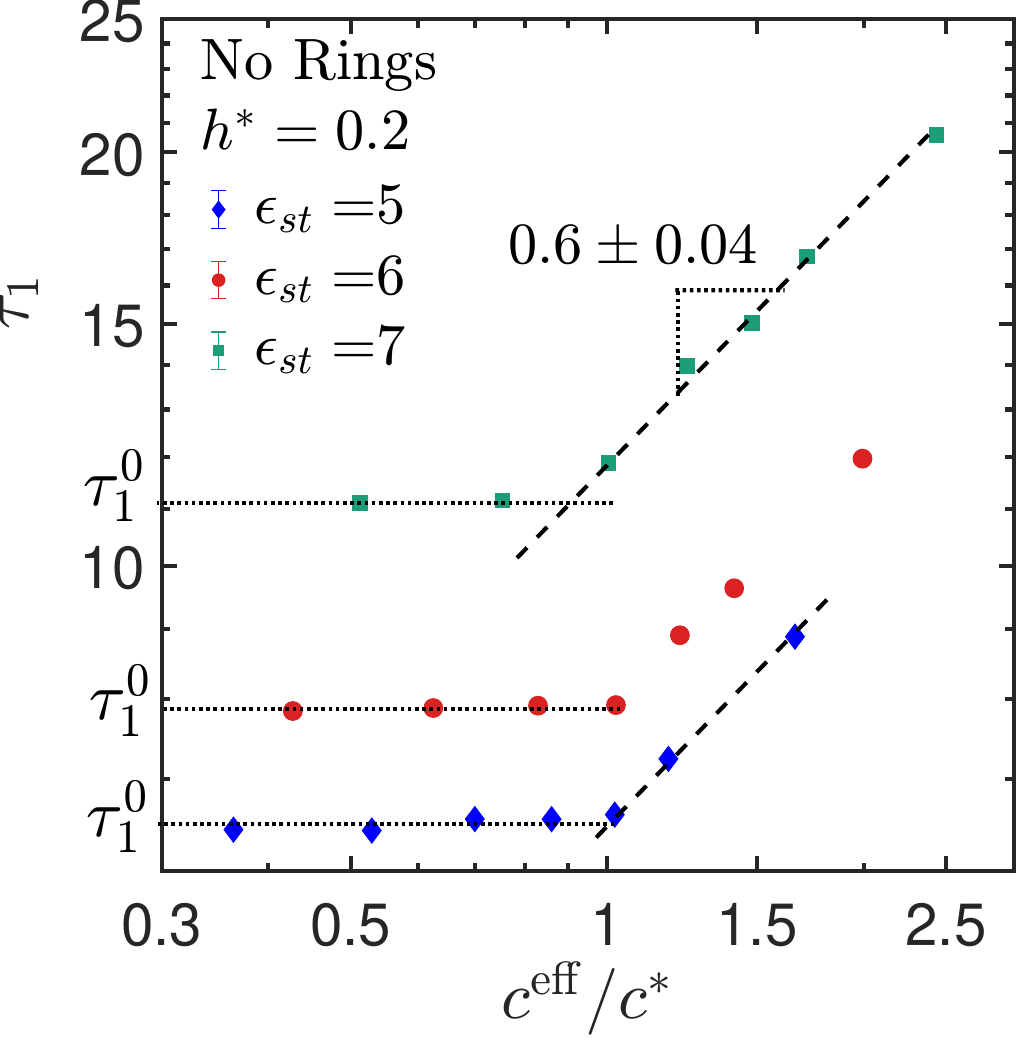} &
\includegraphics[width=8.5cm]{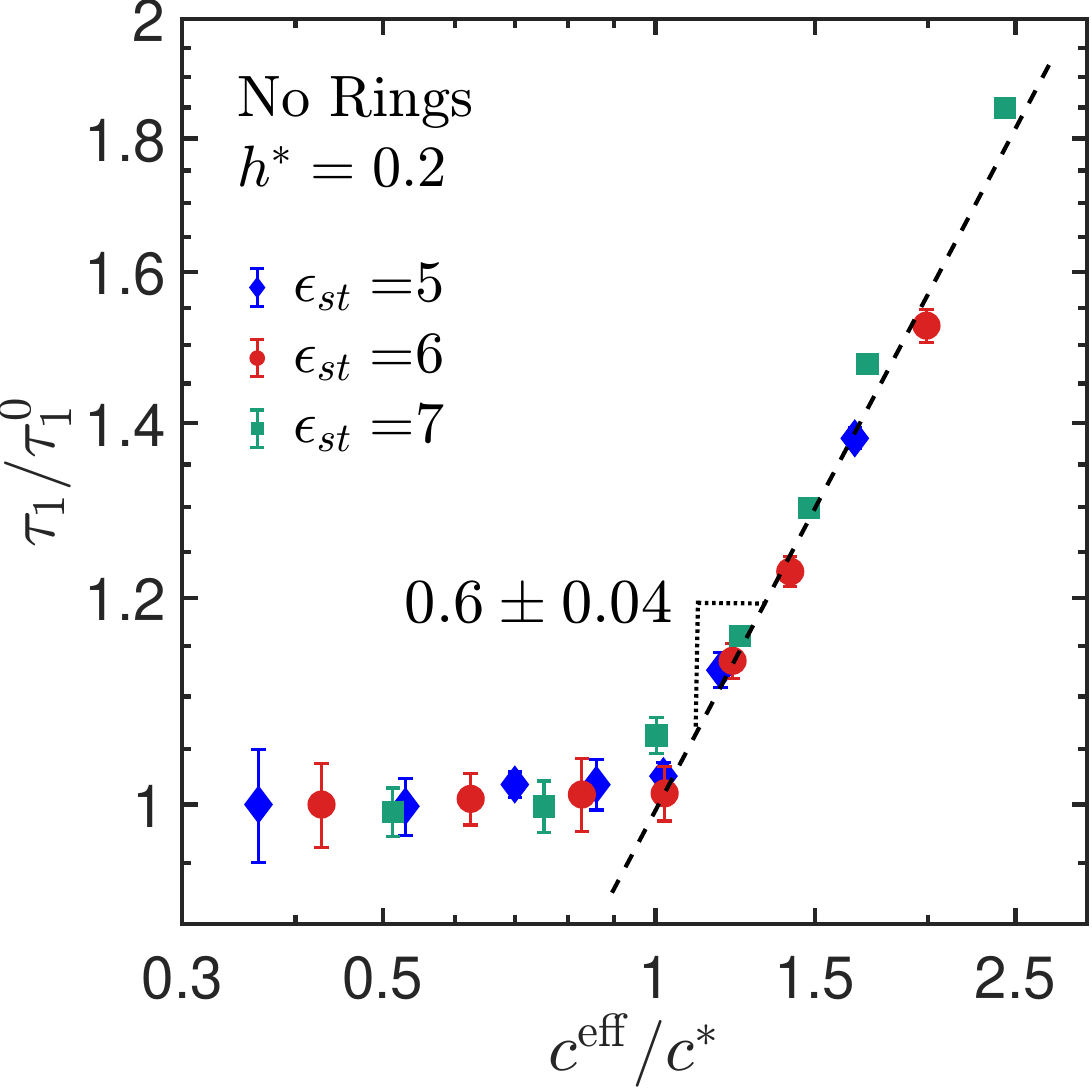} \\[-5pt] 
(a)  & (b) \\
\end{tabular}
\vspace{-10pt}
\end{center}
\caption{(a) Scaling of the longest relaxation time ($\tau_1$) with effective wormlike micelle concentration ($c^{\mathrm{eff}}/c^*$) for solutions with hydrodynamic interactions and in the absence of rings, at different sticker energies ($\epsilon_{st} = 5, 6,$ and $7$). (b) Dependence of the normalized longest relaxation time ($\tau_1/\tau_1^0$) on the effective wormlike micelle concentration ($c^{\mathrm{eff}}/c^*$) under the same conditions. Here, $\tau_1^0$ denotes the longest relaxation time in the dilute limit for each sticker energy, obtained from the intercept of the horizontal plateau (dashed lines) with the $y$-axis in \fref{fig14}(a).}
\label{fig14}
\vspace{-10pt}
\end{figure*}

To the best of our knowledge, the concentration scaling of $\tau_1$ in the semidilute unentangled wormlike micelle regime has not been reported previously. Our results show that $\tau_1$ increases with concentration as a power law with an exponent of $0.6$ (see \fref{fig14}(a)). This scaling differs from that of unentangled homopolymer solutions, where scaling theory \cite{Rubinstein2003} predicts $\tau_1\sim(c/c^*)^{(2-3\nu)/(3\nu-1)} \sim (c/c^*)^{0.25}$ for an athermal solvent with Flory exponent $\nu = 0.6$. Interestingly, when the relaxation times are nondimensionalized by $\tau_1^0$, defined as the dilute-limit value of $\tau_1$ for each sticker energy (obtained from the $y$-axis intercept of the horizontal plateau in \fref{fig14}(a)), the data for all concentrations and sticker energies collapse onto a universal curve, as shown in \fref{fig14}(b).
\begin{figure*}[t]
\begin{center}
\begin{tabular}{cc}
\includegraphics[width=8.5cm]{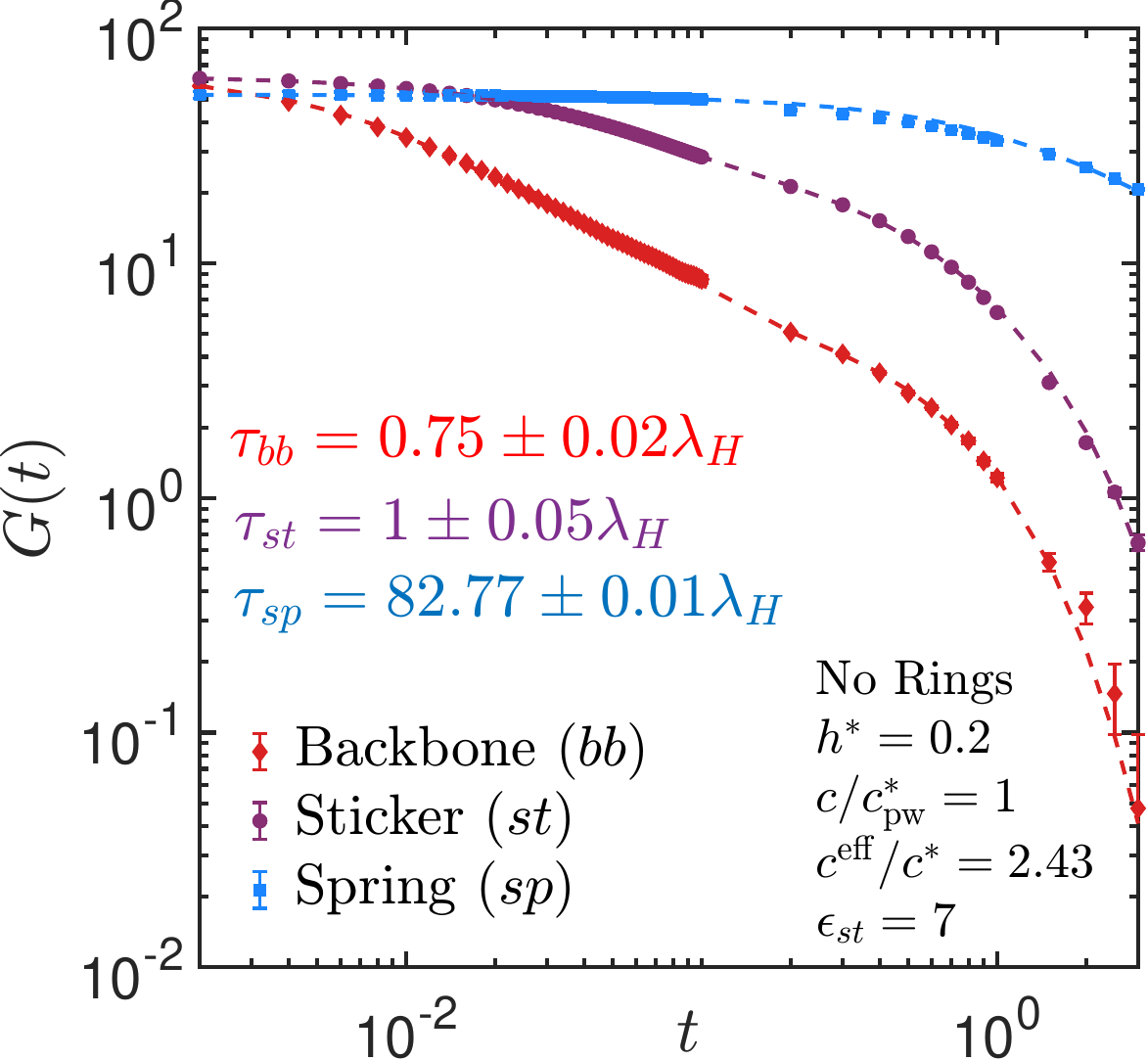} &
\includegraphics[width=8.45cm]{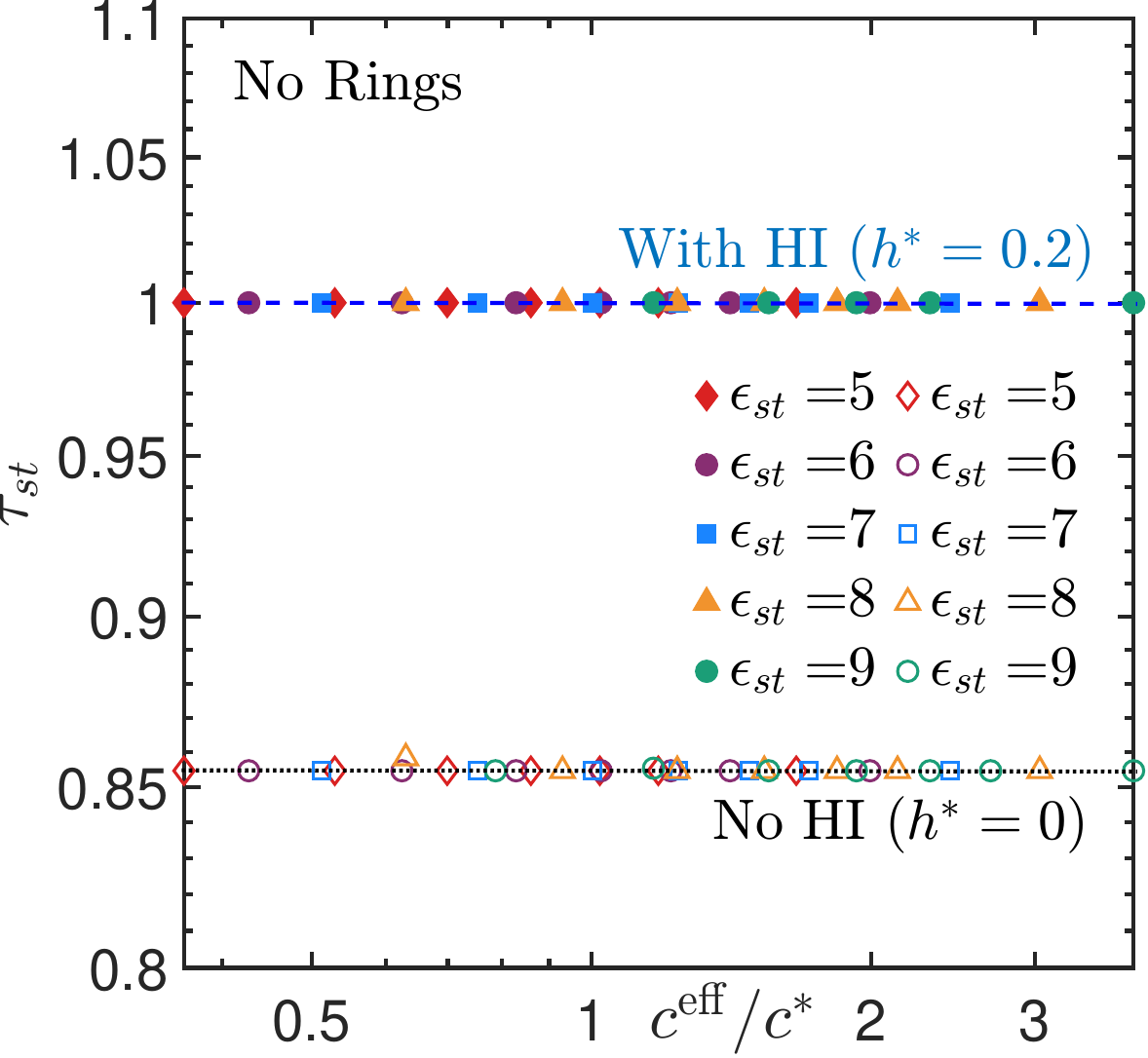} \\
(a)  & (b) \\[10pt]
\includegraphics[width=8.2cm]{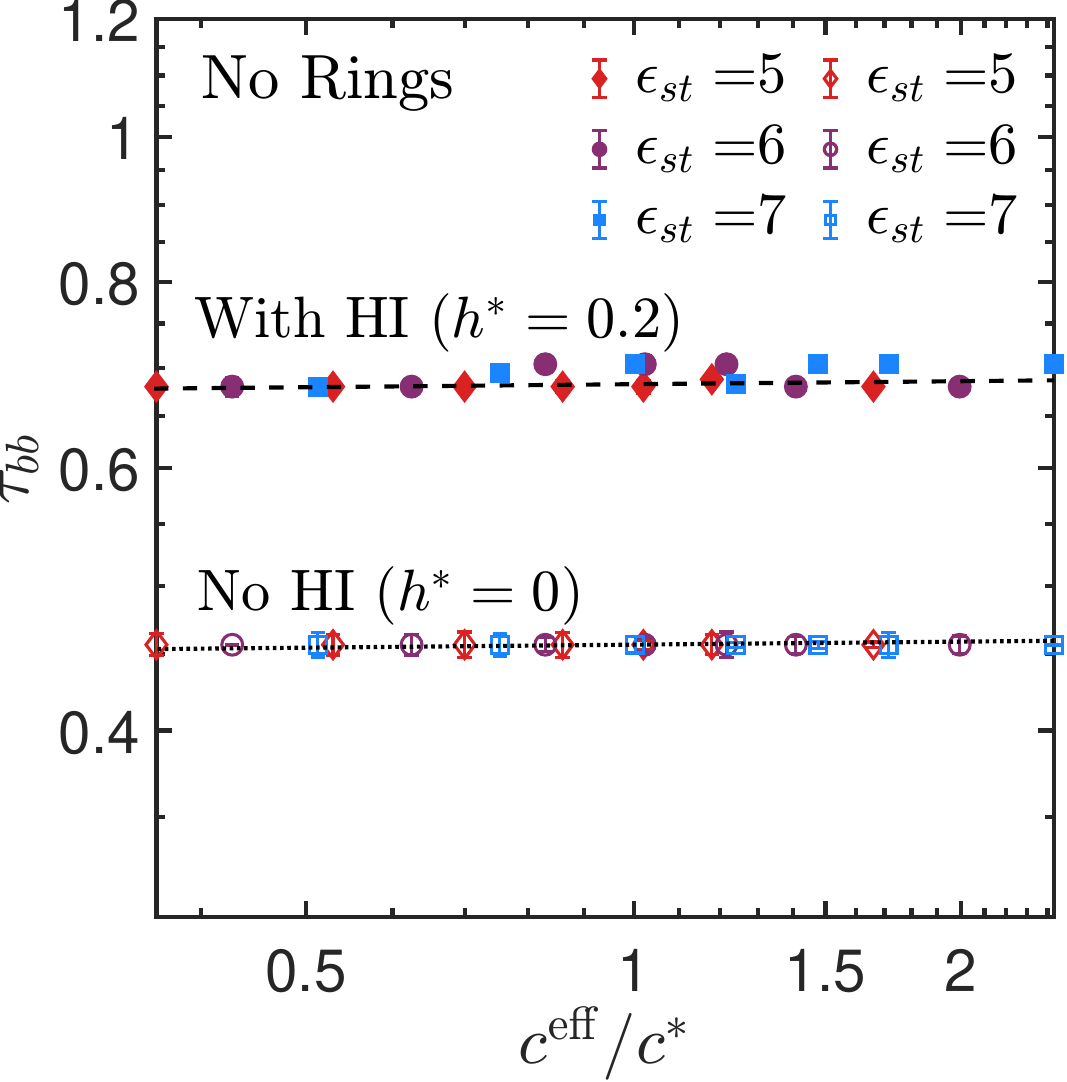} &
\includegraphics[width=8.5cm]{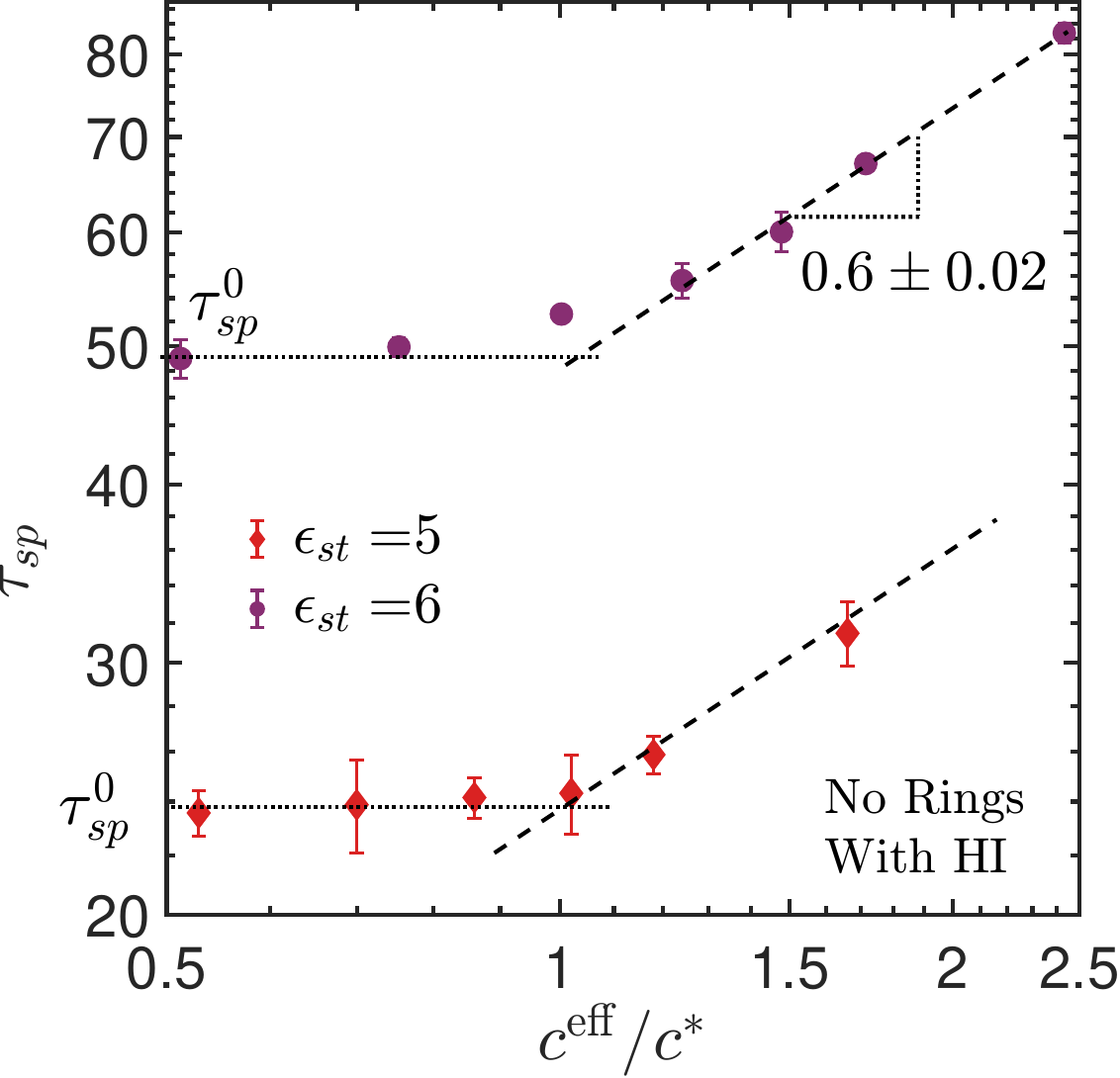} \\
(c)  & (d) 
\end{tabular}
\end{center}
    \vspace{-10pt}
\caption{(a) Stress autocorrelation functions of different contributions in wormlike micellar solutions at $c^{\mathrm{eff}}/c^* = 2.43$ and $\epsilon_{st} = 7$, in the presence of hydrodynamic interactions. (b) Terminal relaxation time of the sticker stress component ($\tau_{st}$) as a function of $c^{\mathrm{eff}}/c^*$, shown in the presence (filled symbols) and absence (hollow symbols) of hydrodynamic interactions for micellar solutions without rings. (c) Terminal relaxation time of the backbone stress component ($\tau_{bb}$) in the presence (filled symbols) and absence (hollow symbols) of hydrodynamic interactions for micellar solutions without rings. (d) Terminal relaxation time of the spring stress component ($\tau_{sp}$) as a function of $c^{\mathrm{eff}}/c^*$ for sticker energies $\epsilon_{st} = 5, 6$, in the presence of hydrodynamic interactions for micellar solutions without rings. Here, $\tau_{sp}^0$ denotes the terminal spring relaxation time in the dilute, unentangled limit, evaluated separately for each sticker energy.}
\label{fig15}
    \vspace{-10pt}
\end{figure*}

The stronger concentration dependence observed in semidilute wormlike micelles, $\tau_1 \sim (c^{\mathrm{eff}})^{0.6}$, compared to homopolymer solutions, $\tau_1 \sim (c^{\mathrm{eff}})^{0.25}$, arises from the dynamic nature of micellar chains. In homopolymer solutions, the chain length is fixed, and the variation in $\tau_1$ with concentration reflects only changes in interchain interactions. In contrast, wormlike micelles are inherently polydisperse, and their chain length is determined by a dynamic equilibrium between scission and recombination. As a result, the mean micellar length increases with both concentration and sticker energy \cite{Wittmer1998,Wittmer2000}, following $\bar{L} \sim (c^{\mathrm{eff}})^{\delta} \exp(\kappa \epsilon_{st})$, and the length distribution shifts toward longer micelles with increasing concentration. Since the longest relaxation time is governed by the slowest-relaxing (longest) micelles in the system, this increase in micellar length introduces an additional length-dependent contribution to $\tau_1$. Consequently, in wormlike micelles, increasing concentration not only enhances intermicellar interactions but also increases the mean micellar length, leading to a stronger dependence of $\tau_1$ on concentration compared to homopolymer solutions.

\subsubsection{\label{sec:taust} Terminal relaxation time of stress components}

From our simulations, we can obtain the different contributions to the relaxation modulus $G(t)$, namely,  $G^{\mathrm{st}}(t)$, the contribution arising from the relaxation of the stress due to sticker–sticker interactions, $G^{\mathrm{bb}}(t)$, the contribution arising from the relaxation of the stress due to backbone–backbone and backbone–sticker interactions, and $G^{\mathrm{sp}}(t)$, the contribution arising from the relaxation of the stress due to spring forces. The terminal relaxation time in each case can be determined by fitting the tail of each of these stress components, which characterizes the decay of the stresses carried by each contribution. The procedure for evaluating the stress components is outlined below.

\begin{figure*}[tbph]
\begin{center}
\begin{tabular}{cc}
\includegraphics[width=8.5cm]{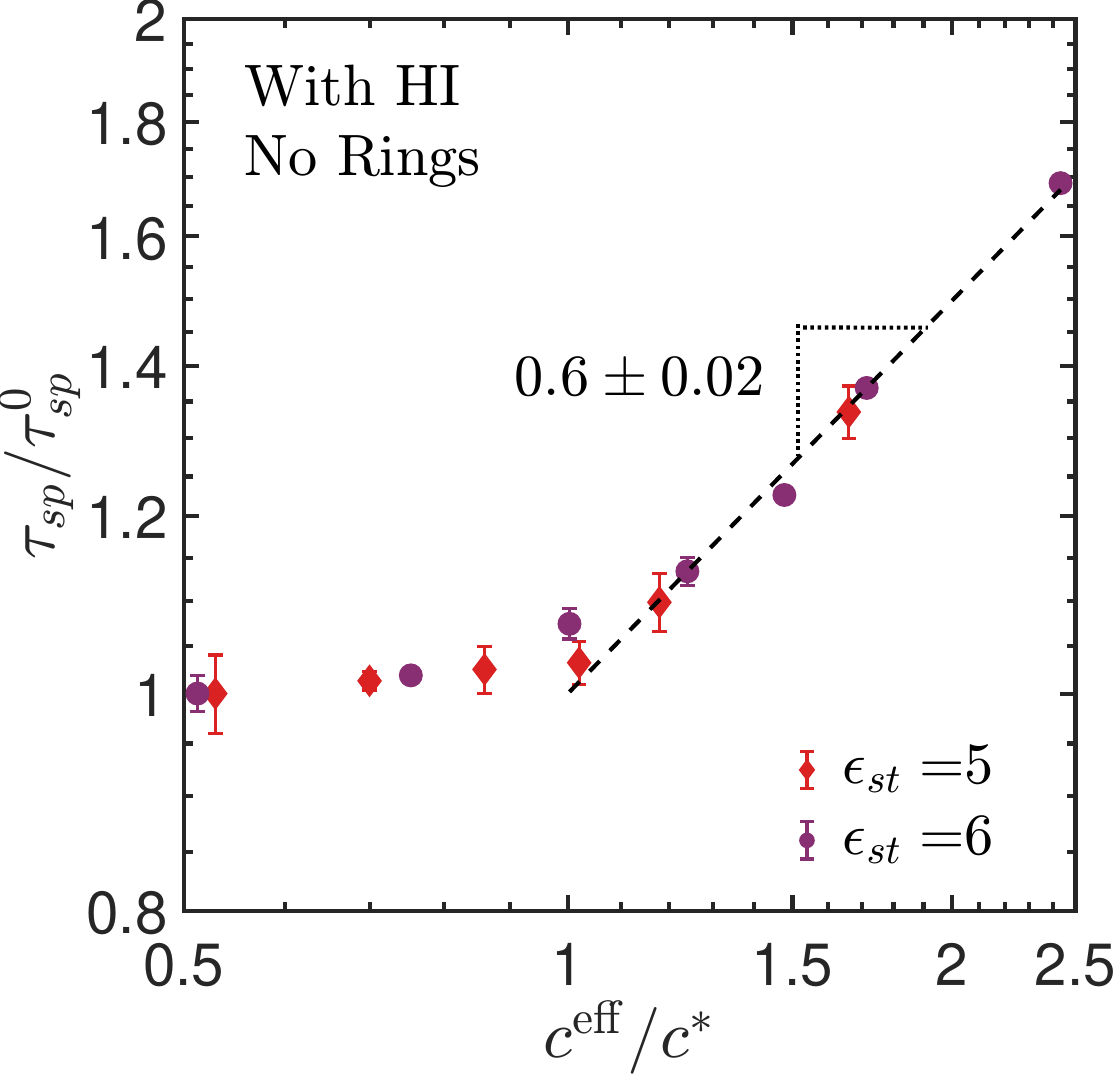} &
\includegraphics[width=9.5cm]{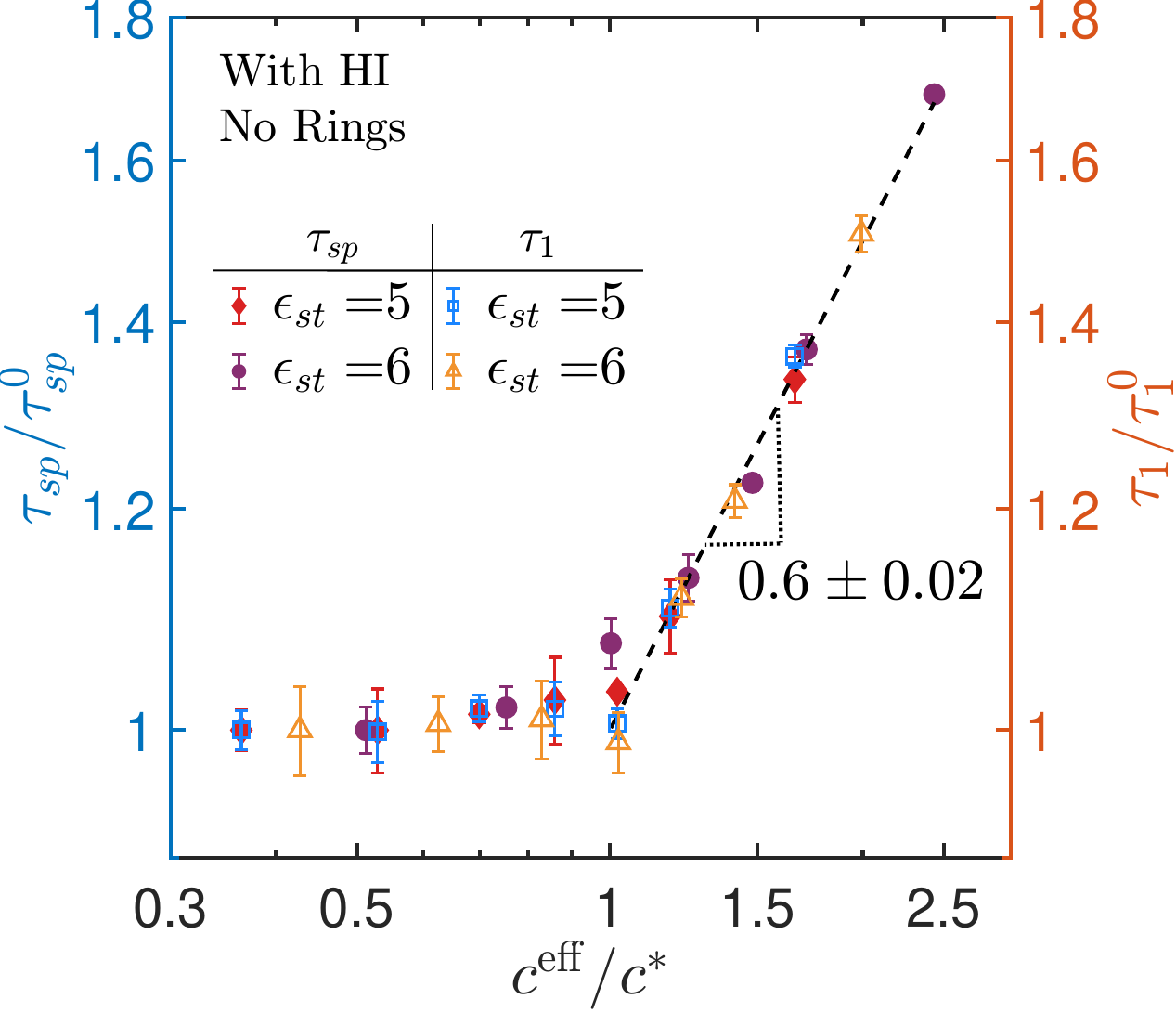} \\[-5pt] 
(a)  & (b) 
\end{tabular}
\vspace{-10pt}
\end{center}
\caption{(a) Scaling of the normalized terminal spring relaxation time ($\tau_{sp}/\tau_{sp}^0$) with the effective wormlike micelle concentration ($c^{\mathrm{eff}}/c^*$) for wormlike micellar solutions in the presence of hydrodynamic interactions and in the absence of rings, at sticker energies $\epsilon_{st} = 5, 6$. (b) Dependence of the normalized longest relaxation time ($\tau_1/\tau_1^0$) (refer \sref{sec:tau1}) and the normalized spring relaxation time ($\tau_{sp}/\tau_{sp}^0$) on the effective micelle concentration ($c^{\mathrm{eff}}/c^*$) under the same conditions.}
\label{fig16}
\end{figure*}

From \eref{netforce}, the net nondimensional force on the $\nu$-th bead in micelle $\alpha$ consists of the spring force along the same wormlike micelle, $\bm{F}_\nu^{\mathrm{sp}}$, and the forces arising from the SDK potential due to pairwise interactions between monomers, within a cut-off radius of bead $\nu$ in micelle $\alpha$, $\bm{F}_{\alpha\nu}^{\text{SDK}}$. For notational simplicity, forces arising from interactions between sticky monomers are denoted with the superscript `$\text{st}$', while those involving all other interactions (backbone-backbone and backbone-sticker monomers) are denoted with `$\text{bb}$', i.e., $\bm{F}_{\alpha\nu}^{\text{SDK}} =  \bm{F}_{\alpha\nu}^{\text{st}}+\bm{F}_{\alpha\nu}^{\text{bb}}$. 

With this notation, the dimensionless polymer stress tensor $\bm{S}$ for a single trajectory defined in \eref{tensor_S} can be rewritten as,
\begin{multline}
\label{tensor_SM}
    \bm{S} = \bm{S}^{st}+\bm{S}^{bb}+\bm{S}^{sp} \\ = \frac{1}{2} \sum_{\nu=1}^{N^{\text{T}}}\sum_{\substack{\mu=1 \\ \mu\ne\nu}}^{N^{\text{T}}} \bm{r}_{\nu\mu} \bm{F}_{\nu\mu}^{\text{st}} + \frac{1}{2} \sum_{\nu=1}^{N^{\text{T}}}\sum_{\substack{\mu=1 \\ \mu\ne\nu}}^{N^{\text{T}}} \bm{r}_{\nu\mu} \bm{F}_{\nu\mu}^{\text{bb}} \\ + \sum_{\alpha=1}^{\mathcal{N}_{\text{wlm}}} \sum_{i=1}^{N_s^{\text{L}}} \bm{Q}_{i}^{(\alpha)} \bm{F}^{{\text{(sp)}}}
    \left( \bm{Q}_{i}^{(\alpha)} \right)
\end{multline}
In accordance with \eref{Gtiso}, the shear relaxation modulus associated with contribution $M$,  under isotropic equilibrium is defined as: $G^{\text{M}}(t) = (1/3) \Big [  G_{xy}^{\text{M}}(t) + G_{xz}^{\text{M}}(t) + G_{yz}^{\text{M}}(t) \Big ]$ with the individual components, $G_{ij}^{\text{M}}(t)$, given by the Green–Kubo relation
\begin{equation}\label{stressauto_M}
    G_{ij}^{\text{M}}(t) =\frac{1}{\mathcal{N}_{\text{wlm}}} \, \Big\langle \! S_{ij}^{\text{M}}(0) \, S_{ij}^{\text{M}}(t) \! \Big\rangle 
\end{equation}
 Here, $M$ is an index representing the stress contributions due to sticker–sticker ($\text{st}$), backbone–backbone and backbone–sticker ($\text{bb}$), and spring ($\text{sp}$) interactions.

Fitting the individual stress components, $G^{\mathrm{st}}(t)$, $G^{\mathrm{bb}}(t)$, and $G^{\mathrm{sp}}(t)$, to a sum of exponentials of the form, $G^{\mathrm{M}}(t) = \sum_{k=1}^{n} a^{\mathrm{M}}_k \exp(-t/\tau^{\mathrm{M}}_k)$ where $a^{\mathrm{M}}_k$ and $\tau^{\mathrm{M}}_k$ are fitting parameters and $n$ denotes the number of exponential modes, allows us to extract the characteristic relaxation times ($\tau^{\mathrm{M}}_k$). The nondimensional terminal relaxation time for each stress component is identified as the largest timescale among $\{\tau^{\mathrm{M}}_k\}$, and is denoted by $\tau_{st}$, $\tau_{bb}$, and $\tau_{sp}$ for the respective contributions.

\fref{fig15}a shows the stress autocorrelation functions corresponding to different force contributions in the absence of rings in the micellar solution and in the presence of hydrodynamic interactions. The stress associated with backbone monomer interactions relaxes rapidly, followed by the stress arising from associative sticker interactions, while the stress arising from spring forces along the wormlike micelle backbone relaxes more slowly.

\fref{fig15}(b) and (c) show the effect of hydrodynamic interactions on $\tau_{st}$ and $\tau_{bb}$ respectively. In the presence of hydrodynamic interactions, bead motions become long-range correlated, which reduces the relative mobility of nearby stickers. Consequently, the stress arising from both backbone monomers and sticker–sticker interactions remains correlated over longer times, leading to slower decay. This results in $\tau_{st} \approx 1 \pm 0.01$ and $\tau_{bb} \approx 0.75 \pm 0.02$ with hydrodynamic interactions, compared to $\tau_{st} \approx 0.85 \pm 0.05$ and $\tau_{bb} \approx 0.54 \pm 0.01$ in their absence. Both $\tau_{st}$ and $\tau_{bb}$ exhibits a weak dependence on sticker energy and micelle concentration. This is because the relaxation of stresses associated with sticker–sticker and backbone–backbone interactions occurs at relatively short timescales, governed primarily by local configurational rearrangements. The presence or absence of rings does not significantly affect these timescales (not shown here).

Once the stresses arising from pairwise backbone and sticker interactions have relaxed, the remaining stress relaxation is governed by micellar scission and non-self recombination of wormlike micelles. These processes control the relaxation of the spring contribution, $G^{\mathrm{sp}}(t)$, which therefore occurs at much longer timescales. As shown in the \fref{fig15}(d), the corresponding terminal relaxation time $\tau_{sp}$ depends strongly on both the concentration of persistent worms and the sticker energy. At higher concentrations and larger sticker energies, the mean micellar length increases, leading to longer relaxation times.

When the terminal relaxation time, $\tau_{sp}$, is normalized by its value in the dilute limit for each sticker energy, $\tau_{sp}^0$ (see \fref{fig15}d), the normalized quantity $\tau_{sp}/\tau_{sp}^0$ collapses onto a master curve that is independent of sticker energy, as shown in \fref{fig16}(a). The data follow a scaling with concentration given by $\tau_{sp}/\tau_{sp}^0 \sim (c^{\mathrm{eff}}/c^*)^{0.6}$, which is consistent with the scaling of the longest relaxation time reported in \sref{sec:tau1} (see \fref{fig16}(b)). This behaviour is consistent with prior studies on associative polymers, where relaxation times obtained from fitting the tail of stress autocorrelation functions and from the inverse characteristic frequency exhibit similar scaling when appropriately nondimensionalized \cite{Robe2024}. In \sref{sec:mapping}, we further demonstrate that $\tau_{sp}$ corresponds to the terminal relaxation regime of wormlike micellar solutions in linear viscoelastic measurements, where the wormlike micellar network has fully relaxed through scission and recombination processes. 

\subsubsection{\label{sec:mapping} Mapping the molecular timescales on the linear viscoelastic curve of wormlike micelles}

\begin{figure}[tbph]
\begin{center}
\begin{tabular}{c}
\includegraphics[width=8.0cm]{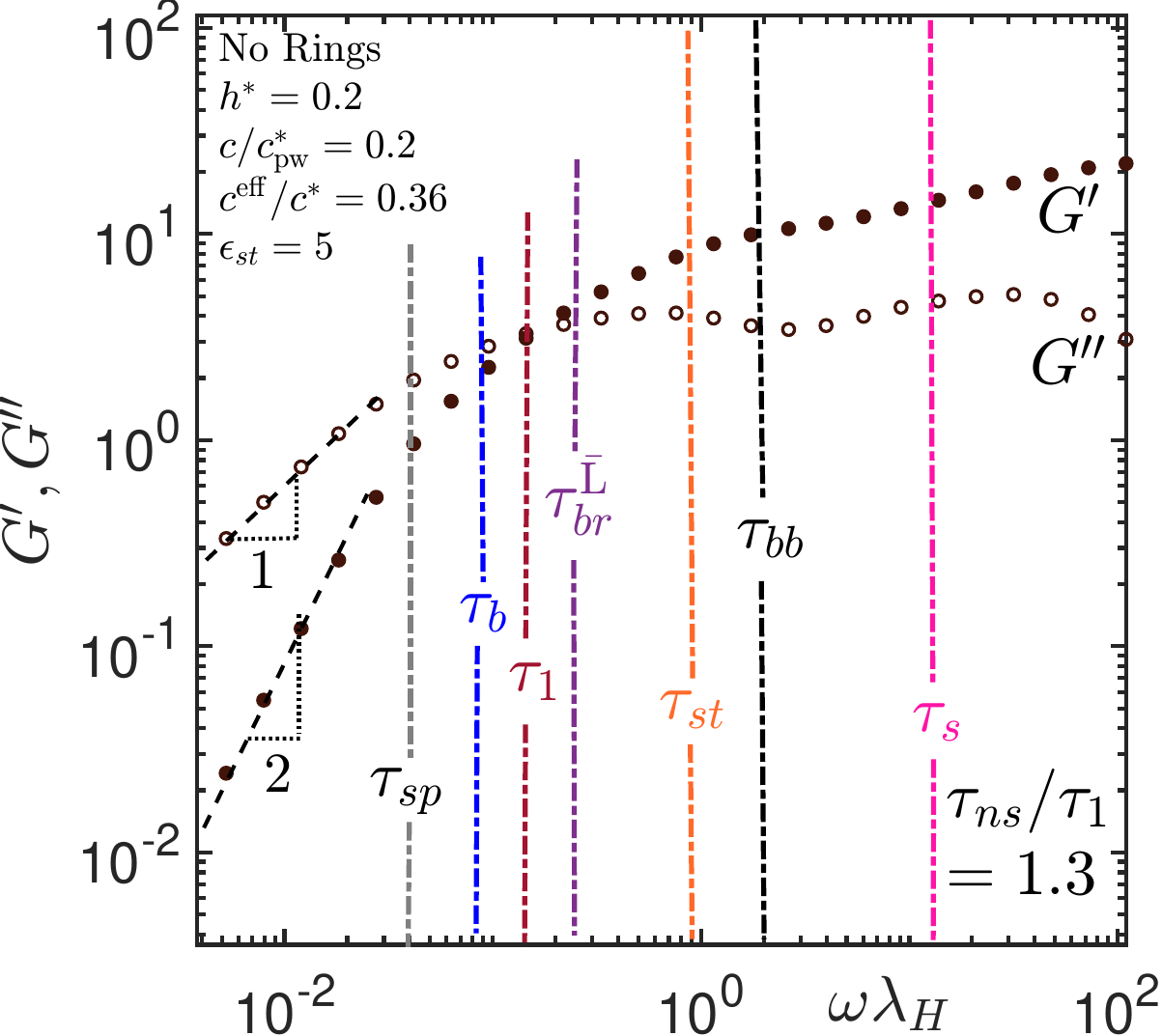} \\
(a)  \\[3pt]
\includegraphics[width=8.0cm]{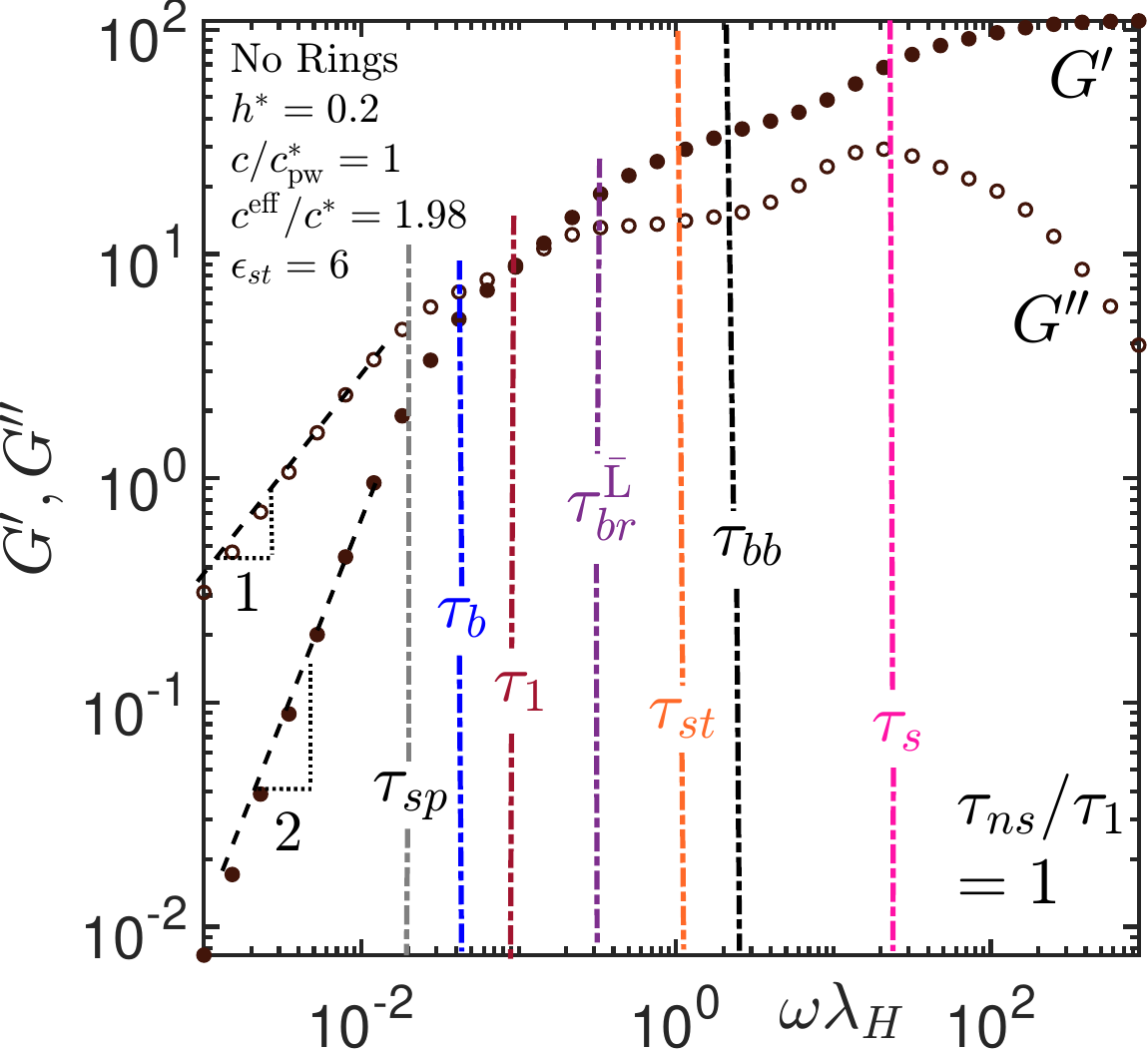} \\
 (b) \\[3pt]
\includegraphics[width=8.0cm]{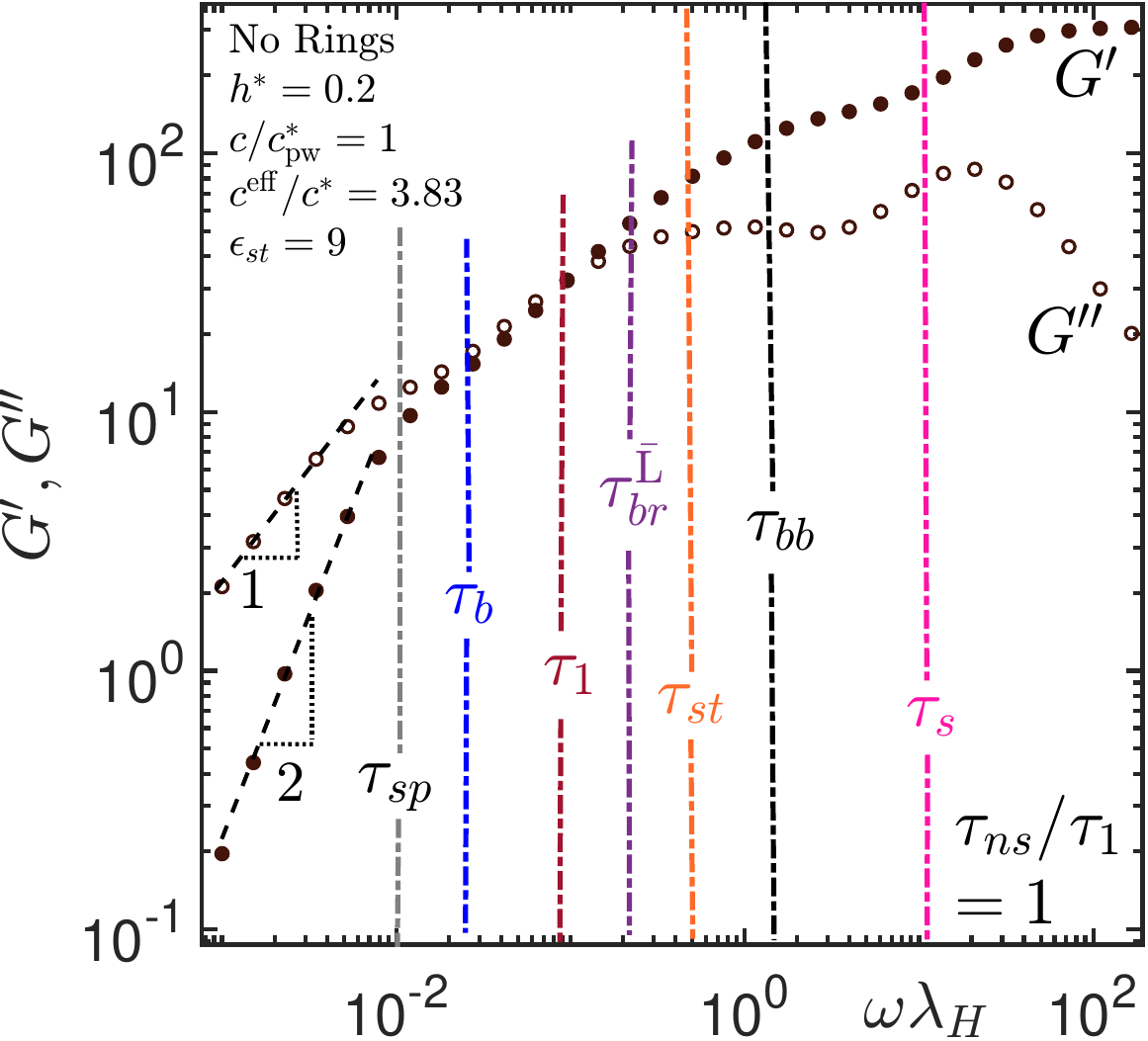} \\[-3pt]
(c) 
\end{tabular}
\vspace{-15pt}
\end{center}
\caption{Mapping of molecular timescales onto the linear viscoelastic response of wormlike micelles in the semidilute regime, in the absence of rings. The storage modulus, $G^{\prime}$, and loss modulus, $G^{\prime\prime}$, are shown as functions of angular frequency, $\omega$, for (a) $c^{\mathrm{eff}}/c^* = 0.36$, $\epsilon_{\mathrm{st}} = 5$; (b) $c^{\mathrm{eff}}/c^* = 1.98$, $\epsilon_{\mathrm{st}} = 6$; and (c) $c^{\mathrm{eff}}/c^* = 3.83$, $\epsilon_{\mathrm{st}} = 9$.}
\label{fig17}
\vspace{-10pt}
\end{figure}

The key molecular timescales extracted from simulations can be directly mapped onto the macroscopic linear viscoelastic response of wormlike micelles, using the relation $\tau = 1/\omega$ between time and frequency. \fref{fig17} illustrates this mapping for semidilute micellar solutions without rings at different effective concentrations and sticker energies.

At high frequencies (short times), the storage modulus $G^{\prime}$ approaches a plateau while the loss modulus $G^{\prime\prime}$ remains small, indicating an almost purely elastic response. In this regime, the transient micellar network does not have time to break or rearrange, so stickers behave as quasi-permanent crosslinks and the stress is stored elastically.

As the frequency decreases, the system begins to probe the short-time recombination dynamics of sticker pairs. On timescales of order the self-recombination time $\tau_s$, stickers repeatedly break and rebind with their original partners within a local volume. This local dynamical behaviour reduces the efficiency with which stickers store elastic stress and leads to the initial decrease of both $G^{\prime}$ and $G^{\prime\prime}$ away from the high-frequency plateau. At slightly longer timescales, the stress associated with backbone monomers relaxes, characterized by the terminal relaxation time $\tau_{bb}$. These processes correspond to configurational rearrangements of the micellar backbone, resulting in a further decay of the storage modulus $G^{\prime}$, while the loss modulus $G^{\prime\prime}$ approaches a plateau. Beyond this regime, the stress contribution arising from sticker–sticker interactions relaxes. Once the corresponding stress correlations decay, the elastic stress decreases further until the system reaches frequencies corresponding to the terminal relaxation time of the sticker stress, $\tau_{st}$.

At intermediate frequencies below $1/\tau_{st}$, micellar breakage becomes dynamically relevant. Breakage events release paired stickers and allow micelles to relax further, while intrachain spring stresses also relax. These processes reduce the stored elastic stress, leading to a continued decay of $G^{\prime}$. In the Cole–Cole representation (III regime in \fref{fig1}a), a pronounced maximum is observed at frequencies corresponding to the breakage time of mean micellar length $\tau_{br}^{\bar{\text{L}}}$. This maximum reflects the timescale at which micelle scission contributes most strongly to viscous dissipation. 

Beyond this regime, both the storage and loss moduli continue to decay as the system explores longer-time relaxation processes. The plateau observed in $G^{\prime\prime}$ terminates near the breakage time associated with the mean micellar length, $\tau_{br}^{\bar{L}}$. At lower frequencies, the loss modulus $G^{\prime\prime}$ decreases more slowly relative to the storage modulus $G^{\prime}$ and eventually crosses it at a frequency corresponding to the longest relaxation time $\tau_1$. We find that $\tau_1$ coincides with the non-self recombination timescale $\tau_{ns}$, indicating that stress relaxation is governed by recombination events in which micelles upon breakage associate with new partners, giving $\tau_1/\tau_{ns} \sim \mathcal{O}(1)$. 

Finally, on timescales longer than the bond lifetime
$\tau_b$, the system enters the terminal flow regime, characterized by $G^{\prime} \sim \omega^2$ and $G^{\prime \prime} \sim \omega$, as expected for viscoelastic fluids. At these timescales, the stress has fully relaxed, corresponding to the terminal relaxation time associated with spring stresses, $\tau_{sp}$.

In the next section, the cumulative effect of all these relaxation processes on the steady zero-shear viscosity is probed. 

\subsubsection{\label{sec:vis_scaling} Zero-shear rate viscosity}

In the linear viscoelastic regime, the dimensional zero-shear viscosity ($\hat{\eta}_0$) is obtained from the time integral of the shear relaxation modulus,
\begin{equation}
\hat{\eta}_0 = \mathcal{N}_{\text{wlm}} \, k_{\text{B}} T \lambda_H \int_0^\infty G(t)\,dt
\end{equation}
where $G(t)$ is the nondimensional stress relaxation modulus defined in \eref{Gtiso}. The corresponding dimensionless viscosity is defined as $\eta_0 = \hat{\eta}_0/\left(\mathcal{N}_{\text{wlm}} \, k_{\text{B}} T\lambda_H\right)$. 

\begin{figure}[tbh]
\centering
\includegraphics[width=8.5cm]{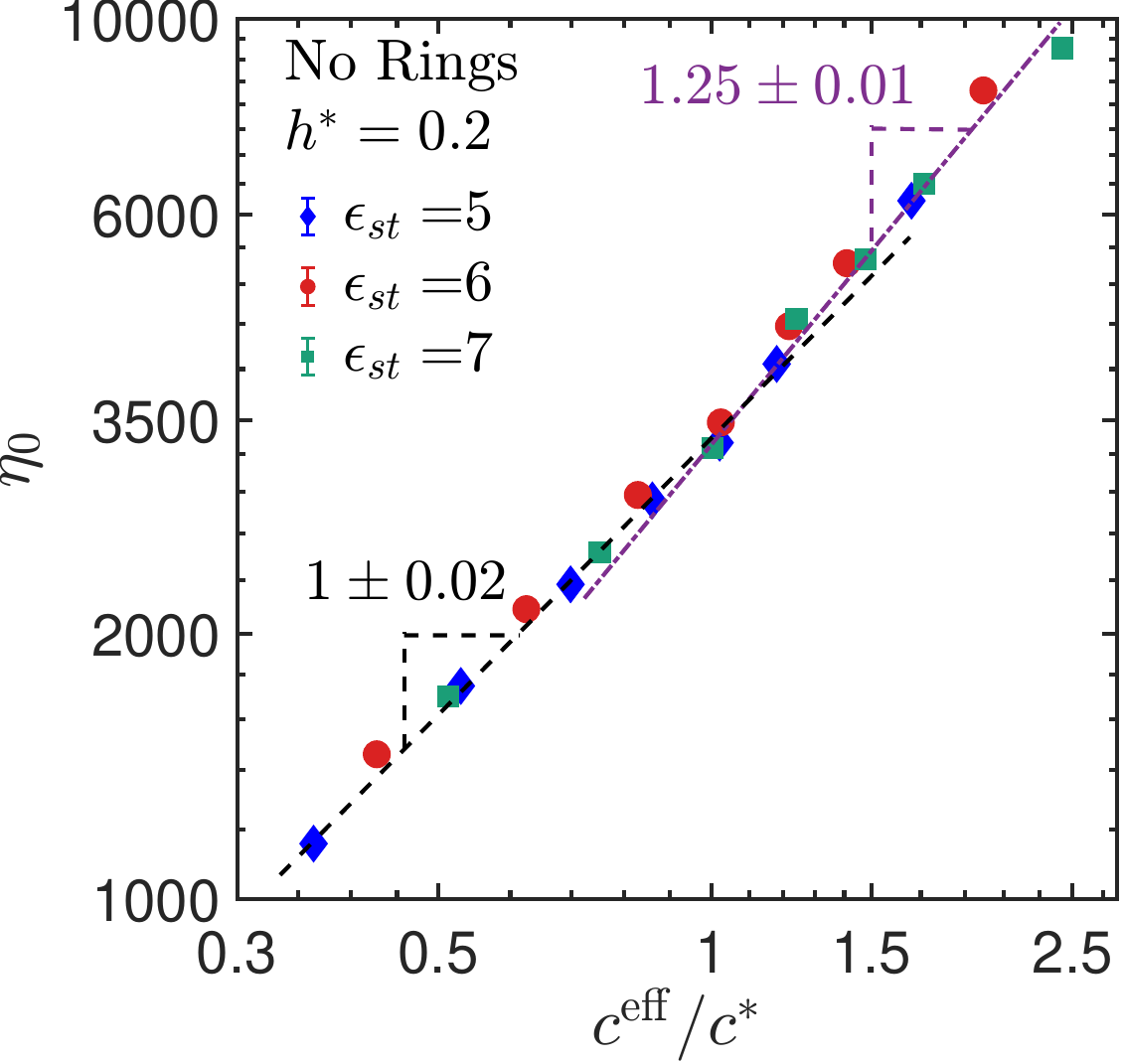}  
\vspace{-10pt}
\caption{Dimensionless zero-shear viscosity ($\eta_0$) as a function of effective micelle concentration, $c^{\mathrm{eff}}/c^*$, for systems with hydrodynamic interactions and no ring structures, at different sticker energies ($\epsilon_{st} = 5,6$ and $7$).}
\label{fig18}
\vspace{-10pt}
\end{figure}

Since $\eta_0$ involves an integral over all timescales, it is dominated by the slowest stress relaxation processes and reflects the long-time relaxation behaviour of the system. \fref{fig18} shows the dimensionless zero-shear viscosity as a function of the effective micelle concentration $c^{\mathrm{eff}}/c^*$ for wormlike micellar solutions with hydrodynamic interactions and in the absence of rings. For all the sticker energies, dimensionless zero shear rate viscosity data collapse onto a single master curve and it follows a power-law dependence on concentration, $\eta_0 \sim (c^{\mathrm{eff}}/c^*)^m$, with two distinct regimes. Below the overlap concentration ($c^{\mathrm{eff}}/c^* \leq 1$), the viscosity increases linearly with concentration ($m \approx 1$), consistent with behaviour in dilute solutions. In the semidilute unentangled regime ($c^{\mathrm{eff}}/c^* \geq 1$), the viscosity exhibits a stronger concentration dependence with an exponent $m \approx 1.25$. Previous experimental and simulation studies have reported values of $m$ ranging from approximately $1$ to $1.7$ for unentangled wormlike micelles \cite{Cates1990,Hoffmann1994,Berret2006,Wang2017}, and the present result lies within this range.

Interestingly, the observed exponent $m \approx 1.25$ is comparable to that reported for semidilute unentangled homopolymer solutions under good solvent ($\nu \approx 0.6$) conditions, for which scaling theory \cite{Rubinstein2003} predicts $m = 1/(3\nu - 1) \approx 1.25$. This similarity reflects a common physical mechanism in the two kinds of solutions at long times. In the unentangled regime, stress relaxation in the wormlike micelles at the longest times is governed by the relaxation of whole micelles of mean contour length $\bar{L}$. By the time these long timescales are reached, all local and intermediate-time relaxation processes associated with reversible sticker interactions have already decayed, and stress relaxes through the longest mode associated with the mean micelle length. This interpretation is further supported by our earlier work \cite{Kumar2025}, where we showed that when the effective concentration of wormlike micelles is matched to that of a monodisperse homopolymer solution and the polymer chain length is chosen equal to the mean micellar length, the longest nondimensional relaxation time is identical in both systems. Consequently, despite their polydispersity, it appears that the steady-state viscosity of unentangled wormlike micellar solutions is governed by the same long-time relaxation physics as that of unentangled polymer solutions, leading to similar concentration scaling of $\eta_0$.

Together, these results establish a clear connection between the microscopic relaxation dynamics of wormlike micelles and their macroscopic steady-state rheology in the unentangled regime.

\section{\label{sec:conclusion} Conclusion}

In this work, we developed and employed a mesoscopic Brownian dynamics framework to establish a direct and quantitative connection between microscopic dynamics and macroscopic linear viscoelastic response of dilute and unentangled semidilute wormlike micellar solutions. Wormlike micelles are modeled as assemblies of persistent worms that reversibly undergo scission and fusion via sticker–sticker interactions, inherently generating polydisperse solutions of linear and ring micelles. Long-ranged hydrodynamic interactions are explicitly incorporated using a Rotne–Prager–Yamakawa tensor, enabling a systematic investigation of the combined effects of reversible bonding, micellar topology, and hydrodynamic coupling on molecular timescales and macroscopic rheology.

A central contribution of this study is the identification and characterization of a hierarchy of molecular timescales governing micellar dynamics: 
\begin{itemize}
    \item  Bond breakage time ($\tau_b$): 
    The bond breakage time characterizes the lifetime of individual sticker–sticker bonds and exhibits an exponential dependence on the sticker energy ($\epsilon_{st}$). The presence of hydrodynamic interactions systematically prolongs bond lifetimes by reducing the relative mobility of bonded stickers through long-ranged hydrodynamic coupling.
    
    \item Self-recombination time ($\tau_s$): The self-recombination time corresponds to short-time recombination of unbound stickers created by the same scission event and is governed by diffusion-controlled kinetics. It is independent of the concentration of wormlike micelles and sticker strength but is sensitive to hydrodynamic interactions. This regime manifests as a power-law decay in the recombination probability distribution at short times.

    \item Non-self recombination time ($\tau_{ns}$): At longer times, unbound stickers lose memory of their original partners and recombine with new partners via mean-field kinetics, characterized by an exponential decay in the recombination probability distribution. The corresponding characteristic timescale, $\tau_{ns}$, represents the average time taken by a free sticker to encounter and recombine with a new partner. The non-self recombination time decreases with increasing concentration and increases with sticker strength, while hydrodynamic interactions further increase $\tau_{ns}$ by slowing relative diffusion. When nondimensionalized by $\tau_b$ and expressed as a function of the mean length of linear micelles $\bar{L}$, $\tau_{ns}$ collapses onto a universal master curve, $\tau_{ns}/\tau_b \sim \bar{L}^{-4/3}$. The presence of rings reduces $\tau_{ns}$ by introducing additional short-range recombination pathways, wherein free end stickers can readily recombine with other free ends on the same micelle.

    \item Breakage time of micelles ($\tau_{br}^{\text{L}}$ and $\tau_{br}^{\bar{\text{L}}}$): Using Kaplan–Meier survival analysis, we quantified the breakage time of micelles of contour length $L$,  $\tau_{br}^{\text{L}}$. Longer micelles exhibit a higher probability of breakage, leading to the scaling $\tau_{br}^{\text{L}} \sim 1/L$, in agreement with mean-field predictions for equilibrium polymers \cite{Cates1987,Wittmer1998, Huang2006}. Increasing sticker strength and the inclusion of hydrodynamic interactions slow micellar breakage and increase $\tau_{br}^{\text{L}}$. However, upon normalization by $\tau_b$, all data collapse onto a universal master curve, indicating that microscopic interaction details enter micellar breakage kinetics primarily through $\tau_b$. Following previous studies \cite{Cates1987,Padding2004}, we estimated the breakage time of micelles of mean length $\bar{L}$ using the established mean-field scaling $\tau_{br}^{\bar{\text{L}}}/\tau_b \sim 1/\bar{L}$.

    \item Terminal relaxation time of sticker-sticker stress component ($\tau_{st}$): By decomposing the stress tensor into distinct microscopic contributions, the relaxation of sticker–sticker stress components is quantified. Fitting the terminal decay of this stress contribution yields the characteristic relaxation time $\tau_{st}$. The timescale $\tau_{st}$ is found to be independent of concentration and sticker strength, reflecting the fact that sticker–sticker stresses relax on relatively short timescales and decorrelate before longer-time network rearrangements, where concentration and sticker energy effects become significant and play a role. Hydrodynamic interactions, however, slightly increase $\tau_{st}$ by slowing the relative motion of stickers and thereby prolonging the time required for stress decorrelation at these short timescales.

   \item Terminal relaxation time of backbone monomers ($\tau_{bb}$) and spring stress component ($\tau_{sp}$): In addition to the terminal relaxation time associated with sticker–sticker interactions, the present study identifies distinct terminal timescales governing the relaxation of backbone ($\tau_{sp}$) and spring ($\tau_{sp}$) force contributions to the stress. The backbone–backbone and backbone–sticker stress components  relax on relatively short timescales, reflecting the rapid decorrelation of excluded-volume interactions between monomers. The timescale $\tau_{bb}$ is weakly dependent on sticker energy and micelle concentration, while the presence of hydrodynamic interactions increases $\tau_{bb}$ due to long-range correlations in bead motion. In contrast, the spring stress component ($\tau_{sp}$) exhibits significantly slower relaxation, arising from configurational rearrangements of the micellar backbone. This relaxation is governed by the intramicellar dynamics of a breakable chain and depends strongly on the mean micellar length as well as hydrodynamic interactions. At long times, the decay of the spring stress dominates the relaxation modulus once sticker-mediated stress contributions have relaxed, thereby contributing to the terminal viscoelastic response. 
    
    \item Longest relaxation time of wormlike micelles ($\tau_1$): The longest relaxation time $\tau_1$ is obtained from the crossover frequency at which the storage and loss moduli intersect, $\tau_1 = 1/\omega_{(G^\prime = G^{\prime\prime})}$. The timescale $\tau_1$ increases with increasing sticker energy and concentration, reflecting the growth of the mean micellar length and the corresponding increase in the time required for relaxation of entire micelles. In the semidilute regime ($c^{\text{eff}}/c^* \ge 1$), $\tau_1$ exhibits a power-law dependence on concentration, $\tau_1 \sim \left(c^{\text{eff}}/c^*\right)^{0.6}$. Upon normalizing $\tau_1$ by its value in the dilute limit for different sticker energies, $\tau_1^0$ (see \fref{fig13}a), the data collapse onto a single master curve for all sticker energies. Furthermore, the spring relaxation time ($\tau_{sp}$) follows a similar scaling as the longest relaxation time, with $\tau_{sp} \sim \left(c^{\text{eff}}/c^*\right)^{0.6}$.

\end{itemize}

The linear viscoelastic response, computed using the Green–Kubo relation, provides a direct macroscopic manifestation of this hierarchy of timescales. Unlike monodisperse homopolymer solutions, wormlike micelles exhibit distinct non-monotonic features in the storage and loss moduli and in Cole–Cole representations. By systematically mapping characteristic nondimensional frequencies ($\omega=1/\tau$) onto these representations, we show that the lowest-frequency (long-time features) are governed by the bond breakage time $\tau_b$ and terminal relaxation time of spring stresses ($\tau_{sp}$). The intermediate-frequency regime is primarily controlled by collective micellar processes, including micellar breakage at the mean length scale and longest relxation time of wormlike micelles, characterized by $\tau_{br}^{\bar{\text{L}}}$ and $\tau_{1}$, respectively. The ratio of the longest relaxation time to the non-self recombination time ($\tau_{ns}$) is of $\mathcal{O}(1)$. At higher frequencies, the response is dominated by short-time processes such as self-recombination $\tau_s$, relaxation of the backbone ($\tau_{bb}$) and sticker stress component $\tau_{st}$ respectively. This mapping shows how reversible scission, recombination, and intramicellar relaxation processes determine the linear viscoelastic response of unentangled wormlike micelles, within a single framework.

The zero-shear rate viscosity follows the scaling $\eta_0 \sim (c^{\mathrm{eff}}/c^*)^m$, exhibiting the same concentration dependence as monodisperse homopolymer solutions under good solvent conditions, with $m = 1$ in the dilute regime and $m = 1.25$ in the semidilute regime. Also, in our earlier work \cite{Kumar2025}, we showed that the longest relaxation time of wormlike micelles, $\tau_1$, is identical to that of a monodisperse homopolymer solution when the polymer chain length and polymer concentration are matched to the mean micellar length and effective micellar concentration, respectively. Together, these observations indicate that, at long timescales, $\tau_1$ is governed by the relaxation of the longest mode associated with the mean micellar length.

Overall, this study clarifies how microscopic scission, recombination, and relaxation processes control the linear viscoelastic response of unentangled wormlike micelles. By explicitly identifying the relevant molecular timescales and their roles in the viscoelastic spectrum, the results provide a clear physical basis for interpreting rheological behaviour in these systems. The framework developed here offers a robust foundation for interpreting experimental rheology in terms of underlying molecular processes and can be readily extended to investigate semiflexibility and nonlinear flow.

\bibliography{wlm}

\begin{thebibliography}{}
\newcommand{\enquote}[1]{``#1''}

\bibitem[Anderson \emph{et~al.}(2020)Anderson, Glaser and Glotzer]{HOOMD2020}
Anderson, J.~A., J.~Glaser and S.~C. Glotzer, \enquote{{HOOMD}-blue: A python
  package for high-performance molecular dynamics and hard particle monte carlo
  simulations,} Comput. Mater. Sci. \textbf{173}, 109363 (2020).

\bibitem[Berret(2006)Berret]{Berret2006}
Berret, J.-F., \enquote{Rheology of wormlike micelles: Equilibrium properties
  and shear banding transitions,} in \emph{Molecular Gels: Materials with
  Self-Assembled Fibrillar Networks}, eds. R.~G. Weiss and P.~Terech, pp.
  667--720, Springer Netherlands, Dordrecht (2006).

\bibitem[Bird \emph{et~al.}(1987)Bird, Curtiss, Armstrong and
  Hassager]{Bird1987}
Bird, R.~B., C.~F. Curtiss, R.~C. Armstrong and O.~Hassager, \emph{Dynamics of
  Polymeric Liquids - Volume 2 : Kinetic Theory}, John Wiley and Sons, New York
  (1987).

\bibitem[Cates(1987)Cates]{Cates1987}
Cates, M., \enquote{Reptation of living polymers: dynamics of entangled
  polymers in the presence of reversible chain-scission reactions,}
  Macromolecules \textbf{20}, 2289--2296 (1987).

\bibitem[Cates(1988)Cates]{Cates1988}
Cates, M., \enquote{Dynamics of living polymers and flexible surfactant
  micelles: scaling laws for dilution,} J. Phys. \textbf{49}, 1593--1600
  (1988).

\bibitem[Cates and Candau(1990)Cates and Candau]{Cates1990}
Cates, M.~E. and S.~J. Candau, \enquote{Statics and dynamics of worm-like
  surfactant micelles,} J. Phys.: Condens. Matter \textbf{2}, 6869 (1990).

\bibitem[Cates and Candau(2001)Cates and Candau]{Cates2001}
Cates, M.~E. and S.~J. Candau, \enquote{Ring-driven shear thickening in
  wormlike micelles?} Europhys. Lett. \textbf{55}, 887 (2001).

\bibitem[Cates and Fielding(2006)Cates and Fielding]{Cates2006}
Cates, M.~E. and S.~M. Fielding, \enquote{Rheology of giant micelles,} Adv.
  Phys. \textbf{55}, 799--879 (2006).

\bibitem[Chu \emph{et~al.}(2013)Chu, Dreiss and Feng]{Dreiss2013}
Chu, Z., C.~A. Dreiss and Y.~Feng, \enquote{Smart wormlike micelles,} Chem.
  Soc. Rev. \textbf{42}, 7174--7203 (2013).

\bibitem[Dreiss(2007)Dreiss]{Dreiss2007}
Dreiss, C.~A., \enquote{Wormlike micelles: where do we stand? {R}ecent
  developments, linear rheology and scattering techniques,} Soft Matter
  \textbf{3}, 956--970 (2007).

\bibitem[Fiore \emph{et~al.}(2017)Fiore, Balboa~Usabiaga, Donev and
  Swan]{PSE2017}
Fiore, A.~M., F.~Balboa~Usabiaga, A.~Donev and J.~W. Swan, \enquote{Rapid
  sampling of stochastic displacements in brownian dynamics simulations,} J.
  Chem. Phys. \textbf{146}, 124116 (2017).

\bibitem[Gowers and Carbone(2015)Gowers and Carbone]{Gowers2015}
Gowers, R.~J. and P.~Carbone, \enquote{A multiscale approach to model hydrogen
  bonding: The case of polyamide,} J. Chem. Phys. \textbf{142}, 224907 (2015).

\bibitem[Hoffmann(1994)Hoffmann]{Hoffmann1994}
Hoffmann, H., \emph{Viscoelastic Surfactant Solutions}, chap.~1, pp. 2--31, ACS
  Symp. Ser. (1994).

\bibitem[Howard \emph{et~al.}(2019)Howard, Statt, Madutsa, Truskett and
  Panagiotopoulos]{Nlist2019}
Howard, M.~P., A.~Statt, F.~Madutsa, T.~M. Truskett and A.~Z. Panagiotopoulos,
  \enquote{Quantized bounding volume hierarchies for neighbor search in
  molecular simulations on graphics processing units,} Comput. Mater. Sci.
  \textbf{164}, 139--146 (2019).

\bibitem[Huang \emph{et~al.}(2009)Huang, Ryckaert and Xu]{Ryckaert2009}
Huang, C.-C., J.-P. Ryckaert and H.~Xu, \enquote{Structure and dynamics of
  cylindrical micelles at equilibrium and under shear flow,} Phys. Rev. E
  \textbf{79}, 041501 (2009).

\bibitem[Huang \emph{et~al.}(2006a)Huang, Xu, Crevel, Wittmer and
  Ryckaert]{Huang2006}
Huang, C.-C., H.~Xu, F.~Crevel, J.~Wittmer and J.-P. Ryckaert,
  \enquote{Reaction kinetics of coarse-grained equilibrium polymers: a brownian
  dynamics study,} in \emph{Computer Simulations in Condensed Matter Systems:
  From Materials to Chemical Biology Volume 2}, pp. 379--418, Springer, Berlin,
  Heidelberg (2006a).

\bibitem[Huang \emph{et~al.}(2006b)Huang, Xu and Ryckaert]{Ryckaert2006}
Huang, C.-C., H.~Xu and J.-P. Ryckaert, \enquote{Kinetics and dynamic
  properties of equilibrium polymers,} J. Chem. Phys. \textbf{125}, 094901
  (2006b).

\bibitem[Israelachvili(2011)Israelachvili]{Israel2011}
Israelachvili, J.~N., \emph{Intermolecular and surface forces}, Academic Press,
  Burlington, MA, 3rd edn. (2011).

\bibitem[Jain \emph{et~al.}(2012)Jain, Sunthar, D{\"u}nweg and
  Prakash]{JainPRE2012}
Jain, A., P.~Sunthar, B.~D{\"u}nweg and J.~R. Prakash, \enquote{Optimization of
  a {B}rownian dynamics algorithm for semidilute polymer solutions,} Phys. Rev.
  E \textbf{85}, 066703 (2012).

\bibitem[Kaplan and Meier(1958)Kaplan and Meier]{Kaplan1958}
Kaplan, E.~L. and P.~Meier, \enquote{Nonparametric estimation from incomplete
  observations,} J. Am. Stat. Assoc. \textbf{53}, 457--481 (1958).

\bibitem[Katashima \emph{et~al.}(2022)Katashima, Kudo, Naito, Nagatoishi,
  Miyata, Chung, Tsumoto and Sakai]{Katashima2022}
Katashima, T., R.~Kudo, M.~Naito, S.~Nagatoishi, K.~Miyata, U.-i. Chung,
  K.~Tsumoto and T.~Sakai, \enquote{Experimental comparison of bond lifetime
  and viscoelastic relaxation in transient networks with well-controlled
  structures,} ACS Macro Lett. \textbf{11}, 753--759 (2022), PMID: 35594190.

\bibitem[Klein(2012)Klein]{Klein2012}
Klein, J., \enquote{The statistical analysis of failure time data,}
  Technometrics \textbf{24} (2012).

\bibitem[Koide(2023)Koide]{Koide2023}
Koide, Y., \enquote{Recombination statistics of nonionic surfactant micelles at
  equilibrium,} J. Chem. Phys. \textbf{159}, 224906 (2023).

\bibitem[Koide and Goto(2022)Koide and Goto]{Koide2022}
Koide, Y. and S.~Goto, \enquote{Flow-induced scission of wormlike micelles in
  nonionic surfactant solutions under shear flow,} J. Chem. Phys. \textbf{157},
  084903 (2022).

\bibitem[Kumar \emph{et~al.}(2025)Kumar, Tabor, Sunthar and
  Ravi~Prakash]{Kumar2025}
Kumar, A., R.~F. Tabor, P.~Sunthar and J.~Ravi~Prakash, \enquote{A mesoscopic
  model for the rheology of dilute and semidilute solutions of wormlike
  micelles,} J. Rheol. \textbf{69}, 873--903 (2025).

\bibitem[Kumar and Saha~Dalal(2022)Kumar and Saha~Dalal]{Indranil2022}
Kumar, P. and I.~Saha~Dalal, \enquote{Fraenkel springs as an efficient
  approximation to rods for brownian dynamics simulations and modeling of
  polymer chains,} Macromol. Theory Simul. \textbf{31}, 2200008 (2022).

\bibitem[{Martin In} \emph{et~al.}(1999){Martin In}, Aguerre-Chariol and
  Zana]{Martin1999}
{Martin In}, O.~Aguerre-Chariol and R.~Zana, \enquote{Closed-looped micelles in
  surfactant tetramer solutions,} J. Phys. Chem. B \textbf{103}, 7747--7750
  (1999).

\bibitem[Mordvinkin \emph{et~al.}(2021)Mordvinkin, D{\"o}hler, Binder, Colby
  and Saalwächter]{Mordvinkin2021}
Mordvinkin, A., D.~D{\"o}hler, W.~H. Binder, R.~H. Colby and K.~Saalwächter,
  \enquote{Rheology, sticky chain, and sticker dynamics of supramolecular
  elastomers based on cluster-forming telechelic linear and star polymers,}
  Macromolecules \textbf{54}, 5065--5076 (2021).

\bibitem[Nicolas-Morgantini(2007)Nicolas-Morgantini]{Nicolas2007}
Nicolas-Morgantini, L., \enquote{Giant micelles and shampoos,} in \emph{Giant
  Micelles}, pp. 493--514, CRC Press, Boca Raton (2007).

\bibitem[Oelschlaeger \emph{et~al.}(2002)Oelschlaeger, Waton, Buhler, Candau
  and Cates]{Oelschlaeger2002}
Oelschlaeger, C., G.~Waton, E.~Buhler, S.~Candau and M.~Cates,
  \enquote{Rheological and light scattering studies of cationic fluorocarbon
  surfactant solutions at low ionic strength,} Langmuir \textbf{18}, 3076--3085
  (2002).

\bibitem[Oelschlaeger \emph{et~al.}(2003)Oelschlaeger, Waton and
  Candau]{Oelschlaeger2003}
Oelschlaeger, C., G.~Waton and S.~J. Candau, \enquote{Rheological behavior of
  locally cylindrical micelles in relation to their overall morphology,}
  Langmuir \textbf{19}, 10495--10500 (2003).

\bibitem[O'Shaughnessy and Yu(1995)O'Shaughnessy and Yu]{Shaughnessy1995}
O'Shaughnessy, B. and J.~Yu, \enquote{Rheology of wormlike micelles: two
  universality classes,} Phys. Rev. Lett. \textbf{74}, 4329 (1995).

\bibitem[Padding and Boek(2004)Padding and Boek]{Padding2004}
Padding, J.~T. and E.~S. Boek, \enquote{Evidence for diffusion-controlled
  recombination kinetics in model wormlike micelles,} Europhys. Lett.
  \textbf{66}, 756 (2004).

\bibitem[Padding \emph{et~al.}(2008)Padding, Boek and Briels]{Padding2008}
Padding, J.~T., E.~S. Boek and W.~J. Briels, \enquote{Dynamics and rheology of
  wormlike micelles emerging from particulate computer simulations,} J. Chem.
  Phys. \textbf{129}, 074903 (2008).

\bibitem[Rehage and Hoffmann(1991)Rehage and Hoffmann]{Rehage1991}
Rehage, H. and H.~Hoffmann, \enquote{Viscoelastic surfactant solutions: model
  systems for rheological research,} Mol. Phys. \textbf{74}, 933--973 (1991).

\bibitem[Robe \emph{et~al.}(2024)Robe, Santra, McKinley and Prakash]{Robe2024}
Robe, D., A.~Santra, G.~H. McKinley and J.~R. Prakash, \enquote{Evanescent
  gels: Competition between sticker dynamics and single-chain relaxation,}
  Macromolecules \textbf{57}, 4220--4235 (2024).

\bibitem[Rubinstein and Colby(2003)Rubinstein and Colby]{Rubinstein2003}
Rubinstein, M. and R.~H. Colby, \emph{Polymer Physics}, Oxford University
  Press, New York (2003).

\bibitem[Shibaev \emph{et~al.}(2015)Shibaev, Molchanov and
  Philippova]{Shibaev2015}
Shibaev, A.~V., V.~S. Molchanov and O.~E. Philippova, \enquote{Rheological
  behavior of oil-swollen wormlike surfactant micelles,} Chem. B \textbf{119},
  15938--15946 (2015).

\bibitem[Soddemann \emph{et~al.}(2001)Soddemann, Duenweg and Kremer]{SDK2001}
Soddemann, T., B.~Duenweg and K.~Kremer, \enquote{A generic computer model for
  amphiphilic systems,} Eur. Phys. J. E \textbf{6} (2001).

\bibitem[Stukalin \emph{et~al.}(2013)Stukalin, Cai, Kumar, Leibler and
  Rubinstein]{Stukalin2013}
Stukalin, E.~B., L.-H. Cai, N.~A. Kumar, L.~Leibler and M.~Rubinstein,
  \enquote{Self-healing of unentangled polymer networks with reversible bonds,}
  Macromolecules \textbf{46}, 7525--7541 (2013).

\bibitem[Sullivan \emph{et~al.}(2007)Sullivan, Nelson, Anderson and
  Hughes]{Sullivan2007}
Sullivan, P., E.~B. Nelson, V.~Anderson and T.~Hughes, \enquote{Oilfield
  applications of giant micelles,} in \emph{Giant Micelles}, pp. 453--472, CRC
  Press, Boca Raton (2007).

\bibitem[Varakhedkar \emph{et~al.}(2026)Varakhedkar, Sunthar and
  Prakash]{Amit2026}
Varakhedkar, A., P.~Sunthar and J.~R. Prakash, \enquote{Linear viscoelasticity
  of semiflexible polymers with hydrodynamic interactions,} Soft Matter
  \textbf{22}, 369--386 (2026).

\bibitem[Wang \emph{et~al.}(2017)Wang, Feng, Agrawal and Raghavan]{Wang2017}
Wang, J., Y.~Feng, N.~R. Agrawal and S.~R. Raghavan, \enquote{Wormlike micelles
  versus water-soluble polymers as rheology-modifiers: similarities and
  differences,} Phys. Chem. Chem. Phys. \textbf{19}, 24458--24466 (2017).

\bibitem[Wittmer \emph{et~al.}(1998)Wittmer, Milchev and Cates]{Wittmer1998}
Wittmer, J., A.~Milchev and M.~Cates, \enquote{Dynamical {M}onte {C}arlo study
  of equilibrium polymers: Static properties,} J. Chem. Phys. \textbf{109},
  834--845 (1998).

\bibitem[Wittmer \emph{et~al.}(2000)Wittmer, van~der Schoot, Milchev and
  Barrat]{Wittmer2000}
Wittmer, J., P.~van~der Schoot, A.~Milchev and J.~Barrat, \enquote{Dynamical
  {M}onte {C}arlo study of equilibrium polymers. ii. {T}he role of rings,} J.
  Chem. Phys. \textbf{113}, 6992--7005 (2000).

\bibitem[Zhu \emph{et~al.}(2004)Zhu, Liao and Jiang]{Zhu2004}
Zhu, J., Y.~Liao and W.~Jiang, \enquote{Ring-shaped morphology of
  “crew-cut” aggregates from {ABA} amphiphilic triblock copolymer in a
  dilute solution,} Langmuir \textbf{20}, 3809--3812 (2004).

\end{thebibliography}
\bibliographystyle{JORnat}
\end{document}